\documentclass[bibyear]{aa}
\usepackage{times}
\usepackage{amsmath,mathrsfs,amssymb,amsbsy}
\usepackage{graphicx}
\usepackage{subfigure}
\usepackage{paralist}
\usepackage{color}
\usepackage{natbib}
\usepackage{aas_macros}
\usepackage[colorlinks=true,linkcolor=blue, citecolor=blue]{hyperref}%
\graphicspath{{plots/}}

\newcommand{\kms}{\,km\,s$^{-1}$} 

\usepackage{tablefootnote}
\usepackage{gensymb} 
\usepackage{soul} 

\newcommand{\teff}{T$_{\rm eff}$}
\newcommand{\logg}{$\log g$}
\newcommand{\meta}{\hbox{[M/H]}}

\newcommand{\feh}{{\rm {[Fe/H]}}}

\def\afe{\rm [\alpha/Fe]}

\def\kms{\,{\rm km~s^{-1}}}

\def\ltsima{$\; \buildrel < \over \sim \;$}
\def\simlt{\lower.5ex\hbox{\ltsima}}
\def\gtsima{$\; \buildrel > \over \sim \;$}
\def\simgt{\lower.5ex\hbox{\gtsima}}

\def\snr{{\rm S/N}}
\def\rsnr{\snr_{\rm resol}}
\def\fwhm{{\rm FWHM}}
\def\fmin{f_{\rm min}}

\def\ltsima{$\; \buildrel < \over \sim \;$}
\def\simlt{\lower.5ex\hbox{\ltsima}}
\def\gtsima{$\; \buildrel > \over \sim \;$}
\def\simgt{\lower.5ex\hbox{\gtsima}}

\providecommand{\orcit}[1]{\protect\href{https://orcid.org/#1}{\protect\includegraphics[width=8pt]{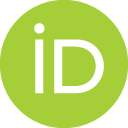}}}                 

\begin{document}

\title{Automatic line selection for abundance determination in  large stellar spectroscopic surveys\thanks{Tables with identified lines from 300 to 1000\,nm, and resolving powers of 3\,000, 6\,000, 20\,000, 40\,000 and 80\,000,  are only available in electronic form
at the CDS via anonymous ftp to \url{cdsarc.cds.unistra.fr} (\url{130.79.128.5})
or via \url{https://cdsarc.cds.unistra.fr/cgi-bin/qcat?J/A+A/}}}
\author{
	Georges~Kordopatis\orcit{0000-0002-9035-3920},\inst{\ref{oca}}\thanks{\email{georges.kordopatis@oca.eu} } 
 	Vanessa~Hill\orcit{0000-0002-3795-0005},\inst{\ref{oca}}
	Karin~Lind\orcit{0000-0002-8892-2573}\inst{\ref{stockholm}} 
	}
\institute{
	Universit\'e C\^ote d'Azur, Observatoire de la C\^ote d'Azur, CNRS, Laboratoire Lagrange, Nice, France\label{oca}
	\and
	Department of Astronomy, Stockholm University, AlbaNova University Centre, SE-106 91 Stockholm, Sweden\label{stockholm}
}
\abstract
{
Over the last  years, new multiplex spectrographs having observed or planning to observe several millions of stars have emerged. The optimisation of these instruments (regarding resolution or wavelength range), their associated surveys (choice of instrumental setup), or their parameterisation pipelines require  methods  that  estimate which wavelengths, or pixels, contain useful information. }
{We propose a method that establishes the usefulness of an atomic spectral line, where usefulness is  defined  by the purity of the line and its detectability. We show two applications of our code: {\it a)} optimising an instrument, by comparing  the number of detected useful lines at a given wavelength range and resolution, and {\it b)} optimising the line-list for a given setup, in the sense of creating a golden subsample, choosing the least blended lines detectable at different signal-to-noise ratios.  }
{The method compares pre-computed normalised synthetic stellar spectra containing all of the elements and molecules with spectra containing the lines of specific elements alone. Then,  the flux ratios between the full spectrum and the element spectrum are computed to estimate the line purities. The method identifies automatically {\it (i)} the line's central wavelength,   {\it (ii)} its detectability based on its depth and a given signal-to-noise threshold and {\it (iii)} its usefulness based on the purity ratio defined above.}
{ We apply this method to compare the three WEAVE high-resolution setups (Blue: $404-465$\,nm, Green: $473-545$\,nm, Red: $595-685$\,nm), and find that the Green+Red setup both 
allows one to measure more  elements  and contains more numerous useful lines. However,  there is a disparity in terms of which elements are detected over each of the setups, which we characterise. 
We also  study the performances of high-resolution ($R\sim20\,000$) and low-resolution ($R\sim6\,000$) spectra covering the entire optical wavelength range.  Assuming a purity threshold of 60 per cent, we find that the high-resolution setup contains a much wealthier selection of lines, for any of the considered elements, whereas the low-resolution has  a "loss" of 50 to 90 per cent of the lines (depending on the nucleosynthetic channel considered) even when the signal-to-noise is increased.
}
{The method presented provides a vital diagnostic of where to focus to get the most out of a spectrograph, and is easy to implement for future instruments that have not decided yet their final configuration, or for pipelines that require line masks.   
}


\titlerunning{Line selection for large surveys}
\authorrunning{G.~Kordopatis et al.}

\maketitle

\section{Introduction}
The relative abundance ratio of atomic elements measured from the stellar photospheres hold key information about multiple fields in modern astrophysics, ranging from galaxy formation \citep[e.g.,][]{Freeman02} to stellar nucleosynthesis \citep[][and references therein]{Burbidge57,Iwamoto99, Nomoto13, Karakas14}, especially if coupled with an estimation of the stellar age \citep[e.g.][]{Kordopatis23}. 
Specifically, by measuring the elemental abundance pattern of a star, it is possible to determine its birthplace and siblings, and/or the star formation history that preceded its formation. Yet, measuring the abundance of specific elements in a stellar spectrum is not straightforward \citep[e.g.][]{Jofre19}. Ultimately, inferring the amount of atoms of a species, present in the photosphere, depends on how easily a specific spectral line is detectable, measurable and transformable into an abundance. 
In other words, this task depends on  the one hand, on the accuracy of the stellar atmosphere and line profile modelling \citep[][]{Gray_spectroscopy} and on the other hand, on how accurately the spectral line can be measured (signal-to-noise ratio, $\snr$, resolution of the spectrum and blending together of several stellar features). The accuracy of the line modelling in turn depends on how accurately and precisely the stellar atmospheric parameters are known (namely, the effective temperature, \teff, surface gravity, \logg, global metallicity, \meta, and $\alpha$-elements enhancement, $\afe$).

As a consequence, the design phase of a spectroscopic survey is a tough negotiation between spectral resolution, exposure time, adopted wavelength range and total number of targets observed by the end of the project \citep[e.g.][]{Feltzing16}. For this purpose, it is important to be able to easily assess, early in the phase of the project, the information available at a given spectral range, resolving power ($R=\lambda/\Delta \lambda$) and signal-to-noise ($\snr$), for specific types of stars. This is intrinsically not trivial as it often requires to have a kind of stellar spectra parameterisation pipeline already available \citep[e.g.,][]{Caffau13,Bedell14, Hansen15}. Yet, such a pipeline often requires a tedious phase of training and/or optimisation \citep[e.g.][]{Recio-Blanco06, Kordopatis11a, Kordopatis13b, Ness15, Piskunov17}, which is therefore incompatible with the timescale or even the scope of the desired tests.
In this context, the recent years have seen the development of codes  that explore in a \emph{quick} way the available information in a spectrograph's configuration, in order to provide answers to the previously raised questions  \citep[e.g.][]{Ruchti16, Ting17,Sandford20}. 

The \emph{Spectral Wavelength Optimization Code} \citep[\emph{SWOC},][]{Ruchti16} requires the user to provide a \emph{predefined}  table containing the central wavelength and the equivalent width (or line-depth) of features that are considered  of particular interest. SWOC then evaluates the quality and the wavelength  distribution of  these features for a considered stellar-type, determines the optimal wavelength coverage based on a defined Figure-of-Merit, and eventually combines this information for different stellar types to ascertain the optimal wavelength coverage for a survey.
This approach therefore relies on already having \emph{a priori} information regarding which lines are of interest. This is not  always the case, especially in wavelength regions that have not yet been commonly used in large surveys in the past.

A different approach has been adopted by \citet[][see also \citealt{Sandford20}]{Ting17}, that employ the Cram\'er-Rao bound metric to quantify the amount of information available in a spectrum of specific wavelength range and resolution, associated with a given label (in this case, elemental abundance). Being based on the so-called gradient spectra, i.e. the variation of the spectrum at a given wavelength associated to a specific label, as well as on the covariance matrix of the spectrum,  the metric sums over the different wavelength pixels, to inform the user about which elements can be detected above a given significance threshold. 
\citet{Ting17} conclude that, given a fixed exposure time and number of pixels (therefore different $\snr$ and wavelength-ranges depending on resolution), low-resolution spectra could provide an equivalent amount of information to high-resolution spectra. 
Yet, this conclusion has been obtained  assuming that resolution and $\snr$ are uniform across the wavelength range and that line-blends are correctly known (and modelled), which is frequently not the case.

The caveats  mentioned in the previous paragraphs motivated the development of a new code, that we present in this paper. Its purpose is to identify  ``useful" lines in a synthetic spectrum, i.e. lines that are visible and not heavily blended at a given spectral resolution and  $\snr$, without any \emph{a priori} knowledge. This information is then stored and can  be used to create either a line-list selection for spectral analysis (e.g. for abundance determination), or to visualise how many lines of a specific element or a nucleosynthetic channel are useful for a given instrumental configuration. 
It therefore has immediate and valuable applications for spectroscopic surveys based on already existing or future spectroscopic facilities or instruments such as
APOGEE \citep{Majewski17}, DESI \citep{Abareshi22},  Gaia-RVS \citep{Gaia, Cropper18}, GALAH \citep{DeSilva15}, LAMOST \citep{Deng12}, 
 4MOST \citep{deJong19},  WEAVE \citep{Jin23}, MOONS \citep{Cirasuolo20}, MSE \citep[][]{MSE19}, PFS \citep{Takada14}, etc.

The paper is structured as follows. In Sect.\,\ref{sect:algorithm} we present the concept of the code: how it runs, which are the required inputs, and which are the  outputs. The synthetic spectral library on which the code  relies on is described in Sect.\,\ref{sec:grid_spectra}. The code is then applied in Sect.\,\ref{sec:Applications} on a handful of examples. In Sect.\,\ref{subsect:GES} a verification of the identified lines based on the line-list established within the Gaia-ESO survey \citep{Randich22, Gilmore22} is performed, and in Sect.\,\ref{sec:Instrument_design} we show an illustration of how an instrument's design can be optimised, by comparing the performances of high- and low-resolution spectrographs for specific types of stars. In Sect.\,\ref{sec:weave_HR_application}, we evaluate the performances of the two WEAVE high-resolution configurations to suggest the setup that best drives Galactic archaeology science. In Sect.\,\ref{sec:linelist_optimisation} we show how our code can be used to create a `golden' line-list for spectral synthesis codes. Finally, Sect.\,\ref{sec:conclusion} concludes.

\section{Description of the code}
\label{sect:algorithm}

\subsection{Description of the algorithm}
\label{sect:algorithm_description}
Let $S_{f,\theta}(\lambda)$  be the normalised synthetic spectrum of a star at a given set of atmospheric parameters $\theta=$\{\teff, \logg, \meta, $\afe$\}. This spectrum is computed at an instrumental resolving power $R=\lambda/\fwhm_{\rm inst}$ (where $\fwhm_{\rm inst}$ is  the full-width at half maximum of the line spread function of the instrument) with a sampling $dx$.  $S_{f,\theta}$ contains the lines and blends of all of the elements and molecules present at the photosphere of the star.  

\begin{figure}[t!]
\begin{center}
\includegraphics[width=0.9\linewidth, angle=0]{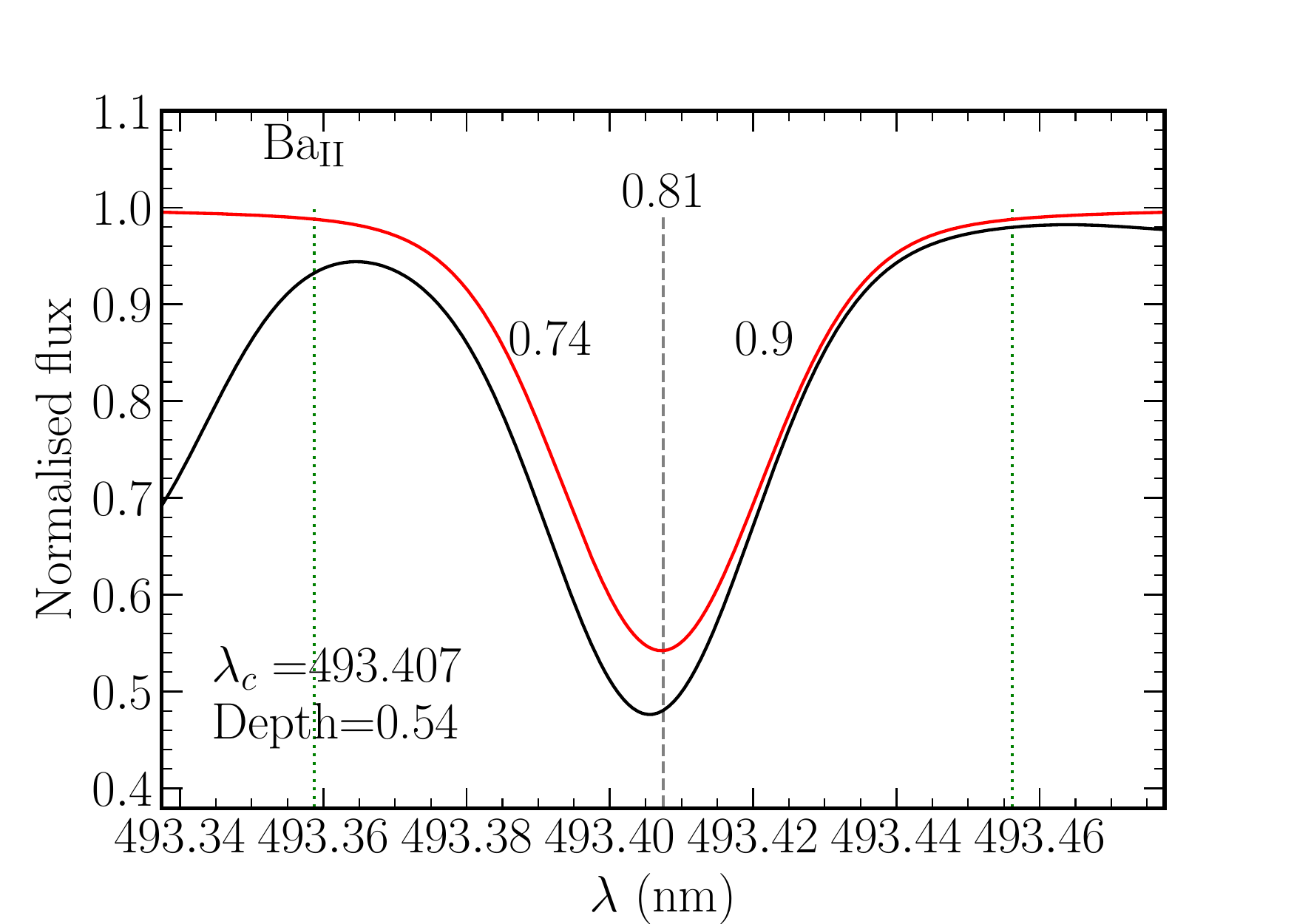}\\
\includegraphics[width=0.9\linewidth, angle=0]{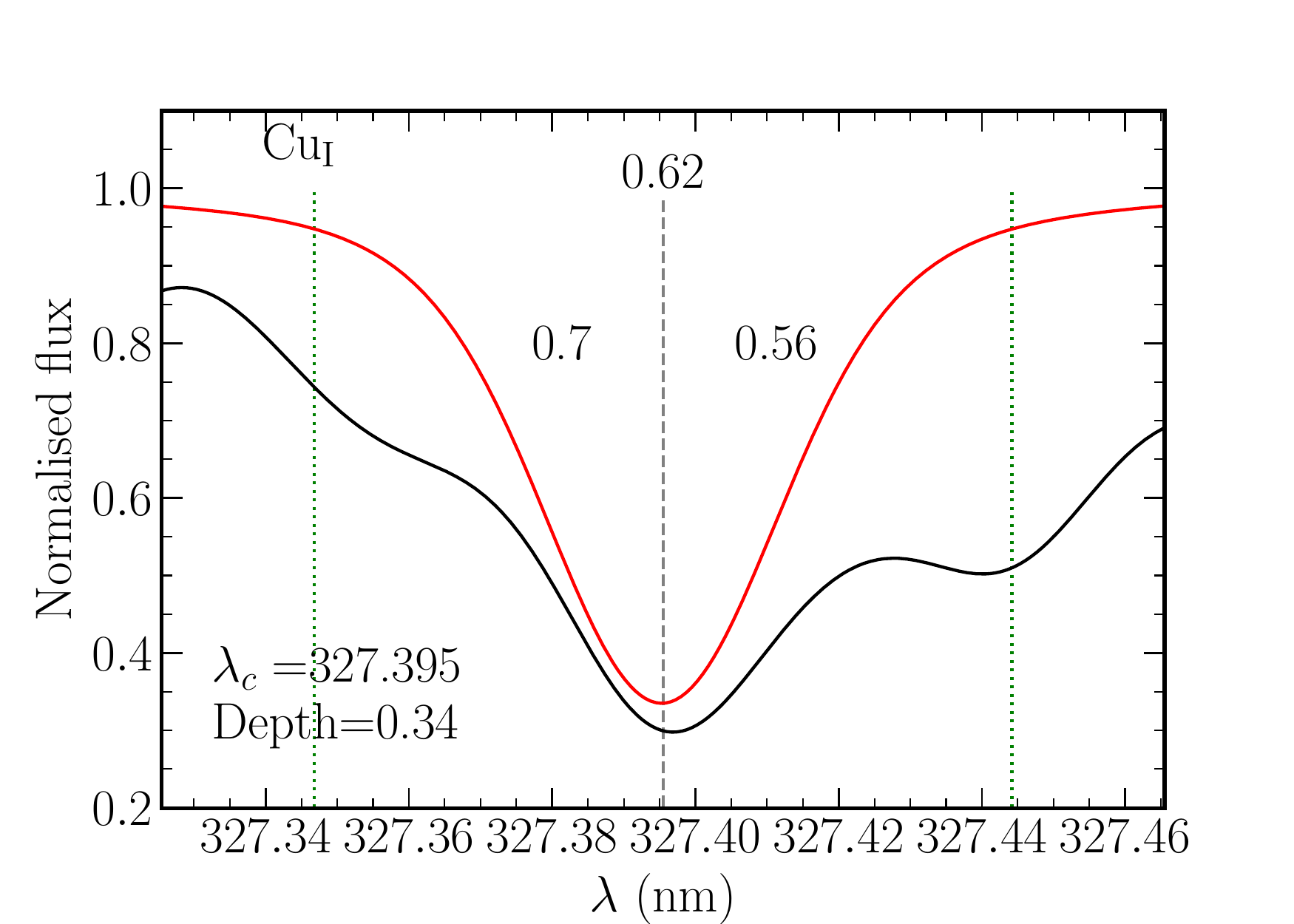}\\
\includegraphics[width=0.9\linewidth, angle=0]{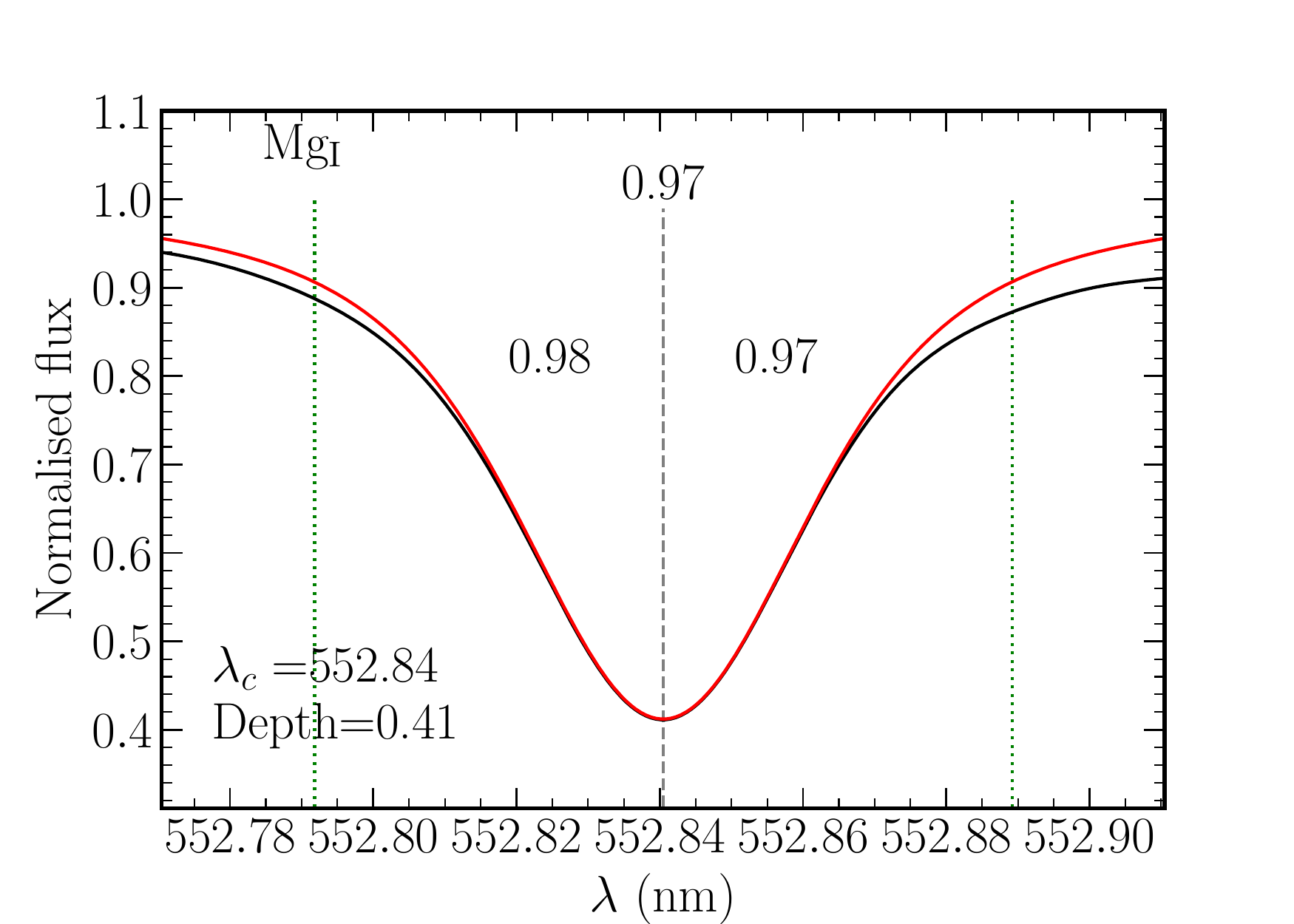}
\caption{Examples of identified lines for a Solar-like spectrum at $R=20\,000$.  The elemental spectrum, i.e. the flux computed with the contribution from atomic lines from only one ionization stage of one element,  is plotted in red and the full spectrum, containing all of the elements and molecules,  in black. The element and ionization stage associated to the line are noted on the upper left corner of each plot. The central wavelength of the identified line, $\lambda_c$, is plotted as a vertical dashed grey line.  The blue-end and the red-end of the line are plotted as vertical dashed green lines. The depth of the line in the element spectrum is indicated at the bottom left corner of each plot. The purity factor for the entire line is written at the middle-top of the plots. The blue-wing and red-wing purities are enclosed within the line at its left and right, respectively.  }
\label{fig:Line_fitting}
\end{center}
\end{figure}

Similarly, let $S_{E,\theta}(\lambda)$ be the normalised stellar spectrum containing only the lines associated to the element $E$, at the same $\theta$ parameters,  resolving power $R$, pixel sampling $dx$ and same continuous opacities as in $S_{f,\theta}(\lambda)$.  Each element $E$ has a reference linelist\footnote{Here, retrieved from
the Vienna Atomic Line Database (VALD), \url{http://vald.astro.uu.se/}} $\{ \lambda_{V,E} \}$ associated to it \citep[][]{Piskunov95,Ryabchikova15}  which is used for the computation of both $S_{f,\theta}(\lambda)$ and $S_{E,\theta}(\lambda)$. For convenience, in what follows we will omit the $\theta$ 
subscript, when implicit. The steps of our algorithm to identify the lines, for a given element $E$, are the following: 

\begin{enumerate}
\item 
We detect all of the lines in $S_E(\lambda)$ by identifying their cores  blindly. To achieve this, we  search for the zero crossings in the derivative of $S_E(\lambda)$, without imposing any threshold in the flux (see, however, below). 
  Let $\{ \lambda_i \}$ be the list containing the wavelengths of all the identified line-cores $i$ of the element $E$. The number of lines in $\{ \lambda_i \}$ is smaller or equal to the number of VALD entries in $\{ \lambda_{V} \}$.

\item
We identify the true central wavelength $\{\lambda_c\}$ of each line in $\{\lambda_i \}$, by cross-matching $\{\lambda_i \}$ with $\{ \lambda_{V} \}$. 
Often, several VALD lines fall within one $dx$ from the considered $ \lambda_i$. 
In this case, we use the Boltzmann equation to evaluate which $\lambda_{V}$  is the most prominent in the considered subset.
In practice, we choose the line that is expected to be the strongest, ranking all candidate lines according to excitation energy and oscillator strength in the following way: 
\begin{equation}
\log(A)\propto -(E_\chi/k T_{\rm eff})+\log(gf),
\end{equation}
where $A$ is the number of atoms,  $E_\chi$ is the excitation potential of the line, $\log(gf)$ is the logarithm of the oscillator-strength times the statistical weight of the parent energy level, and $k$ is the Boltzmann constant\footnote{We note that by doing so, we assume that all of the lines are at the same ionisation level which is not necessarily the case, unless $S_E(\lambda)$ is computed only for a given ionisation level.}.

\item
For each item in  $\{\lambda_c\}$, we evaluate  $S_{E}(\lambda_c)$ and keep the lines that are deep enough to be detected at a given $\snr$. This criterion, derived in Appendix~\ref{sec:Appendix_Careyl}, is defined as:
\begin{equation}
S_{E}(\lambda_c)\leq 1- 3\cdot \frac{1.5}{\snr_{\rm resol}}
\label{eq:min_depth_snr}
\end{equation}
where $\snr_{\rm resol}$ is the signal-to-noise ratio per resolution element (see Appendix~\ref{sec:Appendix_Careyl} for the formula with $\snr$ per pixel). We note that the criterion is applied to the elemental spectrum $S_{E}(\lambda)$ rather than to the observed total spectrum $S_{f}(\lambda)$. The reason for this choice is that we want to impose a criterion on the detectability of the line independently of its blend (or purity, see Eq.~\ref{eq:purity}, below). 

\item
For each $\lambda_c$, we identify the blue-end, $\lambda_b$, and red-end, $\lambda_r$, of the line, defined as the first wavelengths blue-wards and red-wards where :
\begin{equation}
S_E(\lambda) \geq 1- x \cdot S_E(\lambda_c),
\end{equation}
i.e. the wavelengths at which the flux has reached $x$ per cent of the value it had at its core. We limit the search for $\lambda_b$ and $\lambda_r$ to $\lambda_c \pm 1.5\cdot \fwhm_{\rm inst.}$. 
A value of $x=0.02$  (i.e. 2 per cent) has been  empirically adopted. 

 \item
 We define the purity factor $p$ as: 
\begin{equation}
p    =\frac{\sum_{\lambda_{\rm min}}^{\lambda_{\rm max}} 1- S_{e}(\lambda)}{\sum_{\lambda_{\rm min}}^{\lambda_{\rm max}} 1- S_{f}(\lambda)}, 
\label{eq:purity}
\end{equation}
and then compute the purity for the entire line ($p_t$), the blue-half ($p_b$) and the red-half ($p_r$). This is done because the line can be blended differently on its blue or red wing (see, for example, Fig.\,\ref{fig:Line_fitting}) and a line that is free of blend in one of its wings may still be very useful and reliable for abundance determinations  (see how blending can affect the equivalent width measurements in Appendix~\ref{sec:Appendix_blend}). The following ranges, $\lambda_{\rm min}$ and $\lambda_{\rm max}$,  are therefore adopted in Eq.~\ref{eq:purity}:
 
 \[
  (\lambda_{\rm min}, \lambda_{\rm max})=\begin{cases}
              (\lambda_b, \lambda_r)\text{~for~} p_t \\
              (\lambda_b, \lambda_c)\text{~for~}p_b \\
              (\lambda_c, \lambda_r) \text{~for~}p_r
            \end{cases}
\]

 \item 
  Finally,  we evaluate the number of pixels in the blue $N_b$, and in the red $N_r$, that are close to the continuum ($S_{f}(\lambda)>0.9$) within the adopted $ (\lambda_{\rm min}, \lambda_{\rm max})$. This allows us, eventually, to flag the lines that have a purity above a determined arbitrary value but that can  nevertheless be difficult to detect because they are in the wings of stronger lines further away from $\lambda_{\rm min}$ or $\lambda_{\rm max}$.
\end{enumerate}

\subsection{Input/output of the algorithm}
In order to run, the code requires as an input ($i$) a synthetic spectrum that includes all the elements, ($ii$) a set of synthetic spectra with the atomic lines of only one element\footnote{Several ionisation levels can be present in the element spectrum in case the user does not wish to differentiate between them. } each time, computed at the same wavelength range, resolving power and atmospheric parameters as the full spectrum, ($iii$) the reference line-list for each element that has been used to compute the spectra, ($iv$) an arbitrary $\snr_{\rm resol}$ threshold and finally ($v$) the spectral resolving power of the instrument. The latter two parameters are used to evaluate the detectability of a line at a given $\snr$; the spectral  resolving power is applied by convolving the simulated spectra (provided with infinite resolution) with a Gaussian of appropriate FWHM.

The code delivers, for a given element $E$, a table containing the central wavelengths of the lines $\lambda_c$, the blue-ends $\lambda_b$, the red-ends $\lambda_r$, the three purity factors ($p_b$, $p_r$, $p_t$), the depth of the line  in the full spectrum $S_f(\lambda_c)$, the depth of the line  in the element spectrum $S_E(\lambda_c)$, the number of pixels in the blue $N_b$ and the red  $N_r$ that have a flux close to the continuum. This information can later be used as desired to make summary/diagnostic plots or in order to select ``clean'' lines for codes that require such an input.

Figure\,\ref{fig:Line_fitting} shows three cherry-picked  examples of line identifications with our code for a Solar-like spectrum at $R=20\,000$. The total, blue and red purity factors are encapsulated in the figure, together with the central wavelength $\lambda_c$ and the depth of the line for the element spectrum alone.

\section{Grid of synthetic spectra of infinite resolution}
\label{sec:grid_spectra}
 \begin{figure}
\begin{center}
\includegraphics[width=\linewidth, angle=0]{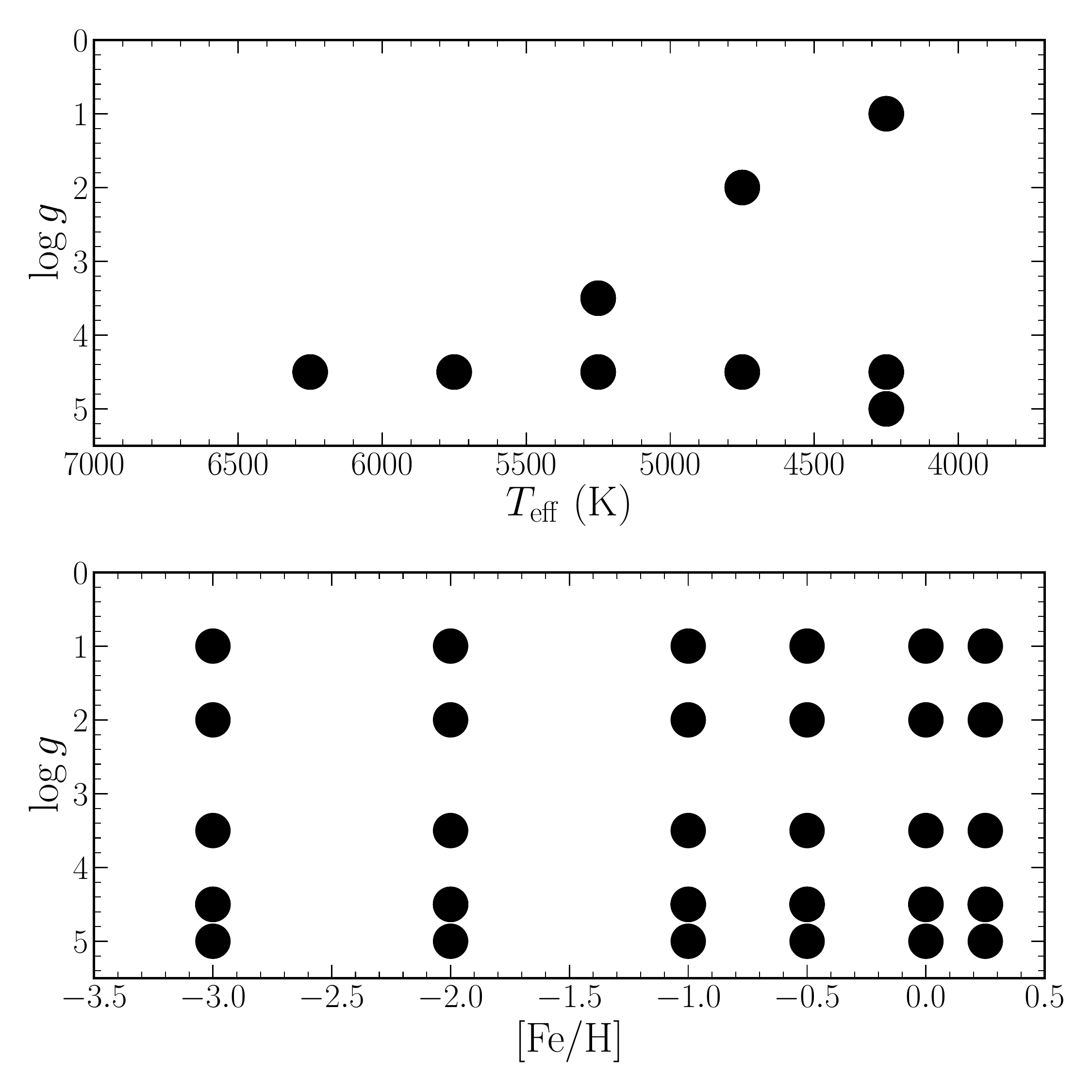}
\caption{Set of atmospheric parameters for which the full and elemental synthetic spectra have been computed.}
\label{fig:input_parameter_spectra}
\end{center}
\end{figure}

We consider  nine stellar types, at different combinations of \teff~and \logg, and six different values of \meta~(see Fig.\,\ref{fig:input_parameter_spectra}), resulting to 54 different templates.

The spectra are computed using PySME v4.10\footnote{\url{https://pypi.org/project/pysme-astro/}} \citep{Wehrhahn22}  and the SME library v5.22\footnote{\url{https://www.stsci.edu/~valenti/sme.html}}   \citep{Valenti96, Piskunov17} together with the 1-dimensional MARCS model atmospheres \citep{Gustafsson08}, assuming  local thermodynamic and hydrostatic equilibrium. The considered total wavelength range is $\lambda=[300-1\,000]$\,nm. The sampling is constant at  $8\cdot10^{-5}$\,nm.  Adopted line-list is from VALD3 database (downloaded in January 2021). The molecular line-list includes CH, CN, C2, TiO, MgH, SiH, CO, and OH. 
 The elemental abundance ratios are the same as for the MARCS model atmospheres i.e. solar scaled with \citet{Grevesse07}  except for Lithium ($A(\mathrm{Li})=2.00$ adopted for all of the stars) and for $\alpha$-elements, for which the abundance is a function of metallicity, as follows: 
 \begin{eqnarray}
 \afe&=&+0.4 \hspace{1cm}\mathrm{for~ \feh\leq-1.0}\\
 \afe&=&-0.4\cdot\feh \hspace{0.5cm}\mathrm{for~ -1\leq \feh< 0.0} \\
 \afe&=&0.0\hspace{1.1cm}\mathrm{for~ \feh\geq0.0}.
 \end{eqnarray}

 We note that the adopted elemental abundances do not necessarily reflect what exists in nature, and that in practice lines could be more easily (in case of over-abundance) or more difficultly detected (in case of under-abundance). 
 
The resolution of the computed spectra is infinite, in the sense that no macro-turbulence, rotational broadening or instrumental broadening have been applied.  
To obtain the spectrum as obtained from a specific instrument\footnote{Assuming that no rapid-rotators are to be observed and that macro-turbulence does not dominate the line-profile.}, one therefore simply needs to convolve the initial spectrum with a Gaussian kernel whose FWHM corresponds to the resolving power of  the considered spectrograph, then crop at the wavelengths the spectrograph observes (see for example Table~\ref{tab:instruments}).

 \begin{table}
\caption{Non-exhaustive list  of spectrographs used for galactic archaeology covering the optical wavelengths at different resolving powers.}
\label{tab:instruments}
\footnotesize
\begin{center}
\begin{tabular}{lcc}
\hline \hline
 Spectrograph-setup & $\lambda$ (nm) & $R$\tablefootmark{(a)} \\ \hline
WEAVE-LR\tablefootmark{(b)} & {\tiny [366;959]} &  $5\,000$  \\
WEAVE-HR (B+R)\tablefootmark{(b)} & {\tiny [404;465] + [595;685]}&   $20\,000$   \\
WEAVE-HR (G+R)\tablefootmark{(b)} & {\tiny [473;545] + [595;685]} &  $20\,000$ \\
4MOST-LR\tablefootmark{(c)} & {\tiny  [370;950]}  &  $6\,500$ \\
4MOST-HR\tablefootmark{(c)} & {\tiny [392.6;435.5] + [516;573] + [610;679]} &  $20\,000$ \\
Gaia-RVS & {\tiny [846;870]}\tablefootmark{(d)} & $11\,500$ \\ 
DESI & {\tiny [360;980]} & $3\,500$ \\ 
HERMES\tablefootmark{(e)} & {\tiny [471.8-490.3] + [564.9-587.3] }& $28\,000$ \\ 
 & {\tiny + [648.1;673.9] +[759.0;789.0]}&  \\ 
 LAMOST-LR\tablefootmark{(f)} & {\tiny [370-900]} &1800 \\

\hline
 \end{tabular}
 \end{center}
 \tablefoot{
 \tablefoottext{a}{Mean resolving power across the considered wavelength range.   }
  \tablefoottext{b}{\citet{Jin23}. }
    \tablefoottext{c}{\citet{deJong19}. }
    \tablefoottext{d}{Wavelengths in the vacuum. }
    \tablefoottext{e}{\citet{Sheinis15}.}
    \tablefoottext{f}{\citet{Zhao12}.}

 }
 \end{table}

The different elements for which individual spectra have been computed  are the following: 
\begin{itemize}
\item
Even-Z elements: 
C, O, Mg, Si, S, Ca, Ti. 

\item Odd-Z elements: 
Li, N, Na, Al, P, K, Sc.

\item
Iron-peak elements: 
V, Cr, Mn, Fe, Co, Ni, Cu, Zn.

\item
Neutron-capture elements $1^{\rm st}$ peak: 
Rb, Sr, Y, Zr, Mo.

\item
Neutron-capture elements $2^{\rm nd}$ peak: 
Ba, La, Ce, Pr, Nd, Sm, Eu.

\end{itemize}

We note that we treat neutral and ionised species separately. Furthermore, whereas molecules are included in the full spectra, molecular lines associated with a given element were not considered for detectability or usefulness.

The VALD line-list used to identify the lines contains 621\,357 unique entries.  It is a merged version coming from two ``extract stellar" requests from the VALD3 database,  for a solar-metallicity giant (\teff$=3800$\,K, $\log g=1.0$) and one solar-metallicity dwarf (\teff$=7000$\,K, $\log g=4.0$), which included hyperfine splitting, a depth detection threshold  set to 0.001 and a micro-turbulence to $1.5\kms$. 


\section{Applications}
\label{sec:Applications}
Below, we show a validation of our code using the Gaia-ESO survey line-list (Sect.\,\ref{subsect:GES}), as well as three different applications/illustrations of it. Section~\ref{sec:Instrument_design} investigates the purity of the lines for different instrument setups (different  resolving powers but similar wavelength range), while Sect.\,\ref{sec:weave_HR_application} compares how two different setups of similar resolving power compare when probing different wavelength regions. Section~\ref{sec:linelist_optimisation} shows how to select a golden sublist of most useful lines, based on the output of our code. 

\subsection{Validation through comparison with the Gaia-ESO line-list}
\label{subsect:GES}
The Gaia-ESO public spectroscopic survey \citep[GES,][]{Randich22, Gilmore22} observed from 2011 to 2018  approximately $10^5$ Milky Way stars using the high-resolution spectrographs UVES ($R\sim47\,000$) and GIRAFFE ($R\sim20\,000$), covering mostly the wavelength regions [480-680] and [850-900]\,nm. The consortium analysed the spectra using more than five different pipelines \citep{Smiljanic14}, based on a variety of methods, ranging from spectral synthesis to equivalent-width measurement, and from model-driven to data-driven parameterisation. In this process, a particular effort has been put into homogeneously selecting lines that were suitable for spectral analysis, both in terms of blending and in terms of reliability of atomic parameters. This effort has been published in \citet{Heiter21}, where the authors provide  blending quality flags (with the keyword {\tt synflag}) based on the visual inspection of  high-resolution spectra ($R\sim 47\,000$) of the Sun and Arcturus. These lines are labelled `Y', for not blended or blended with a line from the same specie for either star, `N' for blended for both  stars, and `U'  for blended for at least one of the stars.

To evaluate the performance of our code, we compared our results for  $R\sim20\,000$ spectra with the ones of GES, 
selecting only the lines that have the {\tt synflag}=`Y'. For that reason, we selected synthetic spectra amongst our templates, with Solar-like and Arcturus-like  parameters  (\teff$=5750$\,K, \logg$=4.5$, \feh=0 and \teff$=4250$\,K, \logg$=1.0$, \feh$=-0.5$, respectively) and ran our code on these with a  signal-to-noise threshold equal to 500 per resolution element and minimum purity equal to 0.2 in order to retrieve as many lines as possible.

\begin{figure}
\begin{center}
\includegraphics[width=\linewidth, angle=0]{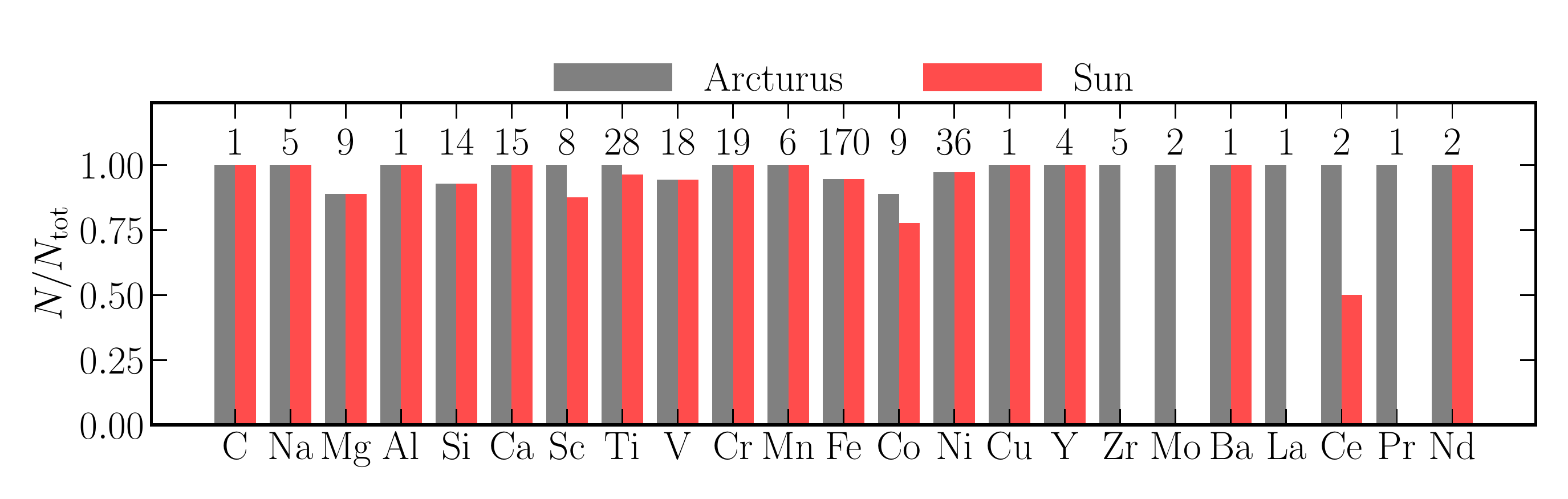}
\caption{ Recovered atomic lines from our code, per element, relative to the list of the Gaia-ESO survey having {\tt synflag}=`Y'. The recovered lines for Arcturus and the Sun are plotted in grey, and red, respectively. We report, at the top of each bar, the number of lines identified in the Gaia-ESO survey for that element. }
\label{fig:Hist_Sun_Arcturus_GES_Lines}
\end{center}
\end{figure}

 Among the 358 lines that GES has identified as reliable\footnote{We do not consider, for this work, the hydrogen lines and we keep only one line per element if within 0.01\,nm from the others.}, we recover 331 of them for the Sun and 344 for Arcturus, i.e. 92.5 per cent and 96 per cent, respectively\footnote{The crossmatch has been performed by rounding the wavelength to 0.01\,nm. }. Figure\,\ref{fig:Hist_Sun_Arcturus_GES_Lines} shows the ratio of recovered lines over the ones available from GES, per element.  For Arcturus, we recover at least a portion of lines for all of the considered elements of \citet{Heiter21}. This is not the case for the Sun, where our code selects none for La, Mo, Pr or Zr,  (\citealt{Heiter21} list contains 1, 2, 1 and 5, respectively).  A deeper investigation of the lines for these elements suggested that our code fails at selecting them because in our synthetic spectra they are too weak (possible disagreement between the model and reality), or too blended. We note, however, that \citet{Heiter21} selection is done on a resolving power which is twice higher than the one considered here and not necessarily in a uniform way for all of the elements, i.e. a {\tt synflag}=`Y' could be assigned to the best line of an element, even if it is rather blended.

\begin{figure*}
\begin{center}
\includegraphics[width=\linewidth, angle=0]{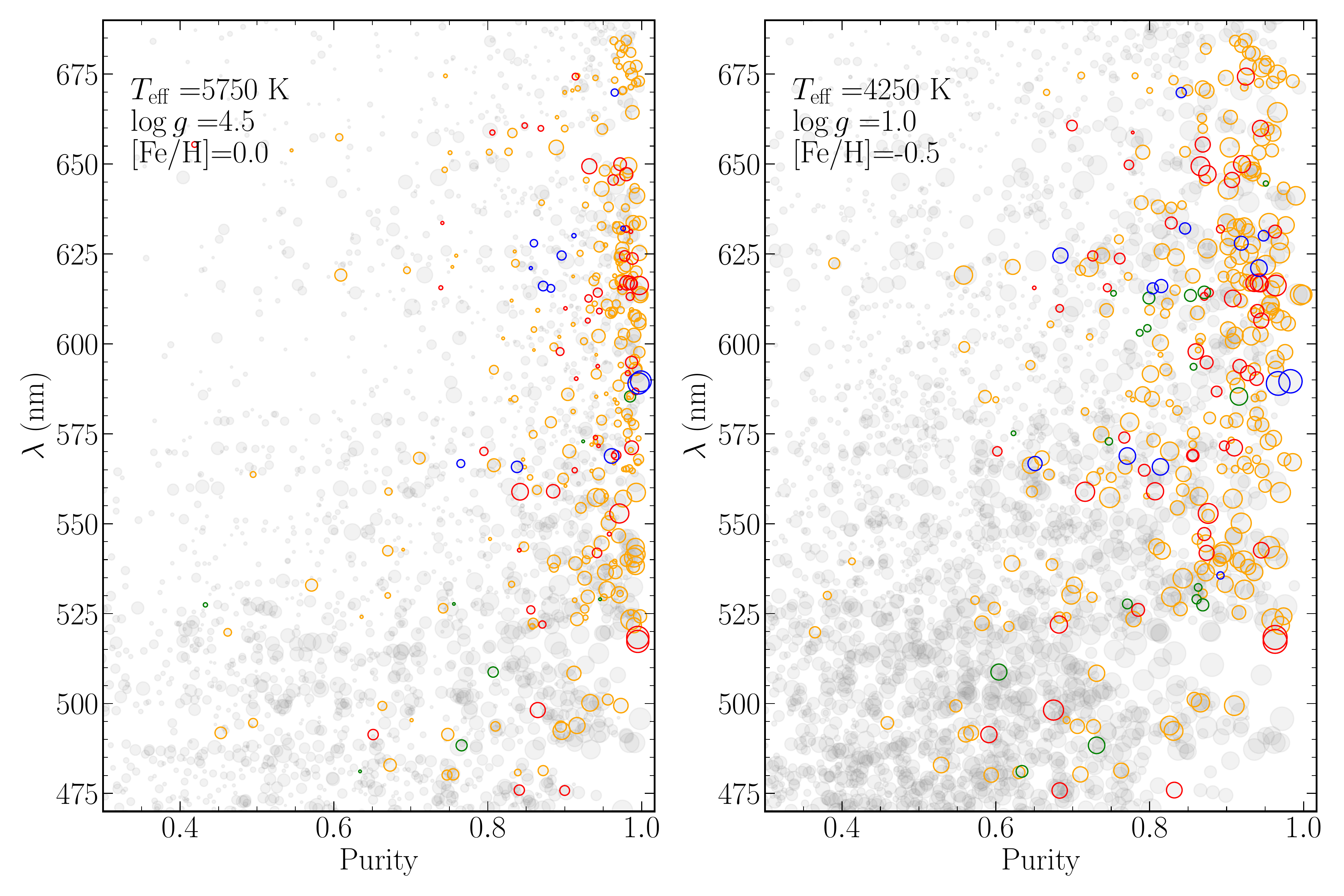}
\caption{ Wavelength versus purity of the atomic lines selected over an arbitrary wavelength range for a Sun-like (left) and an Arcturus-like (right) spectrum at $R=20\,000$ (grey filled circles). The size of the points is  proportional to the strength of the line, i.e. $\sim 1-\mathrm{core_{flux}}$. The subsample of the grey points that are selected as reliable lines for spectral synthesis from \citet{Heiter21} for the Gaia-ESO survey ({\tt synflag}=`Y') are highlighted in colour. Orange circles are associated to iron-peak lines, red  to even-Z elements, blue to odd-Z elements and green to neutron-capture elements. One can see that the GES has selected lines that have on average a purity greater than 0.8, with overall a larger purity for the Sun than for Arcturus.}
\label{fig:Sun_Arcturus_GES_Lines}
\end{center}
\end{figure*}

Figure\,\ref{fig:Sun_Arcturus_GES_Lines} shows the purity of the lines as a function of wavelength, focusing, arbitrarily, on the range [470-690]\,nm. In grey are represented all of the lines we have identified for the Sun or Arcturus, with a purity greater than 0.3 and detectable with a $\rsnr$~less than 500. The lines selected by GES with {\tt synflag}=`Y' that exist in our selection  are circled in coloured solid lines (orange for Fe-peak lines, red for even-Z elements, green for neutron-capture elements and blue for odd-Z elements).

Figure\,\ref{fig:Sun_Arcturus_GES_Lines} illustrates, in a rather unsurprising way, that the lines that are pure for the Sun, are not necessarily of the same purity for Arcturus and vice-versa. Our code, therefore, provides the advantage to visualise immediately the purity of a set of lines for a given set of atmospheric parameters. Furthermore, Fig.\ref{fig:Sun_Arcturus_GES_Lines} validates our code: the lines selected by \citet{Heiter21} are found to be mostly of high purity (mostly above 0.7 for both stars). Finally, the plot indicates that the GES selection is rather conservative and privileging purity for the Solar spectrum. That said, the purity for Arcturus remains rather high, with the majority of the lines having a value greater than 0.8 (as opposed to higher than 0.95 for the Sun). It is beyond the purpose of this paper to discuss the validity and limitation of the GES selection.

\subsection{Instrument design and optimisation}
\label{sec:Instrument_design}
In this section we investigate how lines, selected in a similar way as  in the previous section, compare for a high-resolution ($R\sim20\,000$) and a low-resolution  ($R\sim6\,000$) setup. We  take once again the case of the Sun and Arcturus, with the parameters defined in the previous section, as illustrative of a metal-rich turn-off star and a metal-poor giant.
\begin{figure*}
\begin{center}
\includegraphics[width=1.05\linewidth, angle=0]{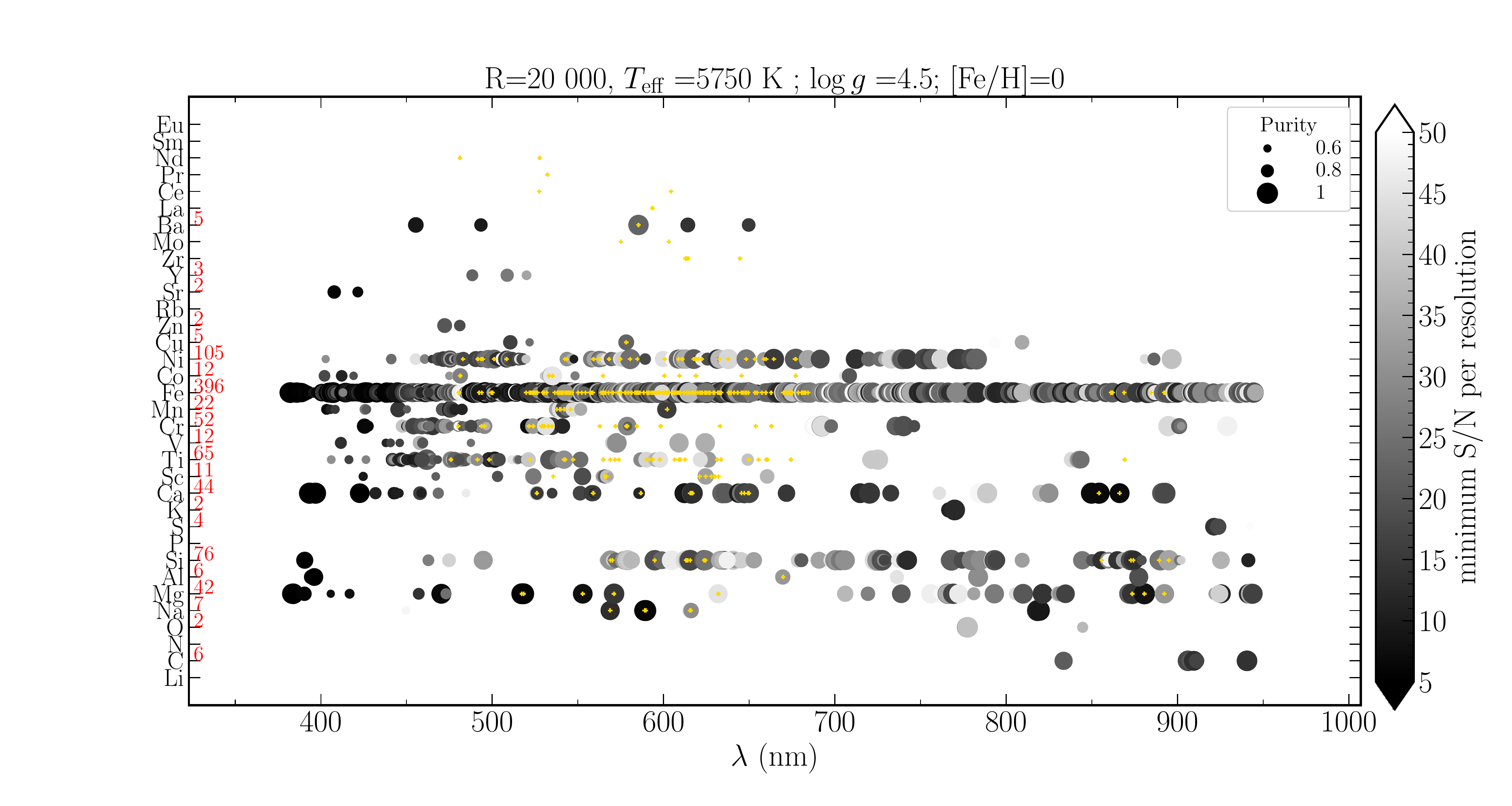}\\
\includegraphics[width=1.05\linewidth, angle=0]{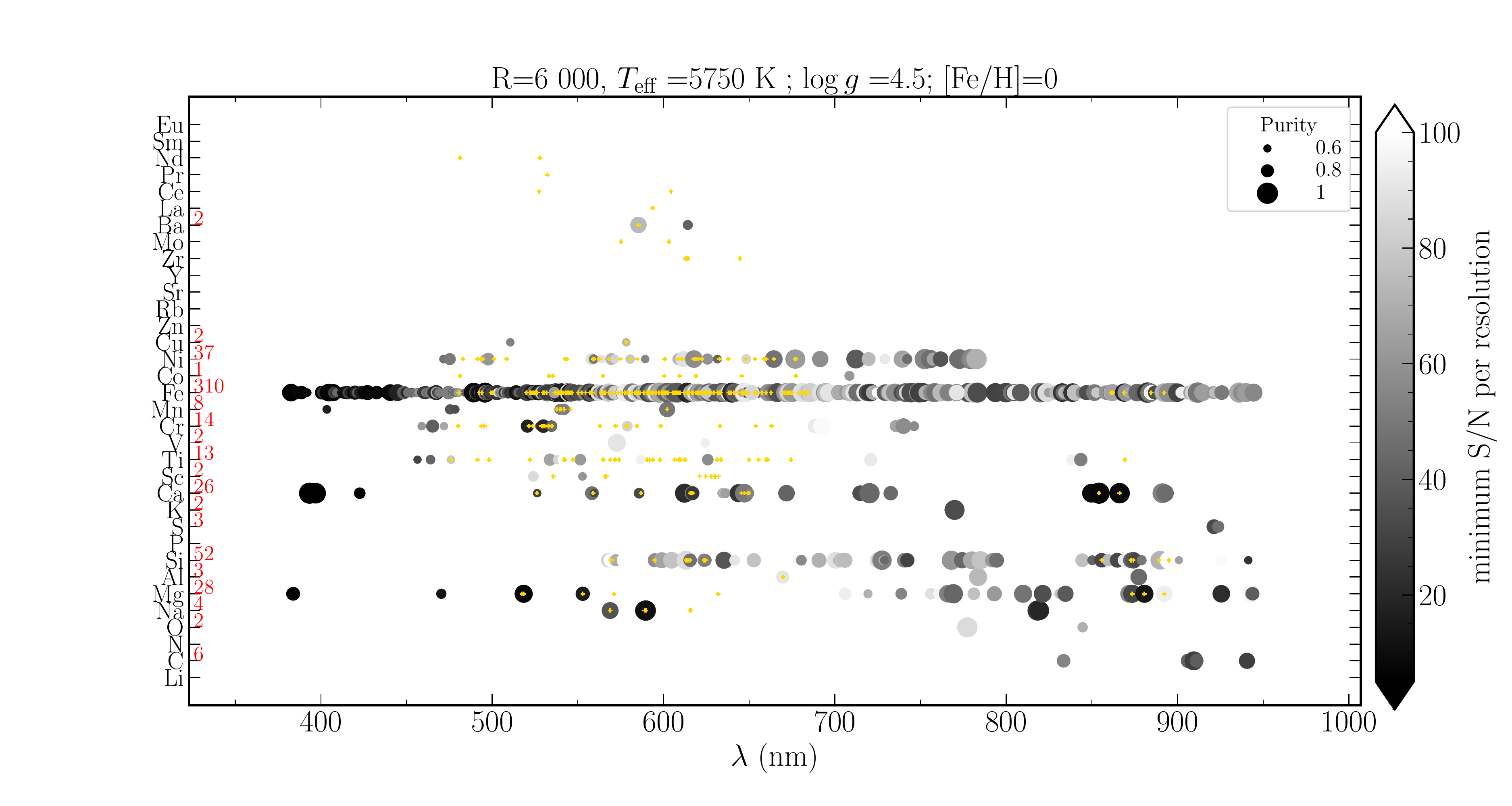}
\caption{  Lines detected for a HR  ($R=20\,000$, top) and a LR ($R=6\,000$, bottom) setup, for a Solar-like star. The names of the elements are indicated on the left-hand side of the plots. The number of identified lines is written in red, next to each element. The points are located at the wavelength where a line is detected. The area  of the circles is proportional to the purity of the line, and their colour to the minimum $\rsnr$ required to detect the line. A simple way to read this plot is the following: if a point is difficult to visualise (small size and white), then the spectral line is difficult to detect and use. The minimum purity plotted is 0.6. Similarly, lines that require a $\rsnr>50$ in HR or $\rsnr>100$ in LR are excluded. Indicatively, yellow '+' symbols are located to the wavelengths at which the Gaia-ESO survey has identified, for $R=40\,000$, lines that are reliable and pure for spectral synthesis \citep[{\tt synflag}=`Y', see][ and Sect.\,\ref{subsect:GES}]{Heiter21}.}
\label{fig:Sun_Lines}
\end{center}
\end{figure*}

\begin{figure*}
\begin{center}
\includegraphics[width=1.05\linewidth, angle=0]{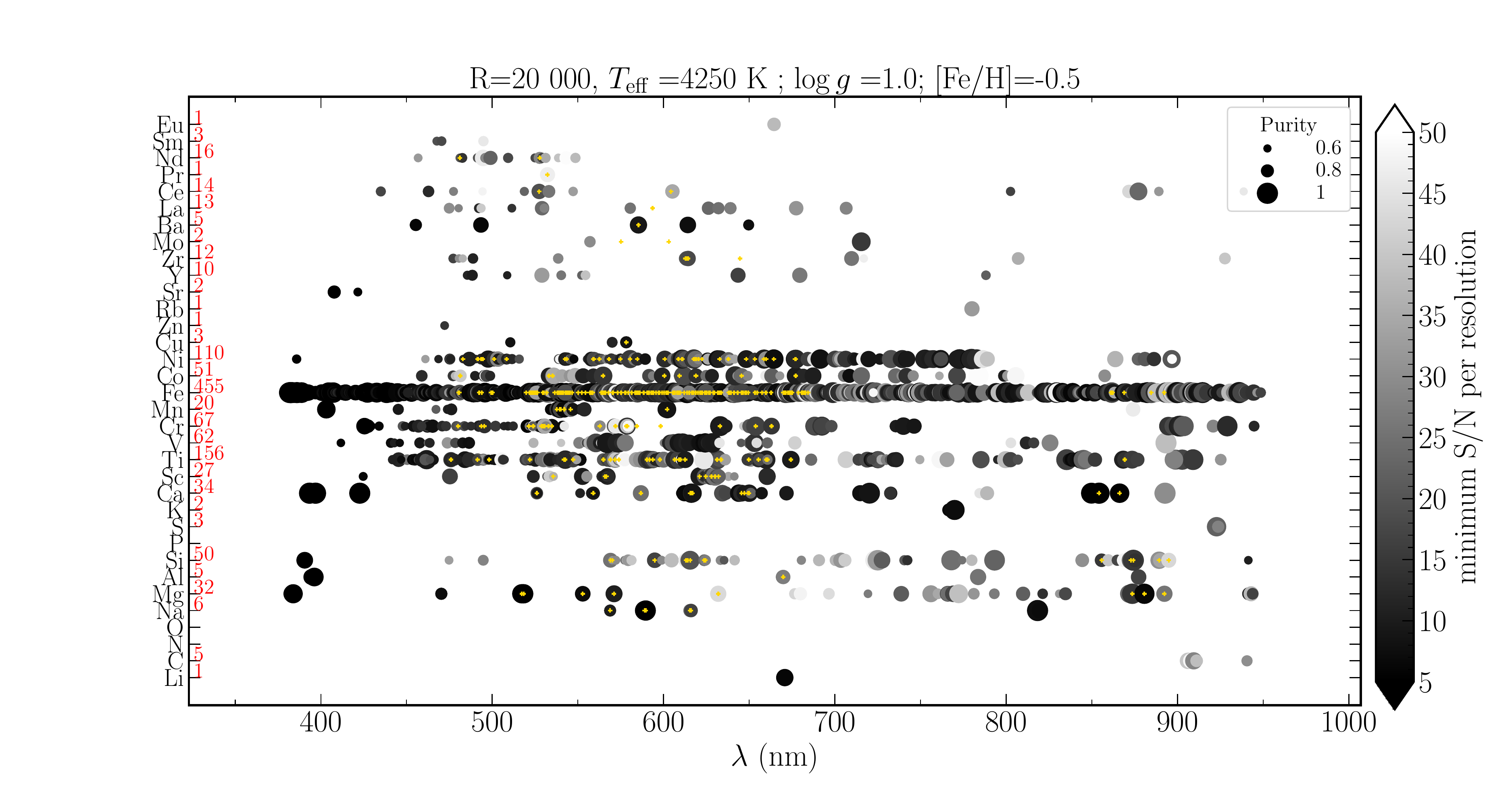}\\
\includegraphics[width=1.05\linewidth, angle=0]{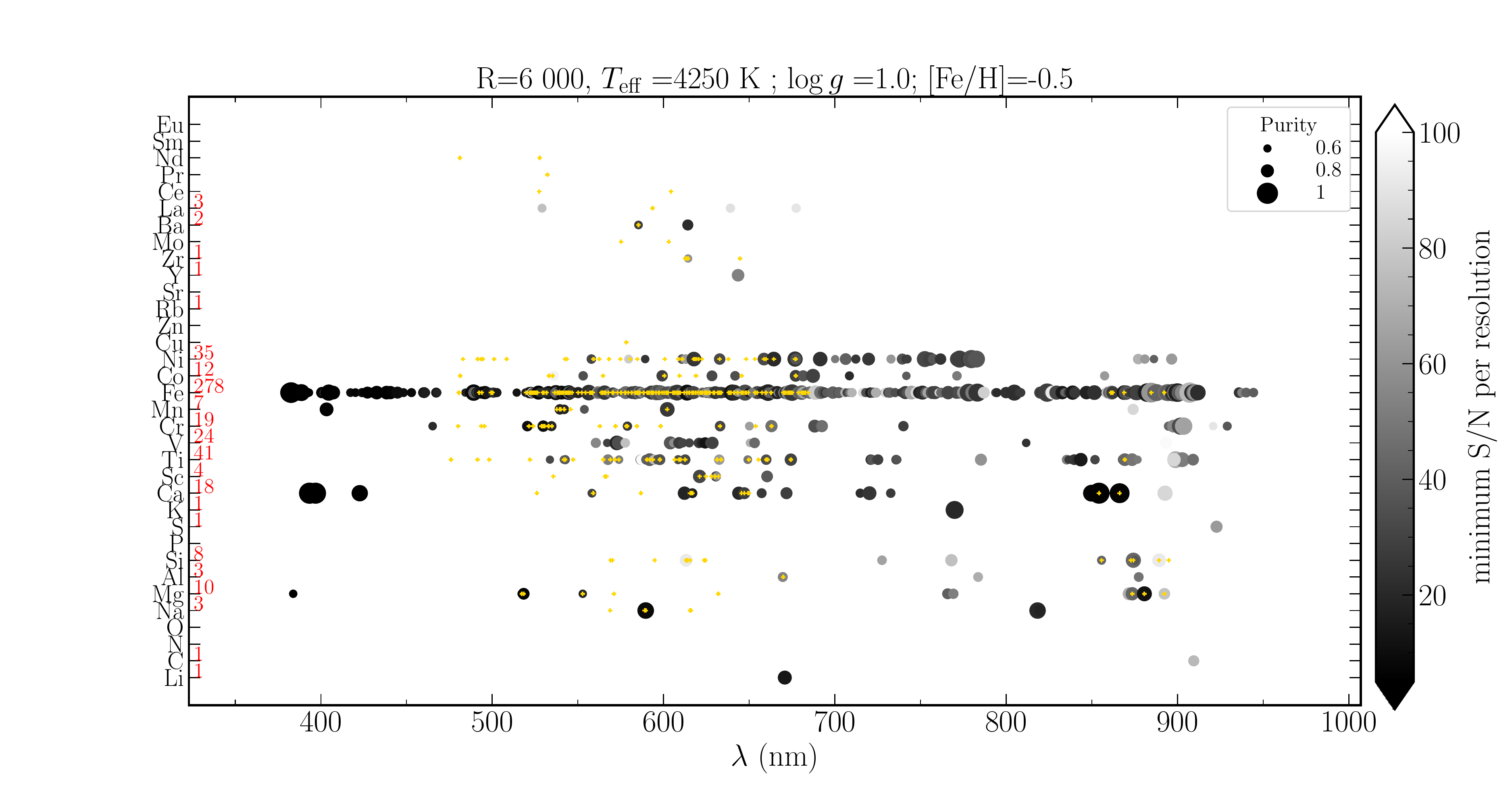}
\caption{ Same as Fig.\,\ref{fig:Sun_Lines}, but for an Arcturus-like giant. }
\label{fig:Arcturus_Lines}
\end{center}
\end{figure*}

Figures~\ref{fig:Sun_Lines} and \ref{fig:Arcturus_Lines} show the lines that are selected for each setup and each star, provided a minimum purity of 0.6 and a maximum $\rsnr=50$  for HR and  $\rsnr=100$ for LR. 
A larger \snr~threshold is adopted for LR, to mimic the fact that one would gain \snr~ by going for LR mode at a fixed exposure time. Note that we assume that the \snr~ is the same across all of the wavelength range, and that the wavelength range is the same for both setups. Neither of these assumptions are true, especially the first one, since noise is in general wavelength dependent (e.g. wavelength dependent efficiency of spectrograph, decreasing optical quality at the borders of the detector, interstellar extinction absorbing preferentially in the blue, ...).

For both of the Sun and Arcturus, Fig.\,\ref{fig:Sun_Lines} and \ref{fig:Arcturus_Lines}, show that the HR setup contains a much wealthier selection of lines, for any of the considered elements. 
Indicatively, 
233 (275) $\alpha$-elements lines, 
606 (769) Fe-peak lines, 
10 (80) neutron-capture lines and
26 (41) odd-Z elements lines are selected for the Sun (Arcturus) in HR, 
compared to 
124 (78), 
374 (375), 
2 (8) and
11 (12) 
in LR, despite the higher \snr~threshold (we recall, however, that the purity threshold is maintained equal to 0.6 in both cases). This corresponds to a "loss" of 50 to 90 per cent of the lines (depending on the nucleosynthetic channel considered). In practice, going for LR implies giving up hopes of detection with a purity greater than 0.6 for Eu, Sm, Nd, Pr, Ce, Mo, Sr, Zn, Cu for Arcturus, while for the Sun the problem is a bit less dramatic, losing only Y, Sr and Zn (due to the fact that many of the aforementioned elements lost in Arcturus LR, are neither detected for the Sun in HR).  
Furthermore, the purity of the lines overall decreases when in LR, as expected due to the blending of the lines. 

This application, illustrates which lines are detectable, for specific spectral types, with what purity, and the required $\rsnr$, given an instrumental  resolving power.  It can be used in order to chose wavelength ranges that contain the most information based on instrumental constraints (e.g. size of the CCD) or observational strategy (e.g. exposure time, target brightness, stellar type).  

In what follows, we will use this information to assess which WEAVE-HR setup performs best per nucleosynthetic channel and per element.

\subsection{Choosing between setups: application to the high-resolution setups of WEAVE}
\label{sec:weave_HR_application}
We now put ourselves in the framework of a survey design, for instance WEAVE. There exist two WEAVE Galactic archaeology (GA) HR surveys,  a HR-chemodynamical survey targeting the thin and thick disc as well as the halo, and an Open Cluster  survey, aiming to target roughly a hundred of young and old open clusters in the disc \citep{Jin23}. WEAVE has the possibility to choose between two HR setups: the first one, dubbed in what follows B+R setup, covers the wavelength ranges [404-465] and [595-685]\,nm. The second one, dubbed G+R setup in what follows, covers the wavelength ranges  [473-545] and [595-685]\,nm. 
The question that we are trying to answer is the following: which setup combination probes the best the different nucleosynthetic channels? In other words, which combination of setups maximises the number of elements  and number of useful lines, per nucleosynthetic channel ($\alpha$-elements, odd-Z elements, Fe-peak elements, neutron-capture elements) across the targeted parameter space of \teff, \logg~and \meta?

 To set this value, we rely on  WEAVE's GA survey plan (WEAVE consortium, private communication) and adopt as a threshold $\snr_{\rm resol}=70$, which is the value of the  expected $\snr$ peak in the blue setup for the typical selection of the WEAVE GA-HR baseline survey. Other setups are expected to have a higher $\snr_{\rm resol}$ value. 
  Using Eqs.~\ref{eq:fmin_appendix1} and \ref{eq:fmin_appendix2}, this corresponds to a minimum required depth   of $\sim 0.94$ for a line to be detected.

We ran our code on a set of metal-poor stars  ($\meta=-2$, representative of the halo),  intermediate-metallicity stars  ($\meta=-0.5$, representative of the thick disc) and metal-rich stars ($\meta=0$, representative of the thin disc and open cluster stars).  The results are shown in Figs.\,\ref{fig:Kiel_alpha_WEAVE_BGR}, \ref{fig:Kiel_irons_WEAVE_BGR} and \ref{fig:Kiel_neutron_WEAVE_BGR}, for $\alpha$-elements, Fe-peak elements and neutron-capture elements, respectively (we have not plotted the results for \teff$=4250$\,K and \logg$=5.0$ for visualisation purposes). They illustrate the number of lines (colour-code) and number of different elements (size of the points) detected for each nucleosynthetic family and each combination of \teff~and \logg.
The purity threshold for the $\alpha$-, Fe-peak and neutron-capture elements has been arbitrarily set at $0.8$, $0.9$ and $0.6$, in order to optimise the number of lines and the purity itself. A detailed view of the detected lines per element across the Kiel diagram is shown in the Appendix, see Figs\,\ref{fig:Summary_WEAVE_BGR_alpha_individual1} to  
\ref{fig:Summary_WEAVE_BGR_neutron_individual3}.

\subsubsection{Even-Z elements}

\begin{figure}
\begin{center}
\includegraphics[width=\linewidth, angle=0]{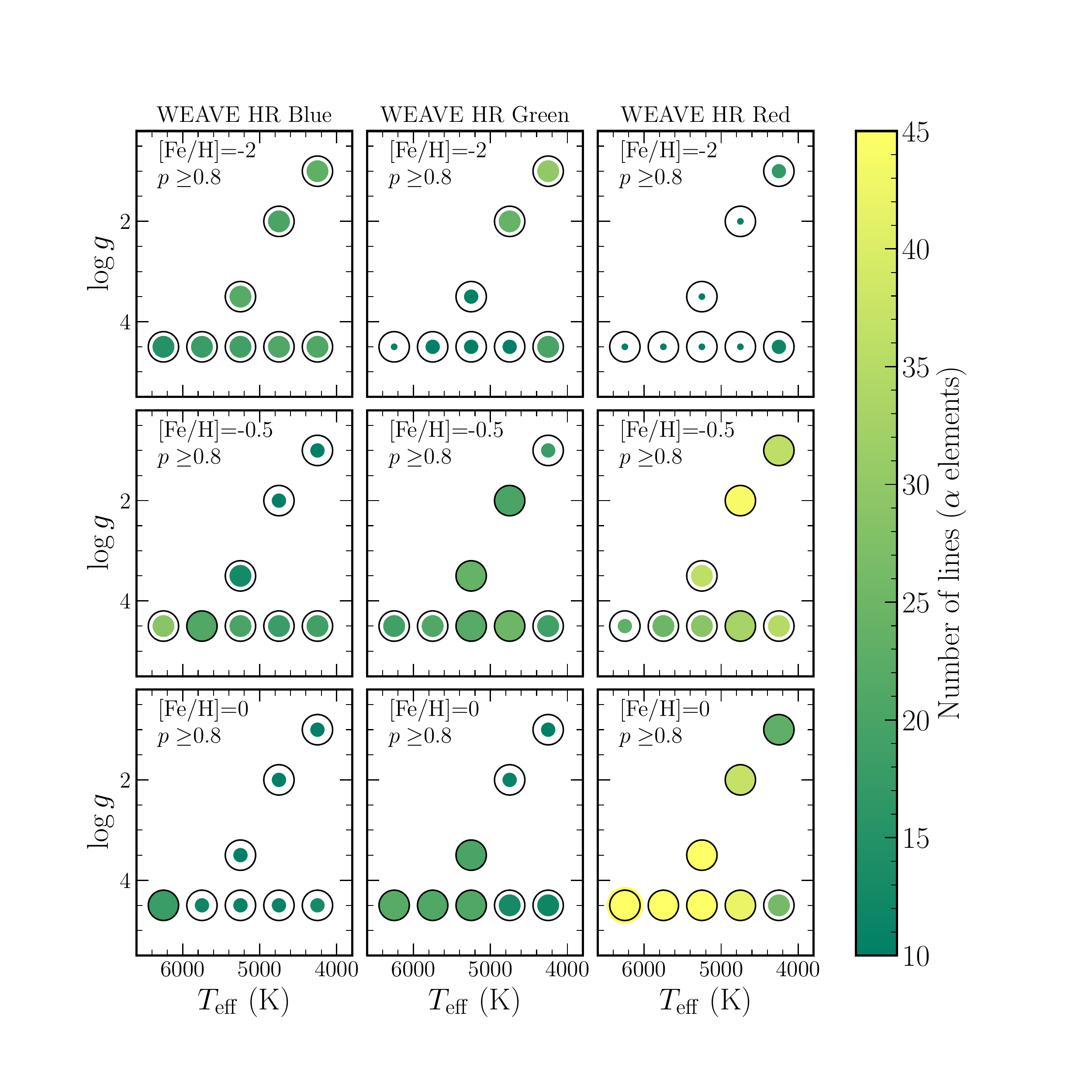}
\caption{Kiel-diagrams for stars with $\feh=-2$ (top), $-0.5$ (middle), $0.0$ (bottom),  and the three different WEAVE HR setups (blue: left, green: middle, red: right). The colour-code represents the number of lines associated with $\alpha$-elements  (O, Mg, Si, S, Ca, Ti) 
having a purity greater than 0.8, and detectable for a $\rsnr<70$.  The size of the points
is proportional to the number of $\alpha$-elements with useful lines at a given combination of \teff, \logg~ and $\meta$. As a reference, the size of a point for which four elements would have been detected is plotted as a black solid  circle at the position of the Kiel diagram at which we have templates. }
\label{fig:Kiel_alpha_WEAVE_BGR}
\end{center}
\end{figure}

As shown in Fig.\,\ref{fig:Kiel_alpha_WEAVE_BGR}, the red setup is the one clearly driving the science for intermediate and high metallicities, with more than $\sim30$ useful lines, throughout the Kiel diagram, and four elements detected with a purity greater than 0.8 (the black solid circle  in Fig.\,\ref{fig:Kiel_alpha_WEAVE_BGR} is proportional to four elements). For metal-poor stars ($\meta=-2$),  the blue setup performs slightly better than the green and red setups, with more elements and more lines detected. The green setup performs slightly better than the blue one for intermediate and high metallicities, a regime, however,  where, as said above, the red setup is the one driving the science for $\alpha$-elements. 

More specifically, based on Figs.\,\ref{fig:Summary_WEAVE_BGR_alpha_individual1} and \ref{fig:Summary_WEAVE_BGR_alpha_individual2}, the following diagnostics can be drawn about individual elements:

\begin{itemize}
\item Carbon (atomic) is seen both in green and red (but not for metal-poor stars) setups, with a purity a bit higher for the green setup ($p\gtrsim0.7-0.8$). It is not detectable in the blue setup.  We note, however, that these are high excitation \ion{C}{I} lines, most readily visible in warmer stars, while C measurements may be achieved using molecular features such as CH in cooler stars.
\item Oxygen is only detectable in the red setup via the $\lambda=630$\,nm line. 
\item Magnesium is detectable in all three setups. The green setup has lines with a very good purity for every metallicity regime (thanks to the \ion{Mg}{I} triplet). The blue setup contains useful lines too, but with a lower purity ($p\lesssim0.7$). 
\item Silicon has many lines detected in the red setup ($\gtrsim20$), as opposed to the green and blue setups which are not optimal for this element (less than 10 lines and $p\lesssim0.7$).
\item Sulphur is only detectable in the red setup, for high and intermediate metallicity stars, with $p\gtrsim0.8$. 
\item  Calcium has many lines detectable (more than fiour at each setup), and its purity is very good in the red ($p\gtrsim0.9$). The blue setup performs better than the green, with  more lines and higher purity. 
\item Titanium has many lines detectable in all setups, with an overall low purity compared to other $\alpha$-elements. For low metallicity stars, the green setup is preferred to the blue one as it shows a higher purity. 
\end{itemize}

\subsubsection{Odd-Z elements}
No global plot combining the odd-Z elements is presented, as these cannot be linked to a specific nucleosynthetic channel. Nevertheless, their abundance determination is of prime importance on many fields of galactic and stellar evolution, and a thorough description on how the setups perform is necessary. Based on Figs.\,\ref{fig:Summary_WEAVE_BGR_odd_individual1} and \ref{fig:Summary_WEAVE_BGR_odd_individual2}, the following diagnostics can be drawn:

\begin{itemize}
\item 
Lithium is detected at all stellar types and metallicities in the red setup, thanks to the  $\lambda= 670.8$\,nm line, and additionally at $\lambda=610.3$\,nm for the most metal-poor giant stars.  We recall, however, that given the adopted Li abundance in the modelled spectra, A(Li)=2.00\,dex, our results are likely overestimated  for giants (for which due to dilution A(Li)$<1$, see however the case of Li-rich giants, e.g. \citealt{Charbonnel00}), and under-estimated  for more metal-rich turn-off stars \citep[see][and references therein]{Karakas14}.
\item Nitrogen (atomic) is not detectable in any of the setups.
\item Sodium is detected in the red setup for all stars, except for the metal-poor regime, with $p\gtrsim0.75$. The green and blue setups perform similarly, each of them providing low-purity lines ($p\lesssim0.7$) that do not allow detection across the whole Kiel diagram at any metallicities.
\item Aluminium is seen only in the red,  for intermediate and high metallicities. The purity is overall high ($p\gtrsim0.8$).
\item Phosphorus is not detectable with either of the setups.
\item Potassium is not detectable in either of the setups.
\item Scandium is detected in the red setup with $p\gtrsim0.85$. Green and blue setups also contain  useful Sc lines, especially at low metallicities, though with a lower purity than in the red setup. The green setup performs better than the blue one both in terms of number of lines and in terms of purity.
\end{itemize}

\begin{figure}
\begin{center}
\includegraphics[width=\linewidth, angle=0]{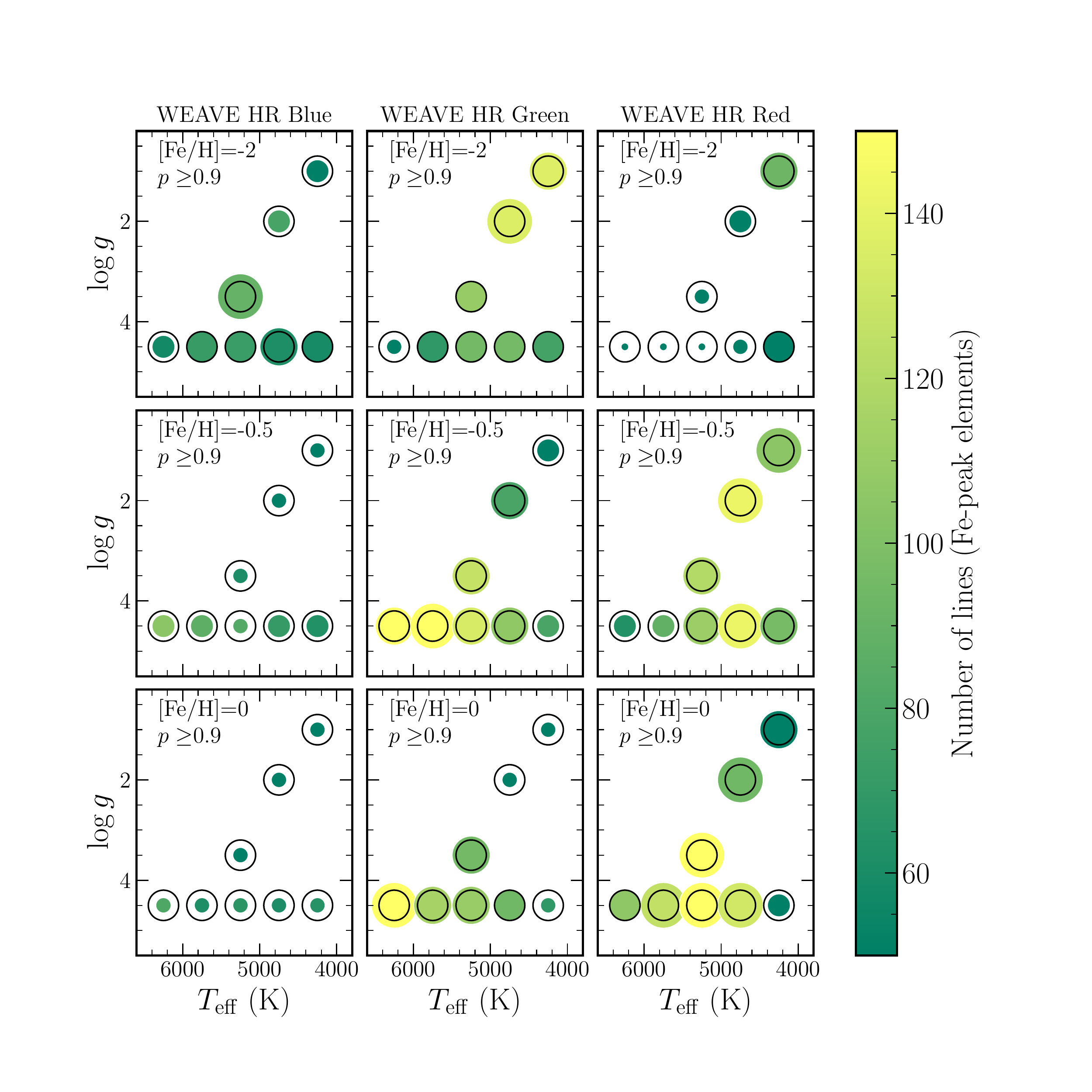}
\caption{Same as Fig.\,\ref{fig:Kiel_alpha_WEAVE_BGR} for the iron-peak elements (V, Cr, Mn, Fe, Co, Ni, Cu, Zn) and a purity greater than 0.9.  Black solid circles  represent the detection of four elements .}
\label{fig:Kiel_irons_WEAVE_BGR}
\end{center}
\end{figure}

\subsubsection{Fe-peak elements}
As shown in Fig.\,\ref{fig:Kiel_irons_WEAVE_BGR}, there exists a plethora of lines to select, with more than 60 lines with a purity greater than 0.9 for any of the setups. Overall, the green setup performs the best for all stars at low metallicity, as well for  main-sequence stars at intermediate metallicity. The red setup is the one driving the science for giants at intermediate metallicity and for all stars at high metallicity. The combination of the green and red setups allows us to get at least seven iron-peak elements at any combination of \teff, \logg~and \meta.  

More specifically, based on Figs.\,\ref{fig:Summary_WEAVE_BGR_iron_individual1} and \ref{fig:Summary_WEAVE_BGR_iron_individual2}, the following diagnostics can be drawn about individual elements: 
\begin{itemize}
\item Vanadium has high purity lines in the red setup ($p\gtrsim0.8$). The blue setup performs better than red or green at low metallicity, with lines detected over the entire Kiel diagram. 
\item Chromium has few high purity lines in the red setup for intermediate and high metallicities. At low metallicity, {both green and blue setup exhibit many lines, with} a marginal advantage of the green setup over the blue one in terms of purer lines. 
\item Manganese has the highest purity lines for intermediate and high metallicity stars in the red setup, which also performs relatively well at low metallicity. Overall, the green setup performs better than blue  the former having purer lines than the latter. 
\item Iron has many  lines that are detectable in all setups, and in fact \ion{Fe}{I} dominates the number counts in Fig.\,\ref{fig:Kiel_irons_WEAVE_BGR}. The red setup has the highest purity ($p\gtrsim0.8$), and the green setup has purer Fe lines than the blue. 
\item Cobalt has the purest lines in the red setup. The blue setup performs better than green at low metallicity, allowing a detectability of Co lines for both giants and main-sequence stars with a purity of $0.7-0.8$.
\item Nickel has the purest lines in the red setup. The green setup performs much  better than the blue,  with more numerous and purer lines.  
\item Copper has lines seen only in the green, with a relatively low purity ($p\lesssim0.7$), except for metal-poor stars, where $p\gtrsim0.8$. 
\item Zinc is not seen in the blue setup. The green setup is the only one that allows us to measure a Zn abundance at low metallicities. 
 \end{itemize}

\begin{figure}
\begin{center}
\includegraphics[width=\linewidth, angle=0]{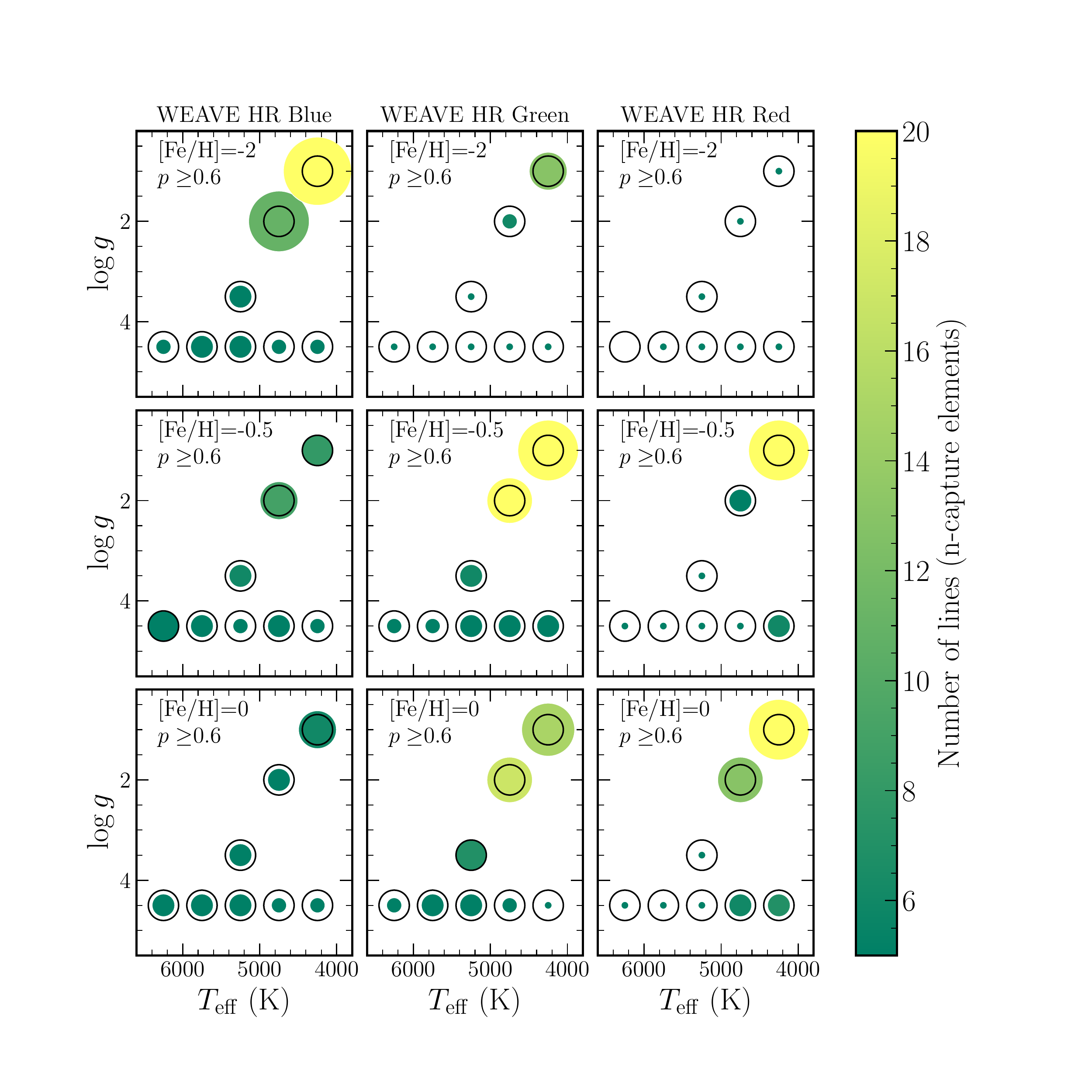}
\caption{ Same as Fig.\,\ref{fig:Kiel_alpha_WEAVE_BGR} for the neutron-capture elements (Rb, Sr, Y, Zr, Mo, Ba, La, Ce, Pr, Nd, Sm, Eu) and a purity greater than 0.6.}
\label{fig:Kiel_neutron_WEAVE_BGR}
\end{center}
\end{figure}

\subsubsection{Neutron-capture elements}
As shown in Fig.\,\ref{fig:Kiel_neutron_WEAVE_BGR},  WEAVE setups contain much less useful neutron-capture element lines than for the $\alpha$- and Fe-peak elements, with at best 20 lines for metal-rich  and intermediate metallicity giants. As far as the turn-off region is concerned, the blue setup is the one performing the best, with more elements being probed compared to the other two setups. 

More specifically, based on Figs.\,\ref{fig:Summary_WEAVE_BGR_neutron_individual1}, \ref{fig:Summary_WEAVE_BGR_neutron_individual2} and \ref{fig:Summary_WEAVE_BGR_neutron_individual3}, the following diagnostics can be drawn about individual elements: 
\begin{itemize}
\item Rubidium is never detected with a purity greater than 0.5 in the considered setups.
\item Strontium is detected only in the blue setup (reaching $p\gtrsim0.8$ for metal-poor stars). 
\item Yttrium has a better purity in the green setup compared to the blue. Yet, for intermediate and high metallicity stars, Y lines are also detected in the red setup.
\item Zirconium is  best detected across the Kiel diagram in the green setup, and sparsely in the blue  for main-sequence and the red setup for cool stars. However, the purity is overall low ($p\lesssim0.7$). 
\item Molybdenum is detected only in the red setup, only for the cooler and  $\feh>-0.5$\,dex stars,  with  $p\gtrsim0.7$.  
\item Barium is detected in all three setups, with the blue one performing slightly better than the green one in terms of purity. 
\item Lanthanum is sparsely detected in all setups for giants with a variety of purities. There is a slight advantage of the green setup over the blue, with purer and more numerous lines. 
\item Cerium is detected  at all evolutionary stages only the blue setup, albeit with a low purity. 
\item Praseodymium is sparsely detected in the blue and green setups. 
\item Neodymium has the most numerous lines detected in the green setup at any evolutionary stage, while having a similar purity similar  than the other setups, at intermediate and high metallicities (or better in the case of metal-poor stars). 
\item Samarium is slightly better detected in the blue setup than in the other two setups. 
\item Europium is only detected in the blue setup for metal-poor stars while only the red setup allows allow the detection of usable lines for intermediate and high metallicity giants. No lines are detected in the green setup.
 \end{itemize}

\subsubsection{Summary}
The diagnostics above were derived for an idealised case of perfectly normalised spectra with white noise (\snr~ that is constant over the wavelength range). In reality, this will not be the case, and the normalisation is expected to be challenging in the blue setup, due to the multiple atomic and molecular lines. Yet,  keeping in mind that WEAVE's HR baseline survey will not target many cool main-sequence stars (WEAVE consortium, private communication), our results seem to slightly privilege the Green+Red setup, both in numbers of elements detected and in terms of number of lines that are useful. We note, however, that there is a disparity in terms of which elements are detected over each of the setups (e.g. Sr is only detectable in the blue setup), and that the final decision needs to be taken according to the elements that the science cases of the considered surveys decide to highlight  and on the expected temperature and metallicity ranges in which those elements need to be detected (i.e. the target selection function). 

\subsection{Line-list optimisation for abundance determination}
\label{sec:linelist_optimisation}
For some abundance determination codes, the masking of a subset of lines of a specific element may be desired, either in order to decrease the computational time and/or in order to improve the precision of the measurement.  The code presented in  Sect.\,\ref{sect:algorithm}, allows one to very simply extract a sub-sample of lines for a given element, provided  some observational (e.g. maximum \snr) and purity constraints. To build such a  "golden line-sublist", the following considerations could be taken into account: 
\begin{itemize}
\item
The purity of the selected lines for a given stellar type and metallicity should be as as high as possible.
\item
The linelist for a given stellar type and metallicity needs to include lines that allow an abundance measurement for both a  high and a low \snr~ (reflecting the range in apparent magnitudes of the survey).
 \item
For a given stellar type and metallicity,  lines on the linear part of the curve-of-growth  (i.e. not strong lines) should be favoured, to maximise the sensitivity of the lines to the elemental abundance \citep{Gray_spectroscopy}.
 \item
When a range of excitation potentials is available for a given species, selecting only the lowest excitation lines should be avoided (typically more prone to non-local thermodynamic equilibrium). In the case of species where many lines are available, spanning a wide range of excitation potentials is desirable to enable checks of the excitation temperature.
\item
The synthetic lines need to reproduce satisfactorily the observed spectra of at least the Sun and Arcturus.
\end{itemize}

\begin{figure*}
\begin{center}
\includegraphics[width=\linewidth, angle=0]{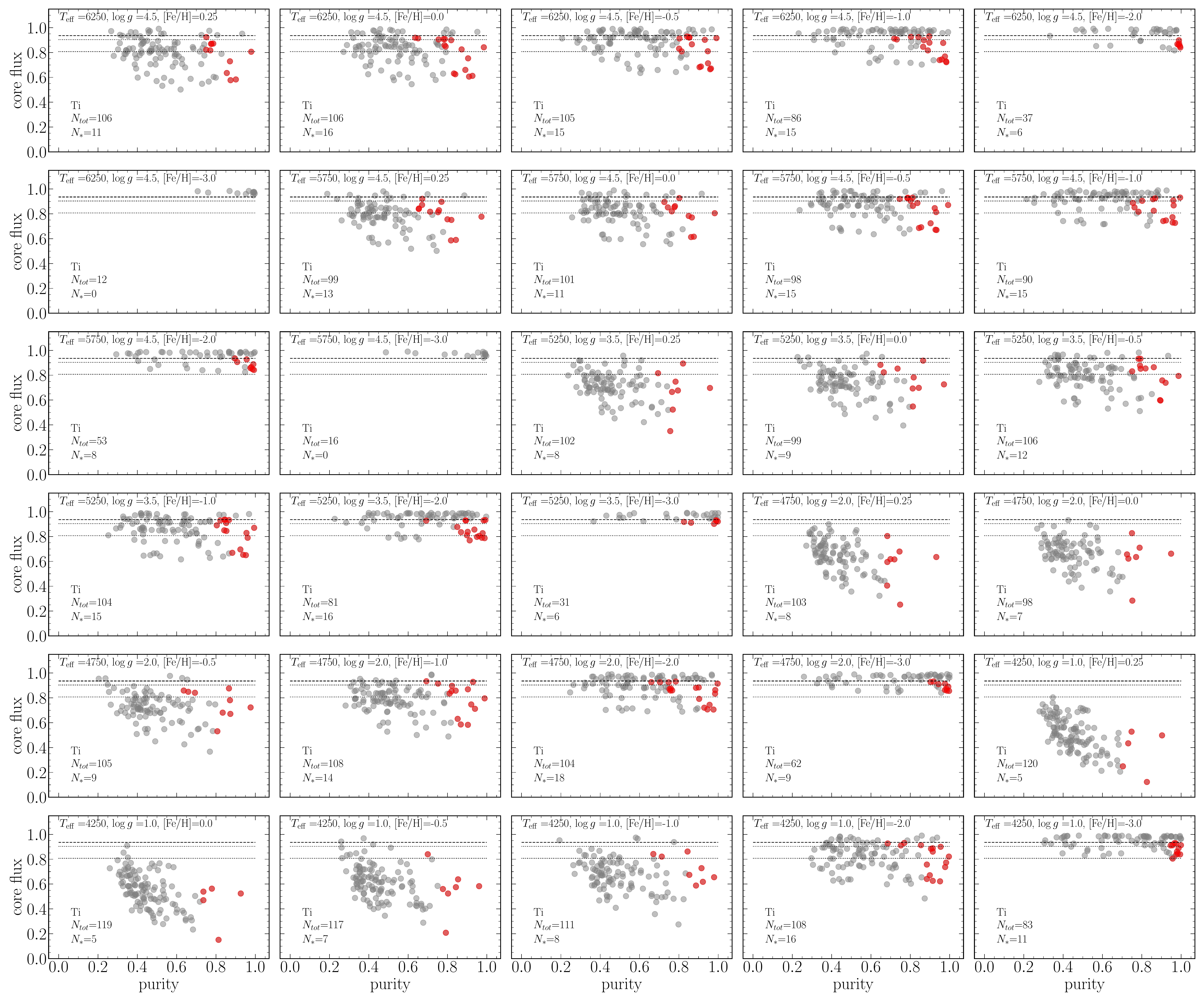}
\caption{Flux at the core of the lines versus purity of the lines for a given element (Ti), and different stellar types (parameters written at the top of each panel). The total available lines, from VALD, are in grey and the selected golden-subsample is in red. The dashed line is located at our $\rsnr$ threshold of 70, i.e. the lines need to be deeper than this in order to be detected. The other two dotted lines split the \snr~ range from 0 to 70 into three, inside which we select at least five lines, provided they have a purity greater than 0.6.  }
\label{fig:Line_selection1_Ti}
\end{center}
\end{figure*}

\begin{figure}
\begin{center}
\includegraphics[width=1.05\linewidth, angle=0]{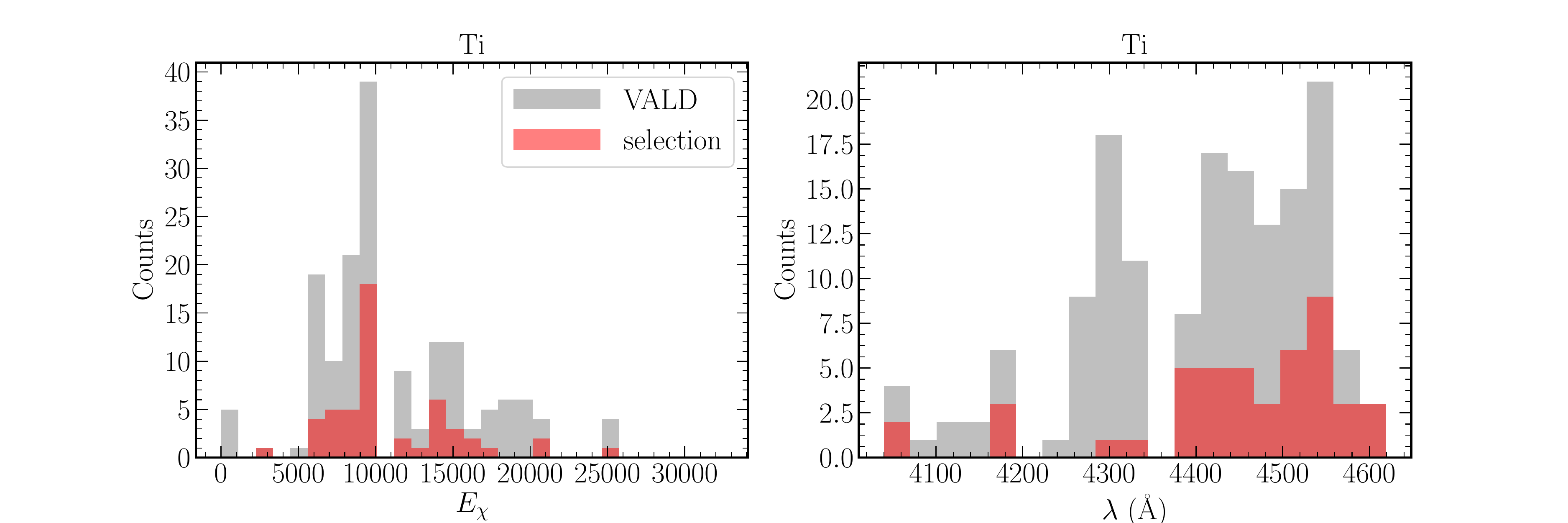}
\caption{Distribution of excitation potential ($E_\chi$, left) and wavelengths (right) of all the Ti lines available in VALD (in grey) and the selected golden sub-sample (in red), the latter being the union of the lines selected in Fig.\,\ref{fig:Line_selection1_Ti} for each stellar type. 
}
\label{fig:Line_selection2_Ti}
\end{center}
\end{figure}

We implemented the above scheme into the creation of a line-list for the blue HR setup of WEAVE.  In practice, we imposed $\snr_{\rm resol,max}=70$ as the maximum $\rsnr$ for the detectability of a line with no purity filter. For each element, we kept all of the available lines if their total number was less than 30 (this number was arbitrarily chosen) when considering all of the set of stellar atmospheric parameters.
When there were more than 30 lines available, each stellar spectrum was investigated automatically, splitting the range $[0,\snr_{\rm resol,max}$] into three bins of equal range, and looking within each of these bins for the lines that had the highest purity. In order to achieve this, we started by imposing a purity of 1 and decreased the latter iteratively by steps of 0.025 until a minimum of five lines was reached while keeping the purity greater than 0.6 (except for Fe, where we imposed a minimum purity of 0.95). Figure\,\ref{fig:Line_selection1_Ti} shows, for Ti, the properties of all the available lines detectable up to  $\rsnr=500$, where we have highlighted in red the ones that we eventually select.

The golden line-sublist for the considered element was then obtained by keeping the union of all of the selected lines across the entire set of atmospheric parameters.   
Figure\,\ref{fig:Line_selection2_Ti} shows a histogram of the excitation potential of all the available Ti lines detectable for  $\snr_{\rm resol,max}$ (in grey) and in red, the sub-sample that we have selected.   One can see that they successfully span all the range of $E_\chi$, with a bias towards lower values, as desired.

    \section{Conclusions}
     \label{sec:conclusion}
    Our automatic line selection for abundance determination code is based on the use of synthetic spectra containing all of the elements and blends available and the comparison with a synthetic spectrum at the same stellar parameters containing only one element at the time. In this sense, a comparison with true, observed, spectra is necessary in order to confirm that the lines that are selected are also representing  nature accurately.  
   Ideally, this comparison should be done with spectra of stars for which both stellar parameters and individual abundances are best known, i.e. the Sun, Arcturus and other benchmark stars \citep[e.g.][]{Blanco-Cuaresma14, Heiter15, Jofre15}. 
  We have not proceeded through this comparison in this work, as results may vary from one  resolving power to the other, yet a simple computation of residuals between the synthetic spectra and the real ones, around the lines that our code selects, should suffice in order to discard lines that are not modelled properly. 
        
    Our code can serve both as an illustration of where the chemical information is present in a stellar spectrum, but most importantly allows one to optimise  {\it i)} observational strategies, such as choosing resolution and spectral windows, as well as {\it 2)} analysis codes, with the application of masks of high quality.
    In particular, direct applications for observations using the WEAVE \citep{Jin23}  and 4MOST \citep{deJong19} facilities (both community and consortium surveys) will benefit largely of the present tool.
    
     The python code allowing to identify and characterise the useful lines can be downloaded on gitlab\footnote{\url{https://gitlab.oca.eu/gkordo/line_selections}}. We also share via CDS the Tables containing the results at five different resolving powers ($R=3\,000, 6\,000, 20\,000, 40\,000$ and $80\,000$) for lines that have a purity greater than 0.4 in at least one of their wings (i.e. $p_b$ or $p_r$, see Sect.~\ref{sect:algorithm_description}), for the entire wavelength range between 300\,nm and 1000\,nm. Results for other resolving powers can be easily computed and provided by contacting the first author of this paper. Finally, the 54 infinite resolution spectra that have been used in this work ($\sim145$\,GB) can be shared upon request.

\section*{Acknowledgments}
We  thank the anonymous referee for their comments that helped  improving the quality of the paper.  
This work has benefited from inspiring and fruitful discussions within the WEAVE and 4MOST consortia, as well as with Michael Hanke. 
Shoko Jin and Scott Trager are warmly thanked for their valuable feedback on early versions of the paper and for discussions that led to the selection of the WEAVE HR wavelength ranges. 
GK and VH gratefully acknowledge support from the french national research agency (ANR) funded project MWDisc (ANR-20-CE31-0004).
KL acknowledges funds from the European Research Council (ERC) under the European Union’s Horizon 2020 research and innovation programme (Grant agreement No. 852977) and funds from the Knut \& Alice Wallenberg foundation.
This work was supported by the Programme National Cosmology et Galaxies (PNCG) of CNRS/INSU with INP and IN2P3, co-funded by CEA and CNES. 
Ansgar Wehrhahn is acknowledged for their contribution to the PySME code. 
This work has made use of the VALD database, operated at Uppsala University, the Institute of Astronomy RAS in Moscow, and the University of Vienna.
as well as the Python packages Numpy \citep{Harris20}, Matplotlib \citep{Hunter07} and Pandas.


\bibliographystyle{aa}
\def\aj{AJ}\def\apj{ApJ}\def\apjl{ApJL}\def\araa{ARA\&A}\def\apss{Ap\&SS}
\def\mnras{MNRAS}\def\aap{A\&A}\def\nat{Nature}
\def\nar{New Astron. Rev.}

\bibliography{master_bib}

\begin{appendix}
\section{Cayrel's formula and minimum depth of a line}
\label{sec:Appendix_Careyl}
Here we derive our approximation on the desired minimum depth of a line, $\fmin$,  in order for it to be detected at a given signal-to-noise ratio, $\snr$.
We start from the standard \citet{Cayrel88} formula, linking the uncertainty $\sigma_{\rm EW}$ on measuring the equivalent width $\rm EW$ of a line, to the $\snr$ per pixel, the full width at half-maximum of the line (assuming it has a Gaussian profile) and the pixel size $dx$ (in wavelength units): 
\begin{equation}
\sigma_{\rm EW} =\frac{1.5}{\snr} \sqrt{\fwhm \cdot dx}. 
\label{eqn:Careyl}
\end{equation}

To a good approximation,  ${\rm EW} \approx \fmin \cdot \fwhm$. One can therefore derive the formula for the uncertainty of the core of the line, $\sigma_{\fmin}$, as:
\begin{eqnarray}
\sigma_{\fmin} &=& \sigma_{\rm EW}/\fwhm \\
		    &=& \frac{1.5}{\snr} \sqrt{dx/\fwhm}. \label{eqn:line_detection}
\end{eqnarray}
where $\fwhm$ and $dx$ are in wavelength units, and $\snr$ is per pixel. 

Equation~\ref{eqn:line_detection} can also be expressed as a function of $\snr$ per resolution element, $\snr_{\rm resol}= \snr_{\rm pix} \cdot \sqrt{\fwhm/dx}$, as follows:  
\begin{eqnarray}
\sigma_{\fmin} &=& \frac{1.5}{\snr_{\rm resol}} \sqrt{\fwhm/dx} \cdot \sqrt{dx/\fwhm} \\
		&=&  \frac{1.5}{\snr_{\rm resol}}. \label{eq:fmin_appendix1}
\end{eqnarray}

The detectability of a spectral absorption line is therefore possible if its intrinsic intensity is deeper than:
\begin{equation} 
\fmin\leq 1-3\cdot \sigma_{\fmin}. 
\label{eq:fmin_appendix2}
\end{equation}

\section{Equivalent width uncertainties in presence of a blend}
\label{sec:Appendix_blend}
We consider $EW_0$ the measured EW of a line, which we assume to be a combination of the real EW of the line alone,  $EW_R$, and a {fractional contribution of a blend} to the line, $\rm blend$. One can hence write:
\begin{equation}
 EW_0=EW_R+ \mathrm{blend} \cdot EW_R. 
\end{equation}
The contribution of the blending to the error on $EW_0$ can be written as:
\begin{eqnarray}
\frac{dEW_0}{EW_0} &=& \frac{dEW_0}{EW_R+ \mathrm{blend} \cdot EW_R}  \\
		&=&  \frac{\mathrm{blend} \cdot EW_R}{EW_R(1+\mathrm{blend})} \\
		&=& \frac{\mathrm{blend}}{1+\mathrm{blend}}.
\end{eqnarray}

In order to have an error  on $EW_0$ smaller than 10 per cent (corresponding to an abundance uncertainty of $\sim0.05$\,dex if the line is in the linear part of the curve of growth), one therefore needs : 
\begin{equation}
\rm
\frac{blend}{1+blend} \leq 0.1 
\end{equation}
and hence: 
$\rm 0.9 \cdot blend \leq 0.1  \implies blend\leq0.11$.

 Similarly, assuming that the blend is known by a factor of $a$, one can write: 
\begin{equation}
 EW_0=EW_R+a\cdot  \mathrm{blend}\cdot EW_R.
\end{equation}

 Following the previous steps, in order to have an error smaller than 10 per cent on $EW_0$, one therefore needs : 
\begin{equation}
\frac{a\cdot  \mathrm{blend}}{1+ \mathrm{blend}} \leq 0.1 
\end{equation}
and hence: $(a-0.1)  \cdot  \mathrm{blend} \leq 0.1 \implies  \mathrm{blend}\leq\frac{0.1}{a-0.1}$. 
So that if $a=0.5$, then $\mathrm{blend}  \leq 0.25$.

\section{Purities and detectability of elements for WEAVE high-resolution setups}
\label{sec:Appendix_weave_BGR}
\begin{figure*}
   \centering
    \renewcommand{\arraystretch}{0.01} 
\begin{tabular}{cc}

\includegraphics[width=0.49\linewidth]{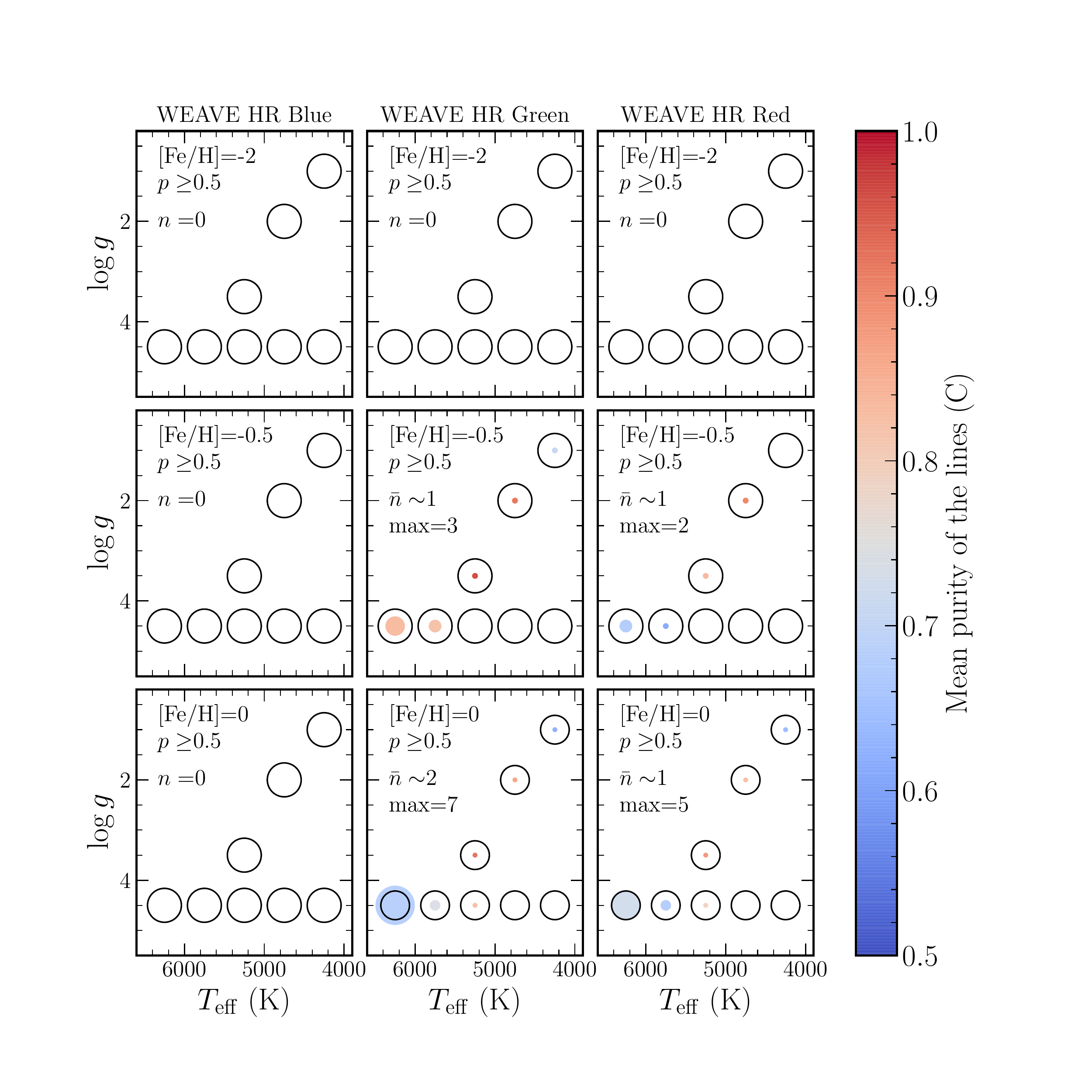}&
\includegraphics[width=0.49\linewidth]{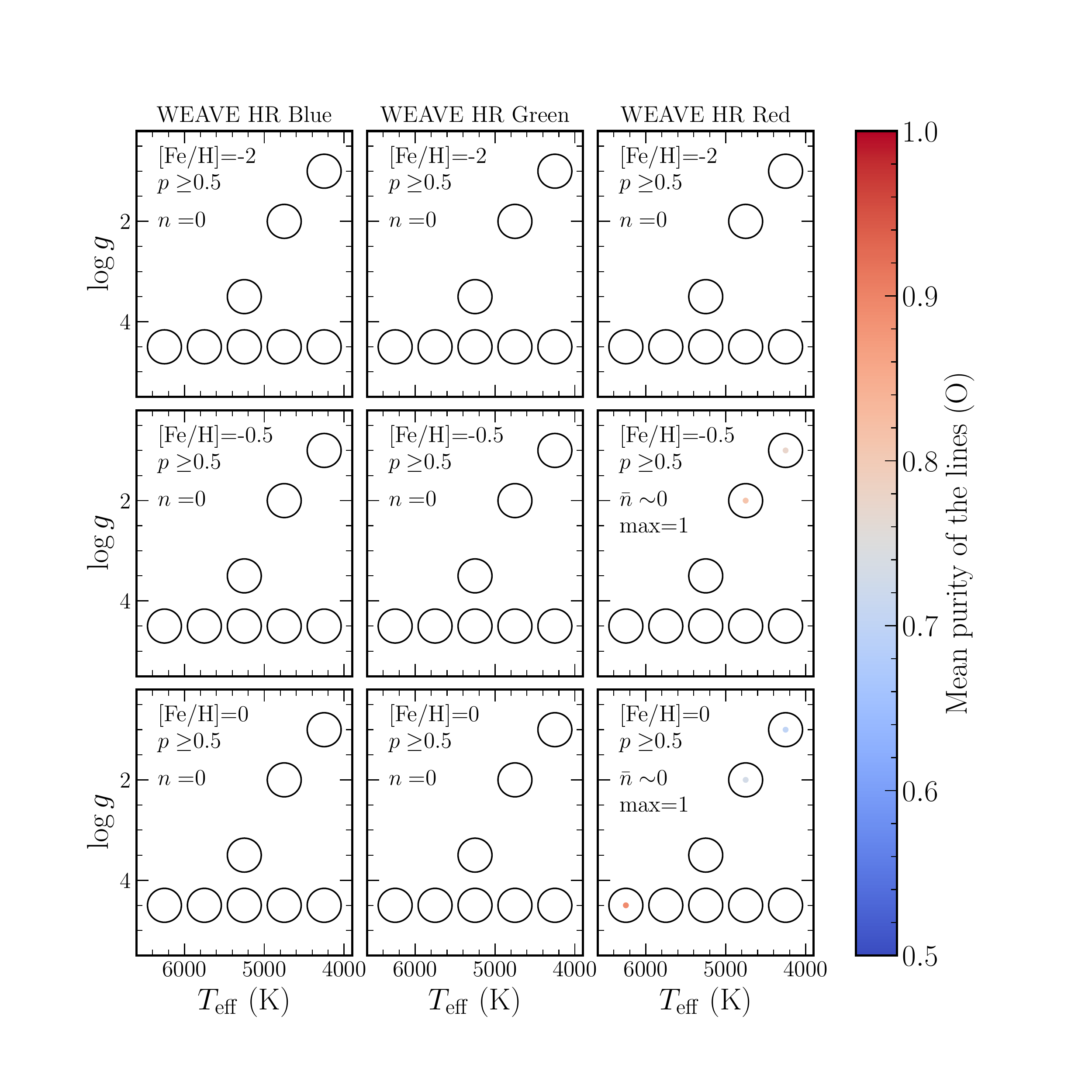}\\
\includegraphics[width=0.49\linewidth]{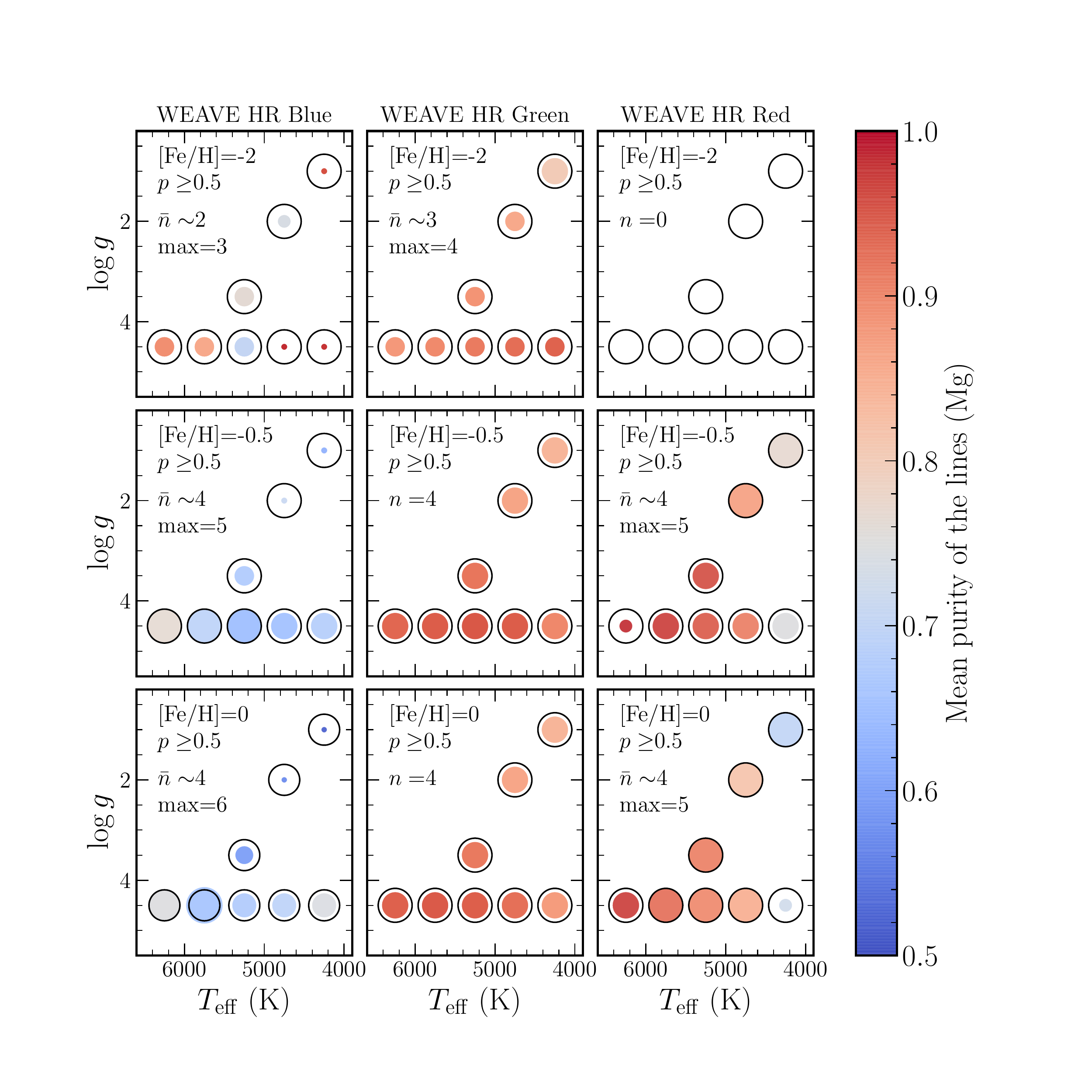}&
\includegraphics[width=0.49\linewidth]{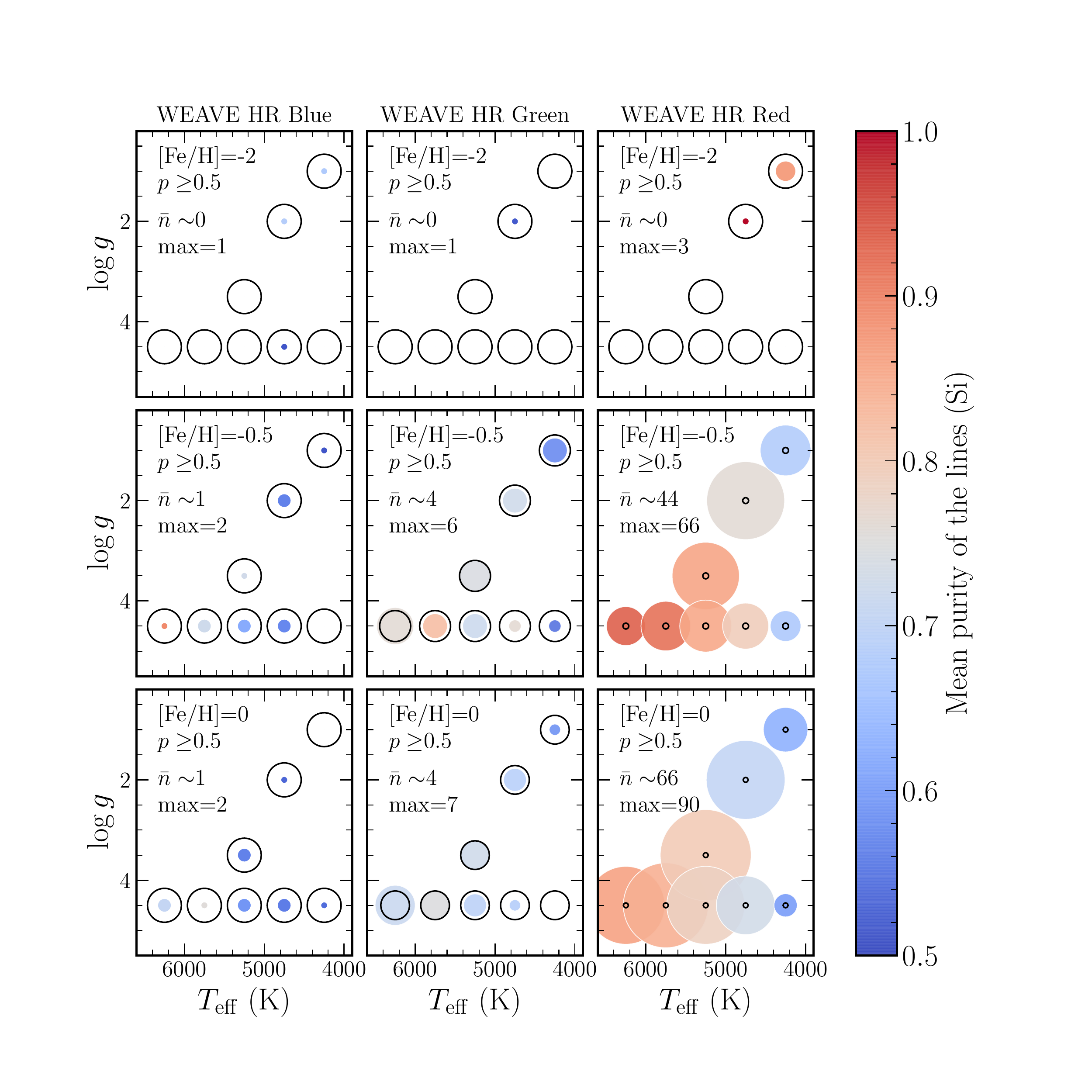}
\end{tabular}
\caption{Number of identified lines (size of the points) and average purity (colour-coded)  for the even-Z elements  
C, O, Mg and Si, at different combinations of \teff, \logg~and \feh, for the different WEAVE HR setups.  To guide the eye, black circles proportional to the identification of five lines are also plotted for each combination of \teff$-$\logg~ (note that the relative size of the black circles and hence of the coloured points change from one frame to the other, for visualisation purposes). The absence of coloured points implies the non-detection of lines. A minimum purity threshold of $0.5$ has been set.  The average ($\bar{n}$) and maximum ($\rm max$, reported only if different than $\bar{n}$) number of identified  lines across the selected models are also indicated within each frame.}

\label{fig:Summary_WEAVE_BGR_alpha_individual1}
\end{figure*}

\begin{figure*}
   \centering
    \renewcommand{\arraystretch}{0.01} 
\begin{tabular}{cc}
\includegraphics[width=0.49\linewidth]{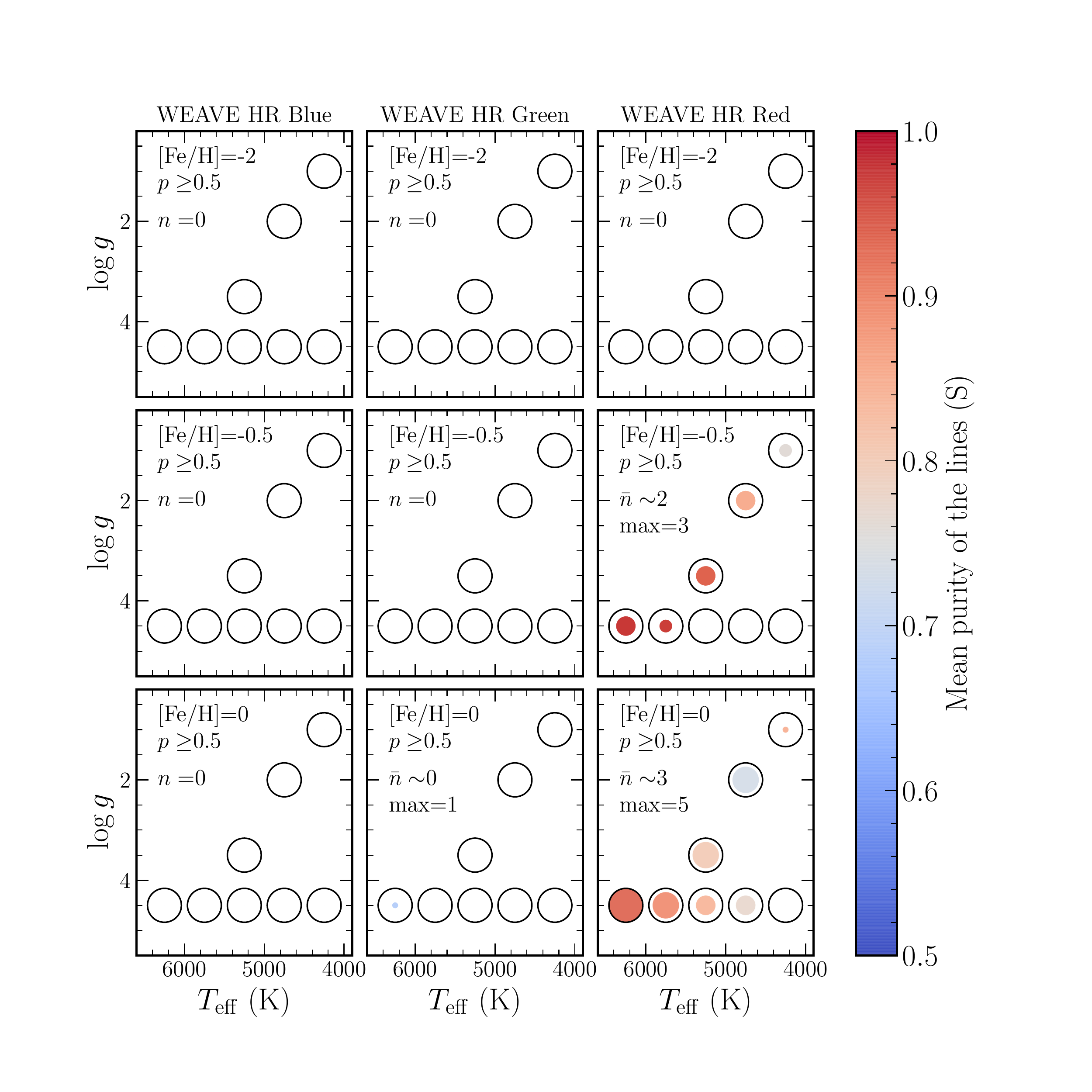}&
\includegraphics[width=0.49\linewidth]{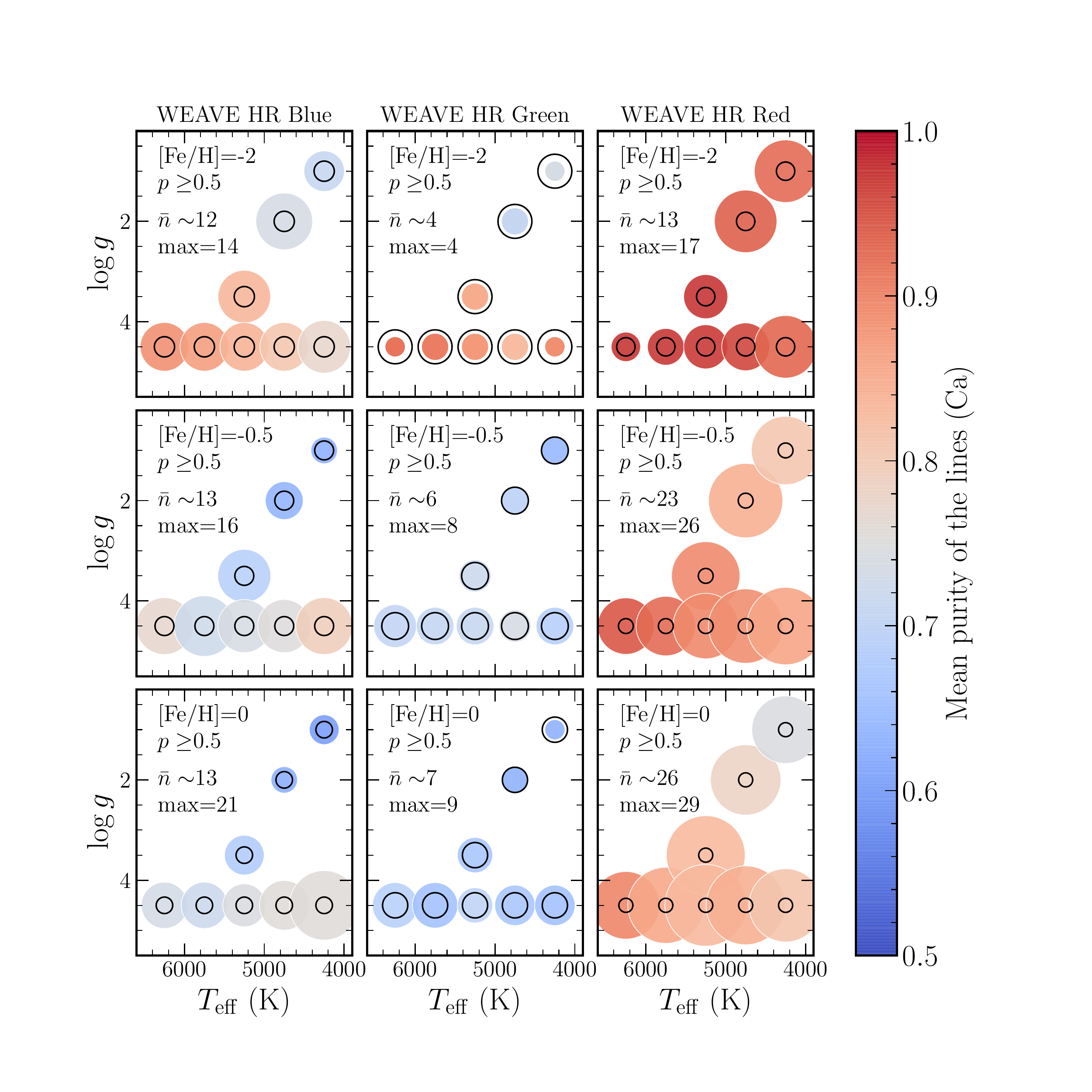}\\
\includegraphics[width=0.49\linewidth]{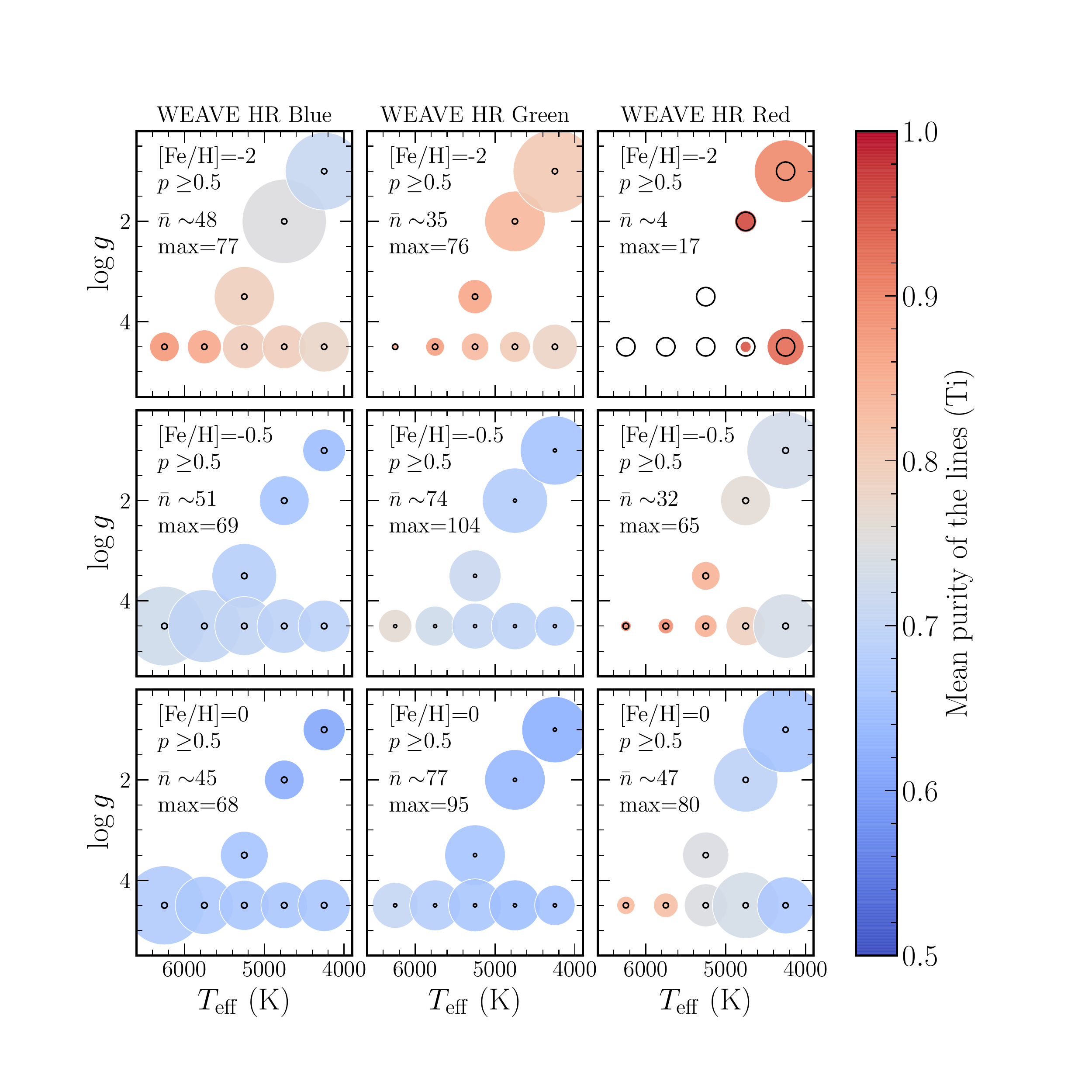}

\end{tabular}
\caption{Same as Fig.\,\ref{fig:Summary_WEAVE_BGR_alpha_individual1} but for the even-Z elements 
S, Ca, Ti. }
\label{fig:Summary_WEAVE_BGR_alpha_individual2}
\end{figure*}

\begin{figure*}
   \centering
    \renewcommand{\arraystretch}{0.01} 
\begin{tabular}{cc}
\includegraphics[width=0.49\linewidth]{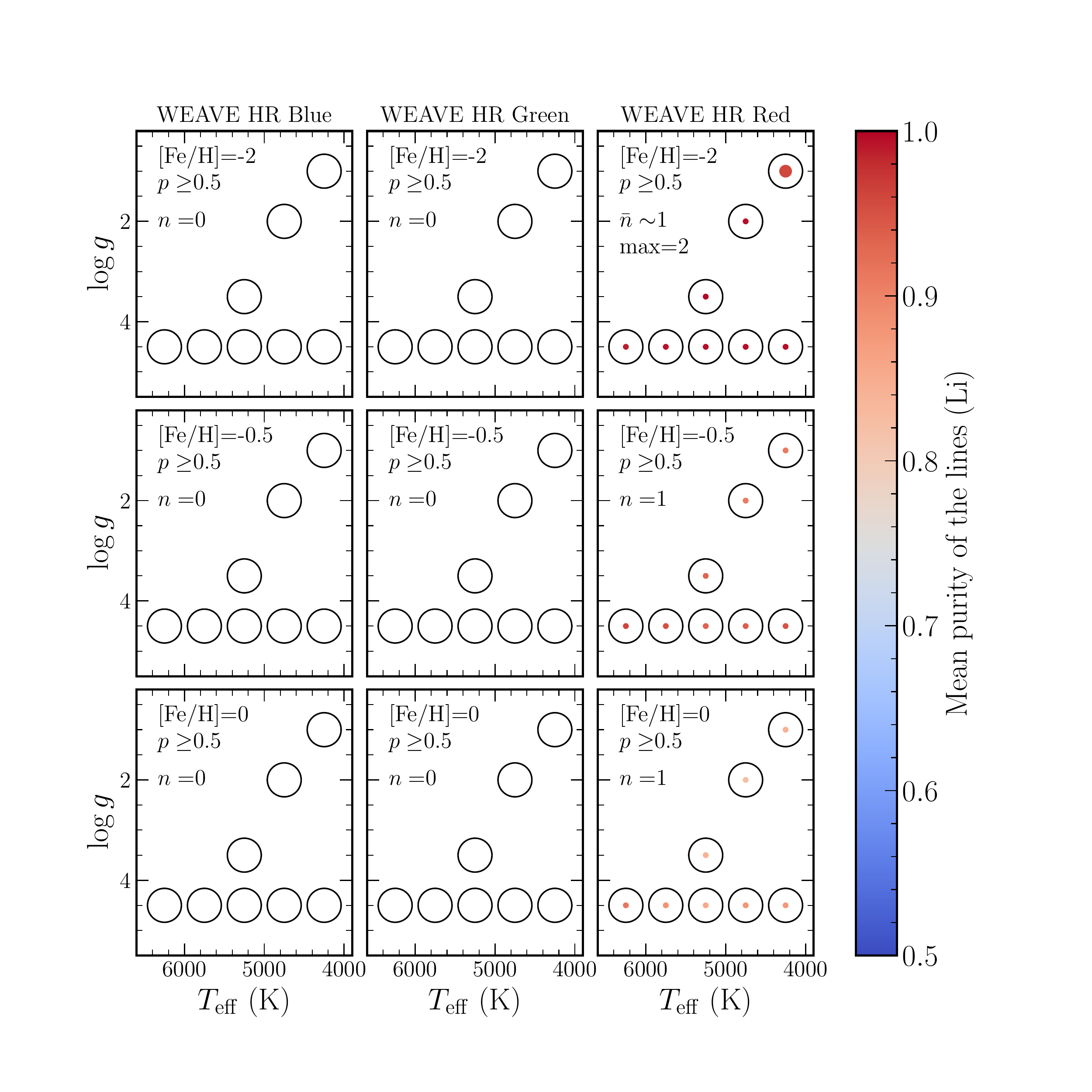}&
\includegraphics[width=0.49\linewidth]{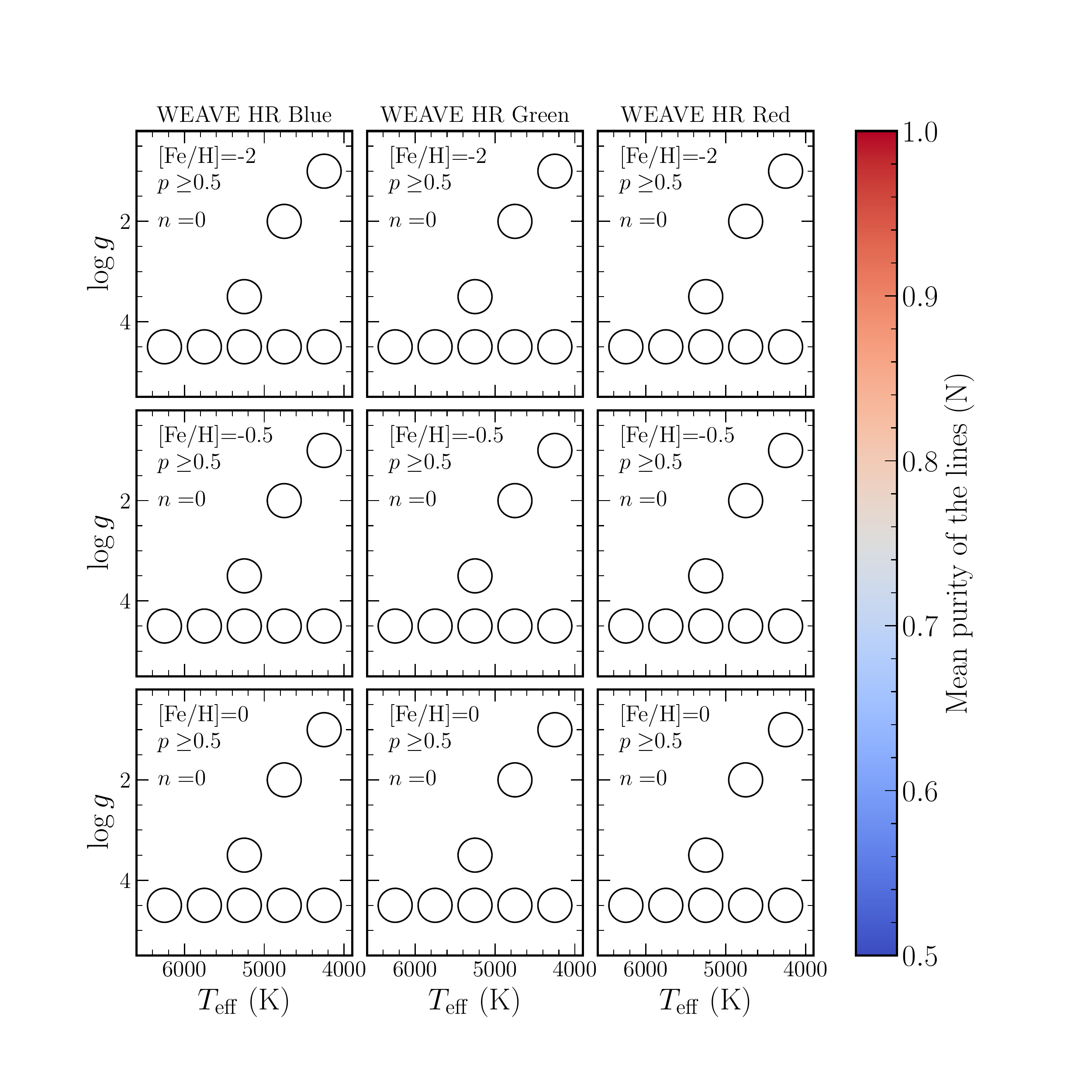}\\
\includegraphics[width=0.49\linewidth]{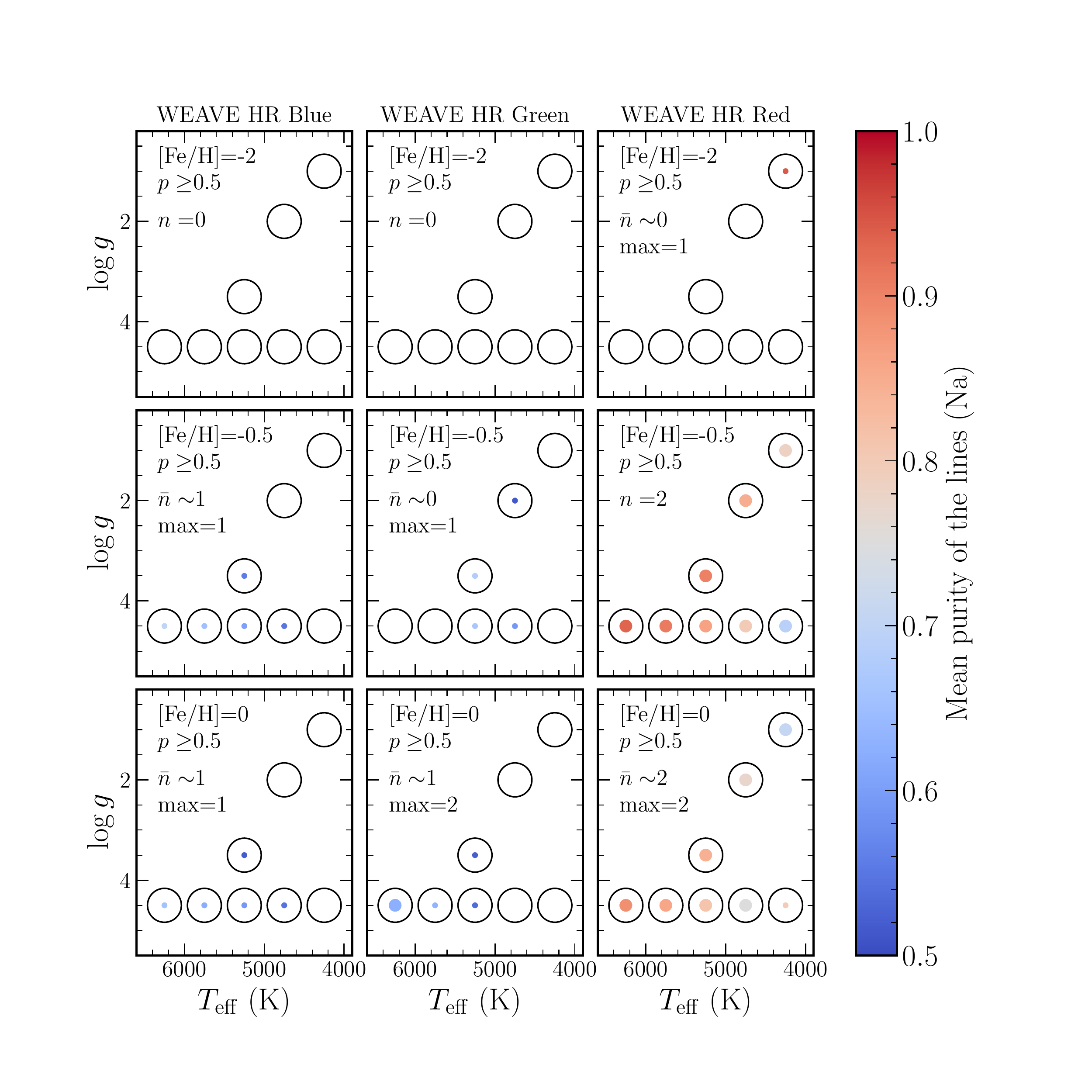}&
\includegraphics[width=0.49\linewidth]{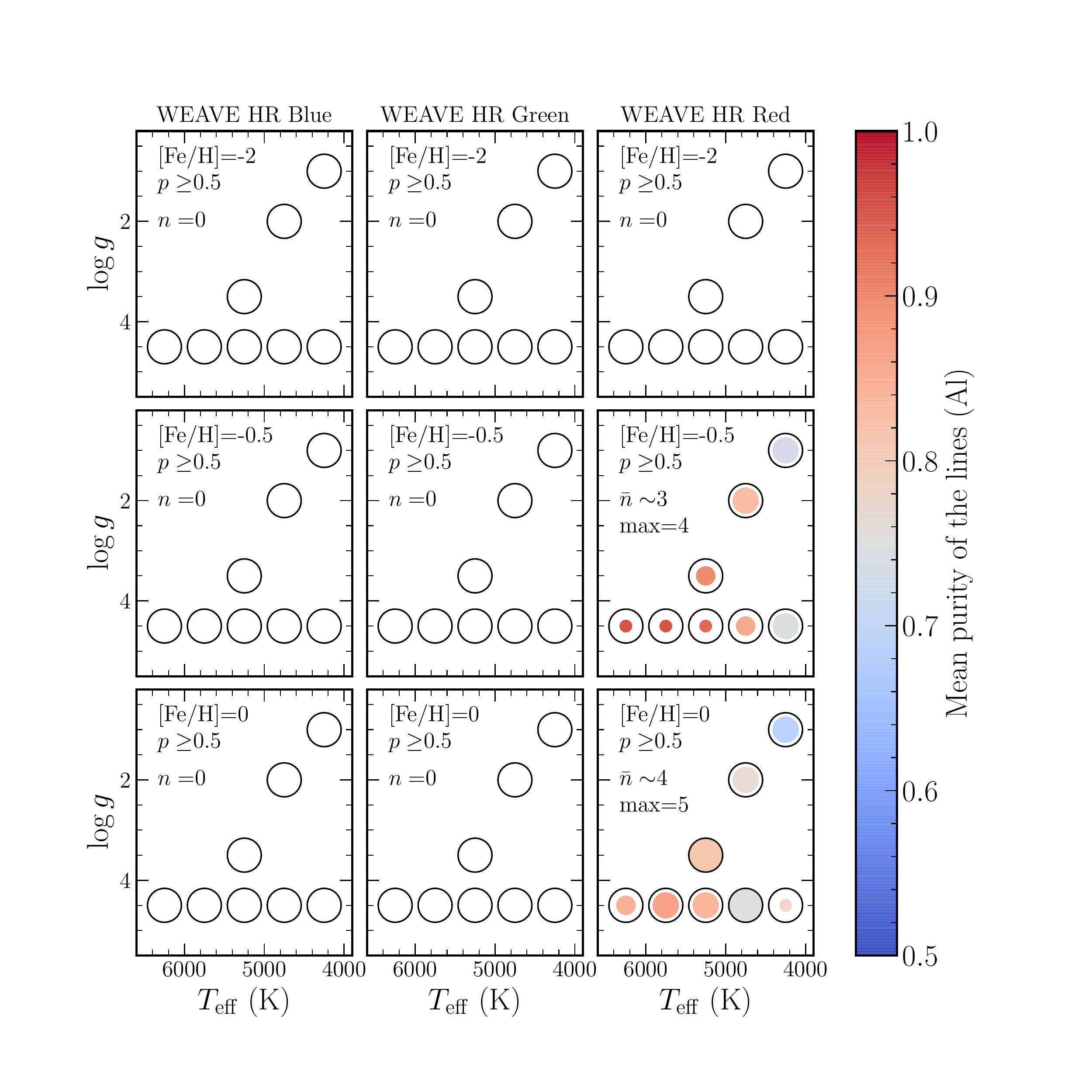}
\end{tabular}
\caption{Same as  Fig.\,\ref{fig:Summary_WEAVE_BGR_alpha_individual1}, for the odd-Z elements Li, N, Na, Al.}
\label{fig:Summary_WEAVE_BGR_odd_individual1}
\end{figure*}

\begin{figure*}
   \centering
    \renewcommand{\arraystretch}{0.01} 
\begin{tabular}{cc}
\includegraphics[width=0.49\linewidth]{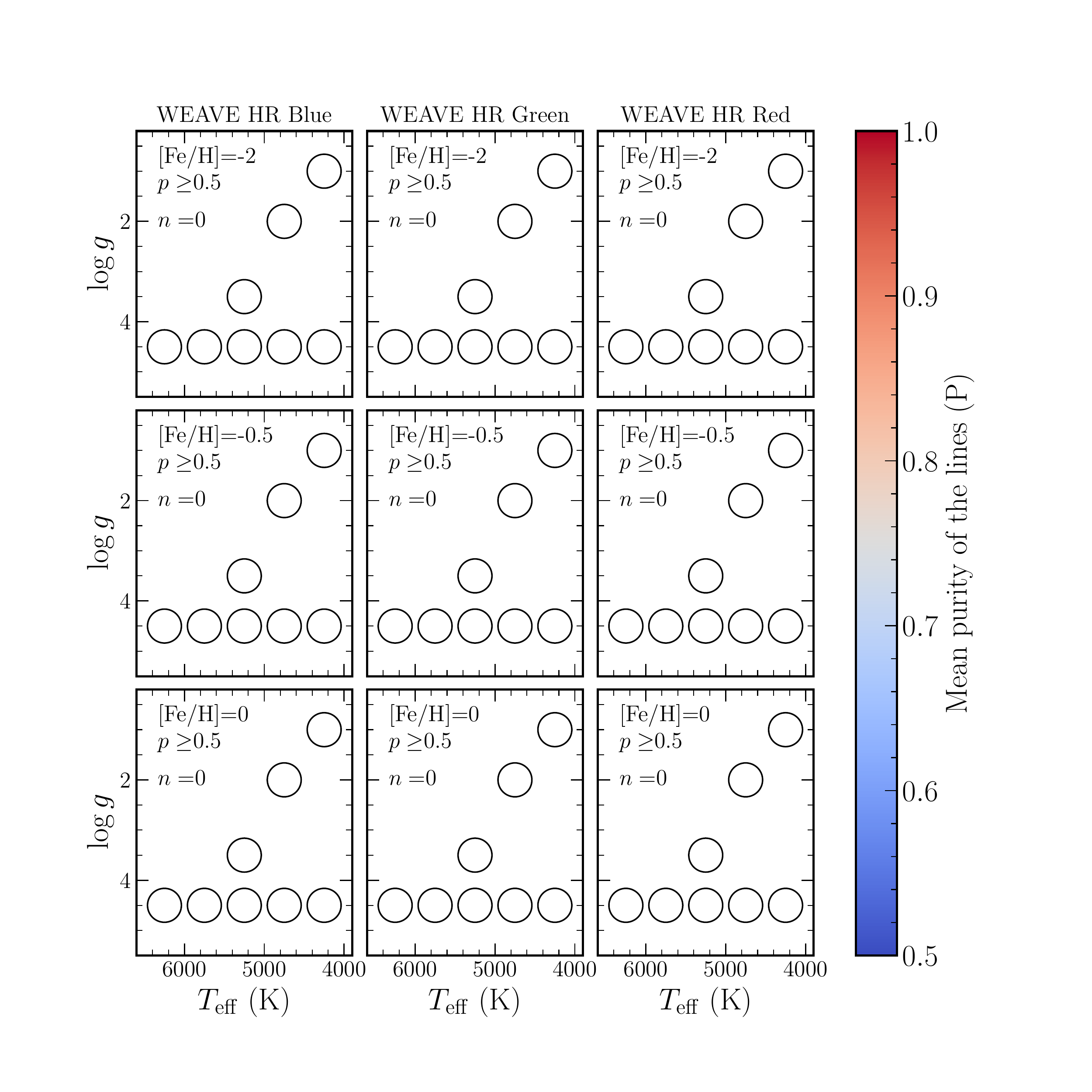}&
\includegraphics[width=0.49\linewidth]{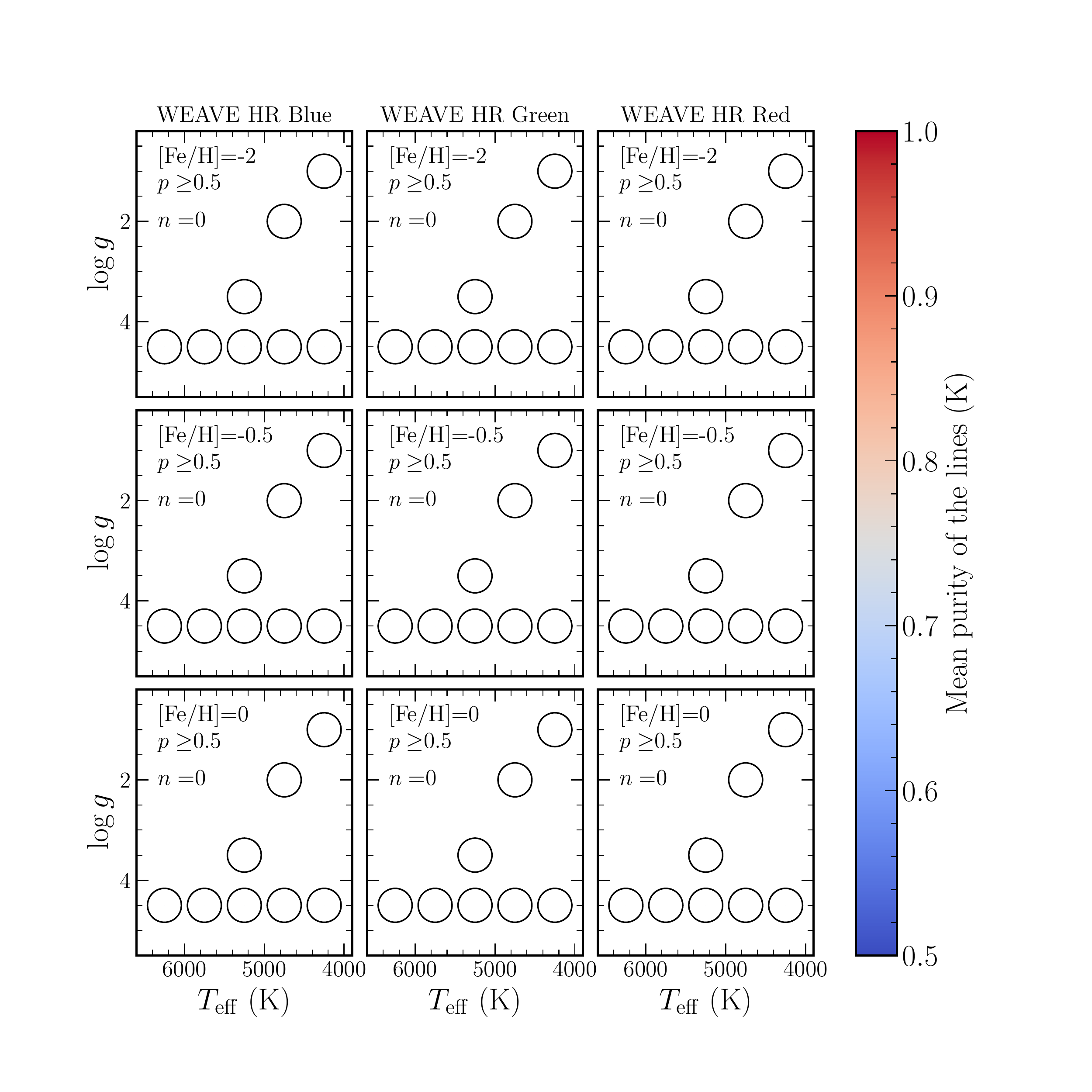}\\
\includegraphics[width=0.49\linewidth]{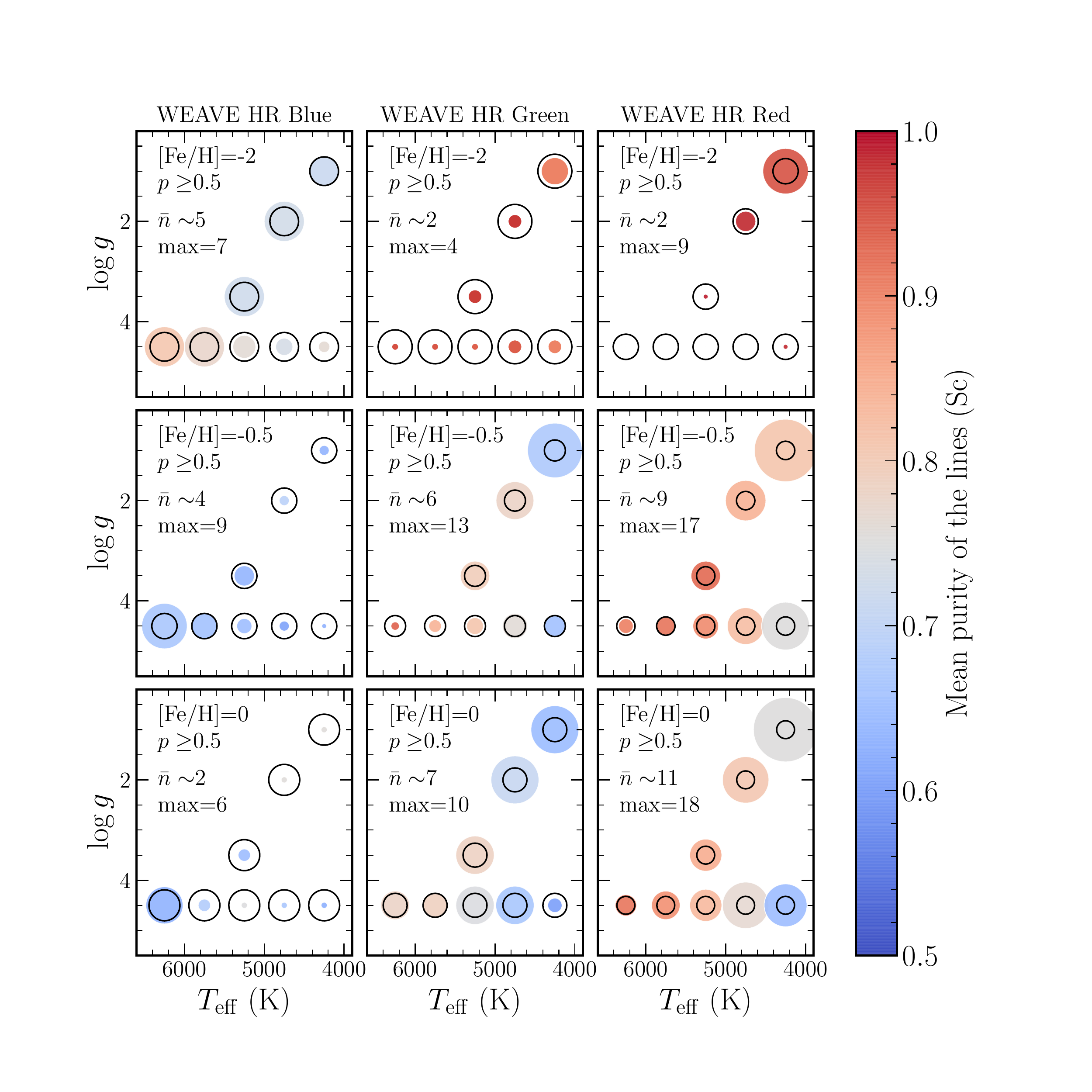}
\end{tabular}
\caption{Same as  Fig.\,\ref{fig:Summary_WEAVE_BGR_alpha_individual1}, for the odd-Z elements P, K, Sc.}
\label{fig:Summary_WEAVE_BGR_odd_individual2}
\end{figure*}

\begin{figure*}
   \centering
    \renewcommand{\arraystretch}{0.01} 
\begin{tabular}{cc}
\includegraphics[width=0.49\linewidth]{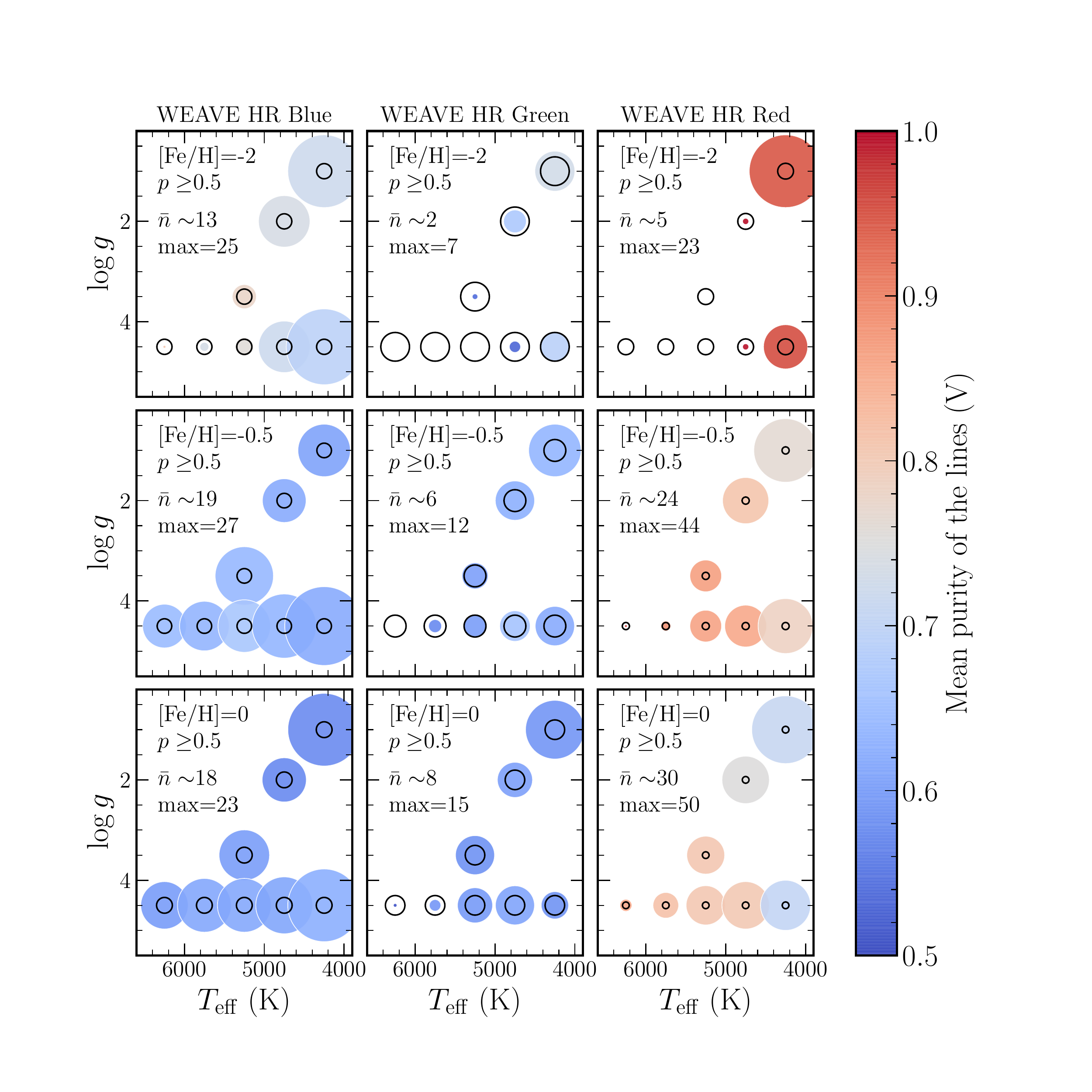}&
\includegraphics[width=0.49\linewidth]{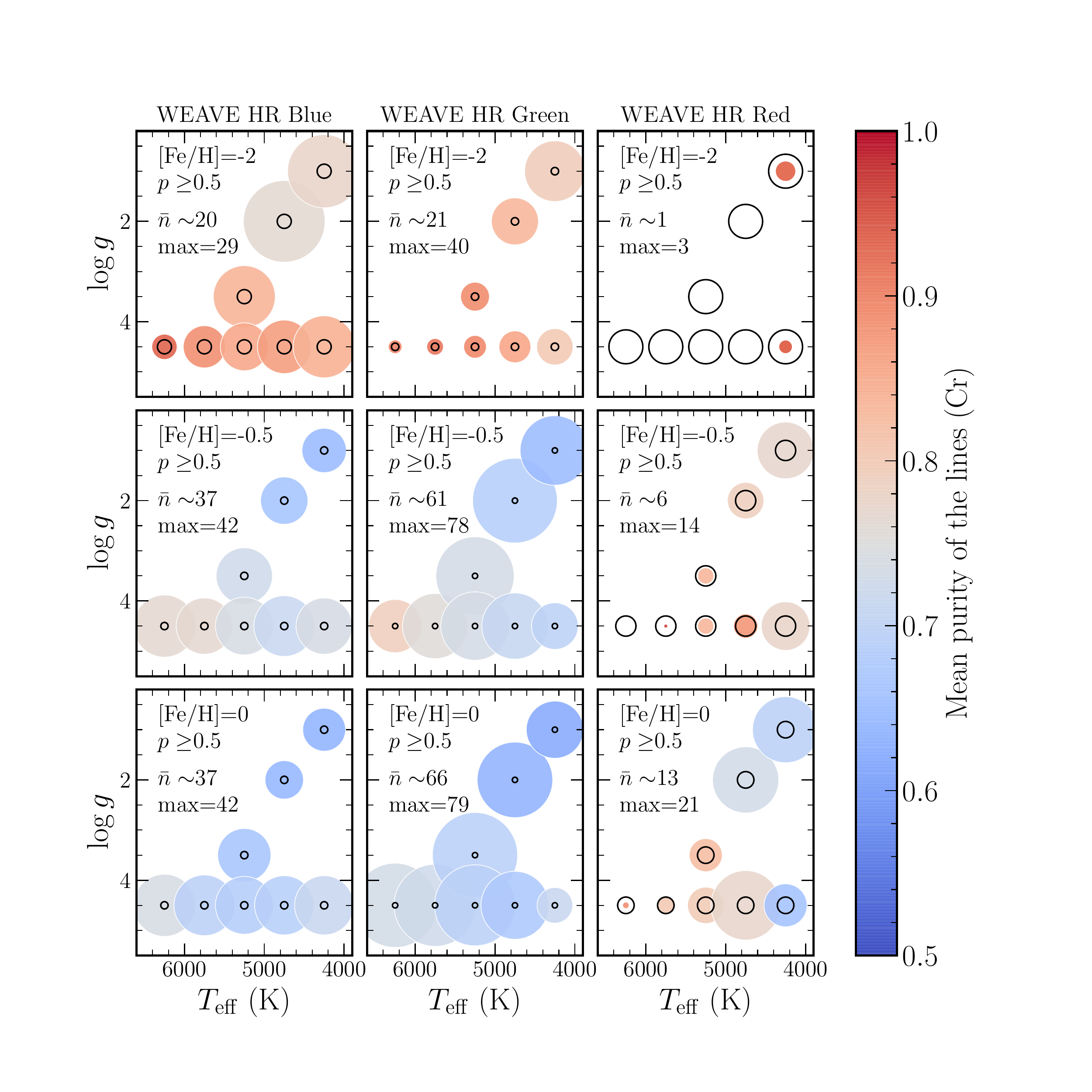}\\
\includegraphics[width=0.49\linewidth]{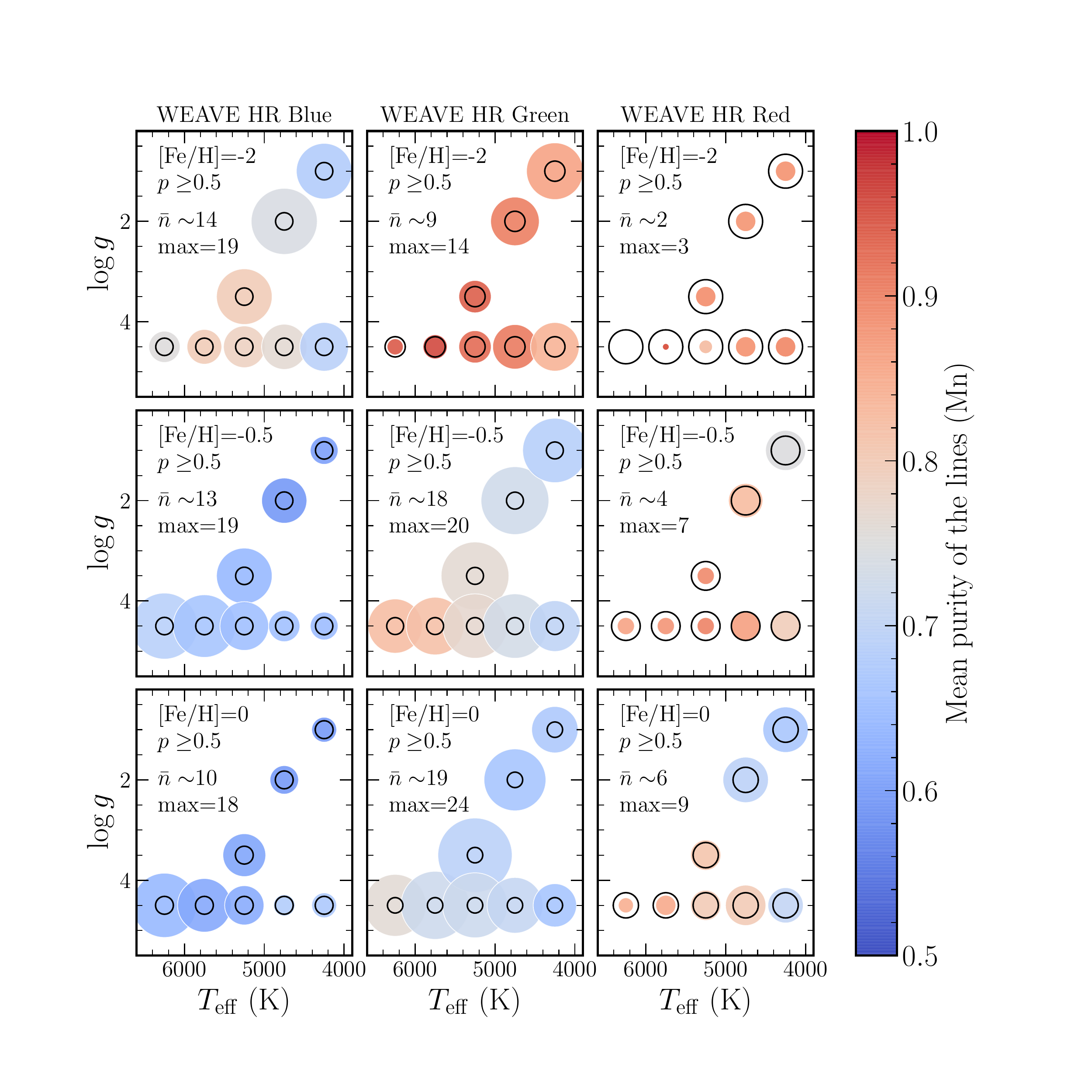}&
\includegraphics[width=0.49\linewidth]{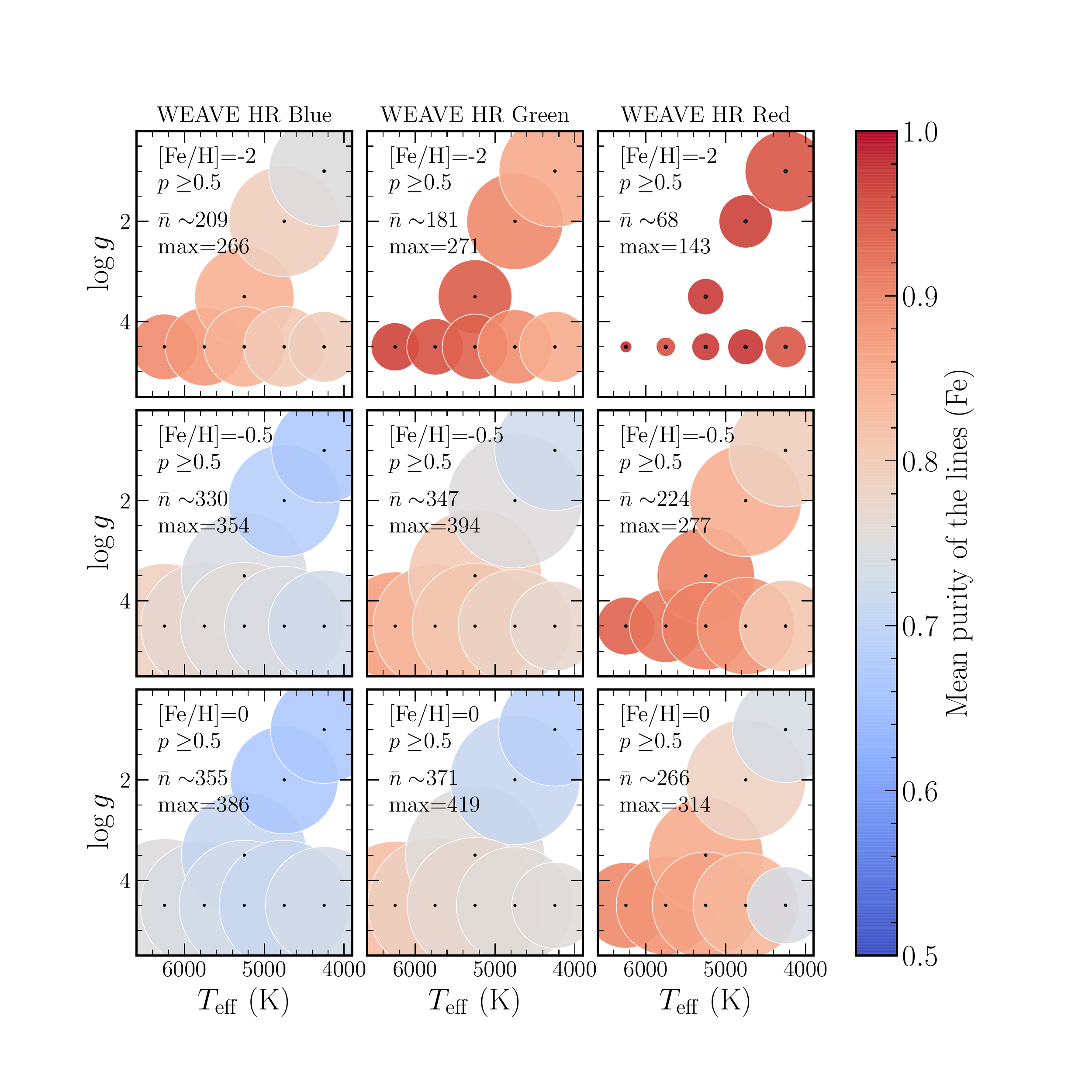}
\end{tabular}
\caption{Same as  Fig.\,\ref{fig:Summary_WEAVE_BGR_alpha_individual1}, for the Fe-peak elements V, Cr, Mn, Fe.}
\label{fig:Summary_WEAVE_BGR_iron_individual1}
\end{figure*}

\begin{figure*}
   \centering
    \renewcommand{\arraystretch}{0.01} 
\begin{tabular}{cc}
\includegraphics[width=0.49\linewidth]{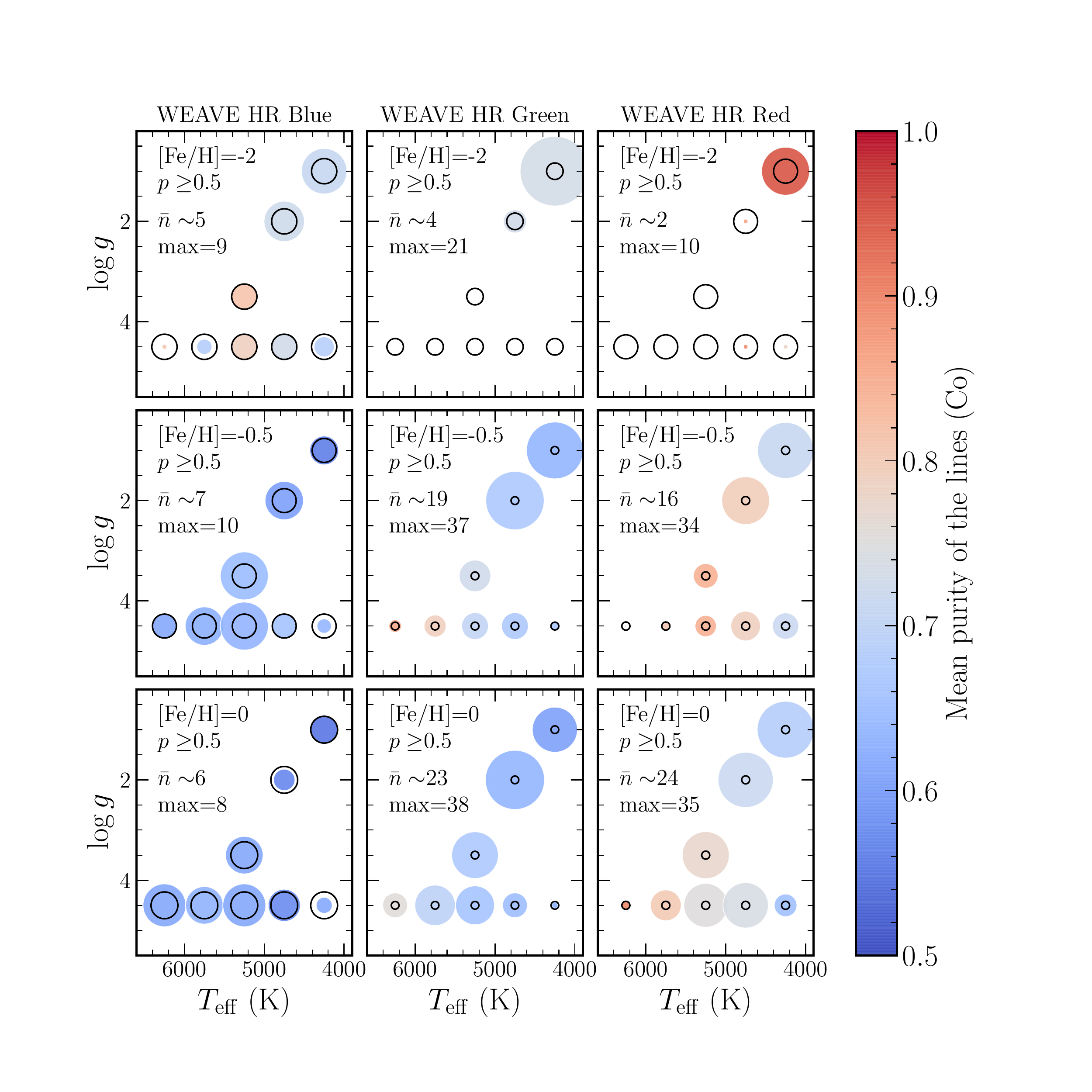}&
\includegraphics[width=0.49\linewidth]{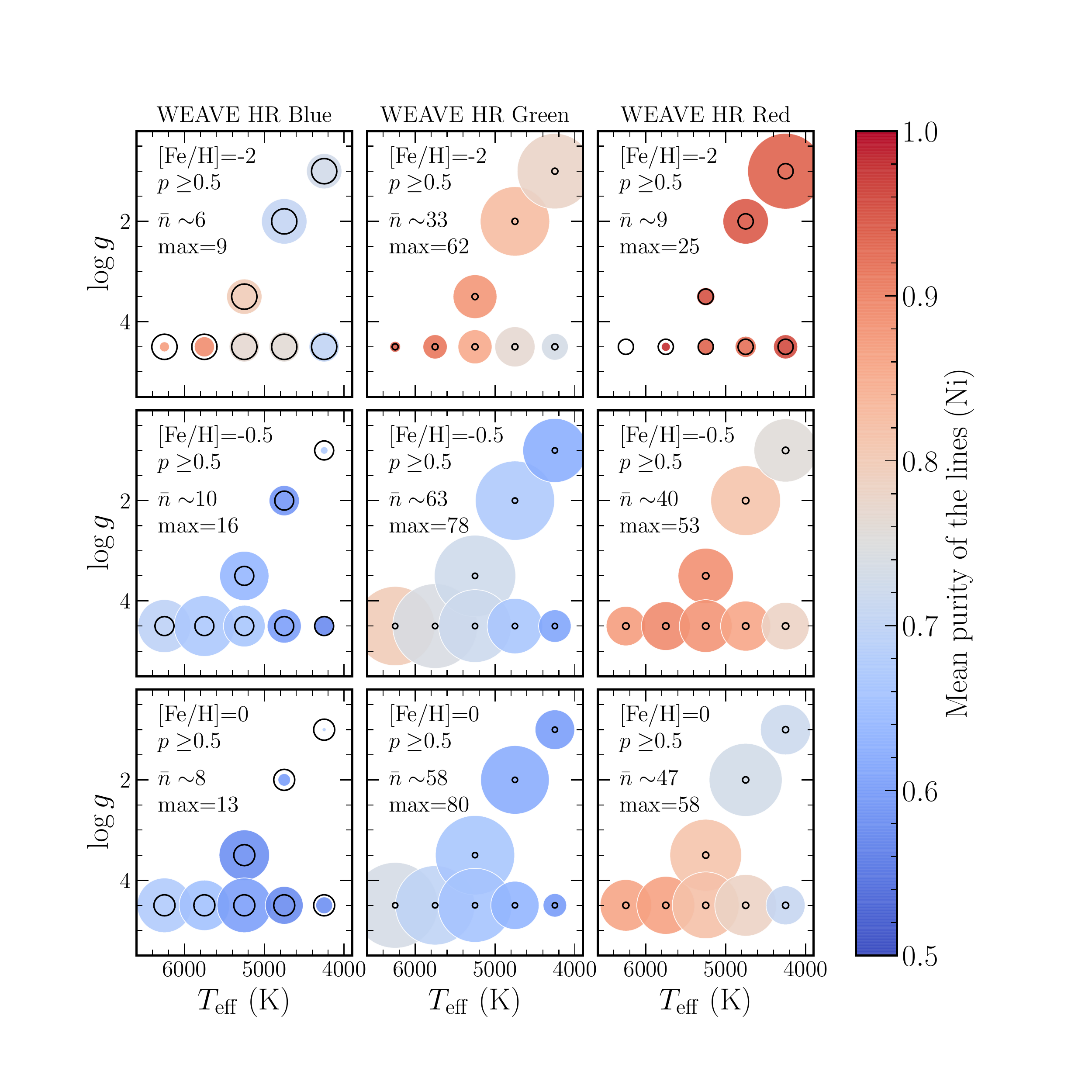}\\
\includegraphics[width=0.49\linewidth]{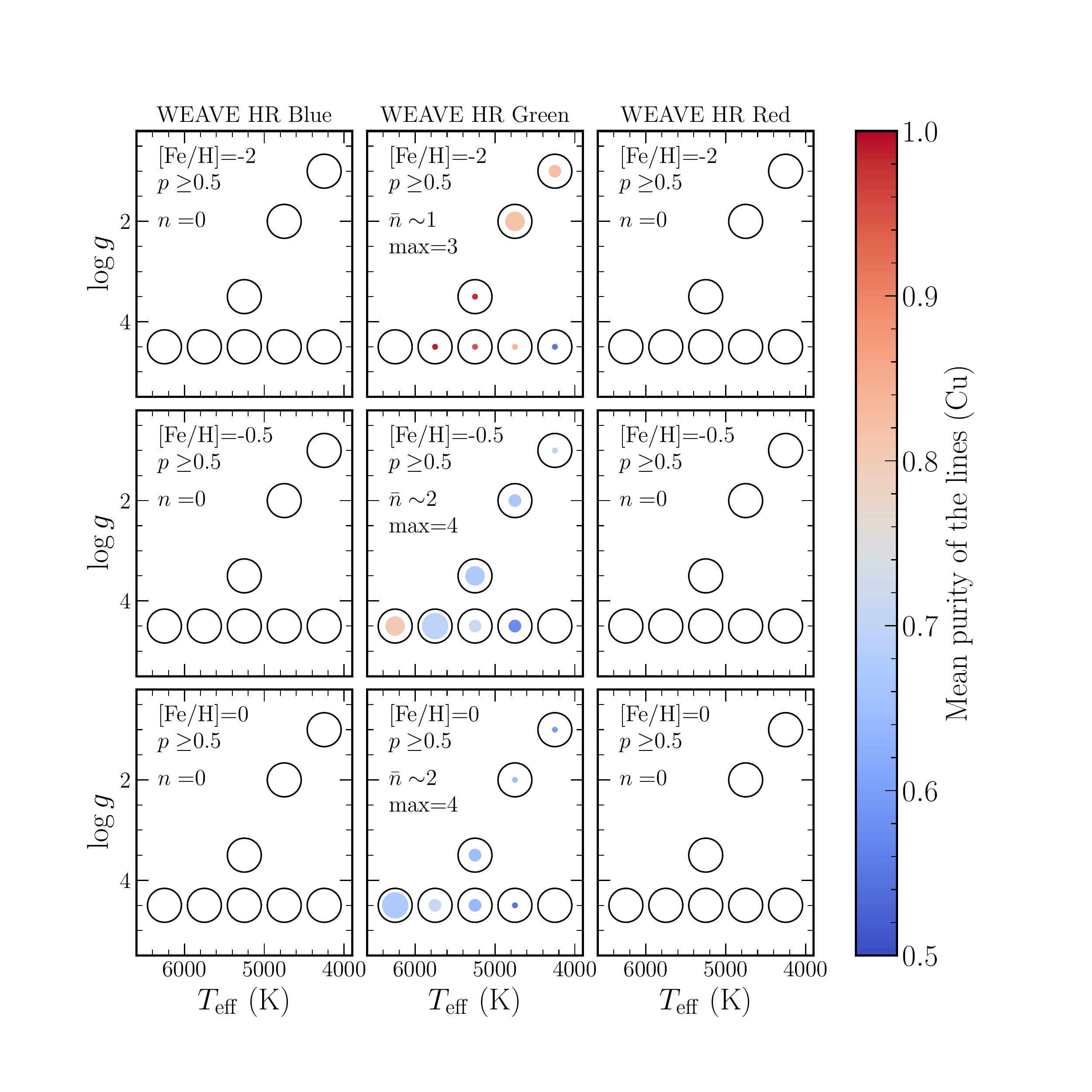}&
\includegraphics[width=0.49\linewidth]{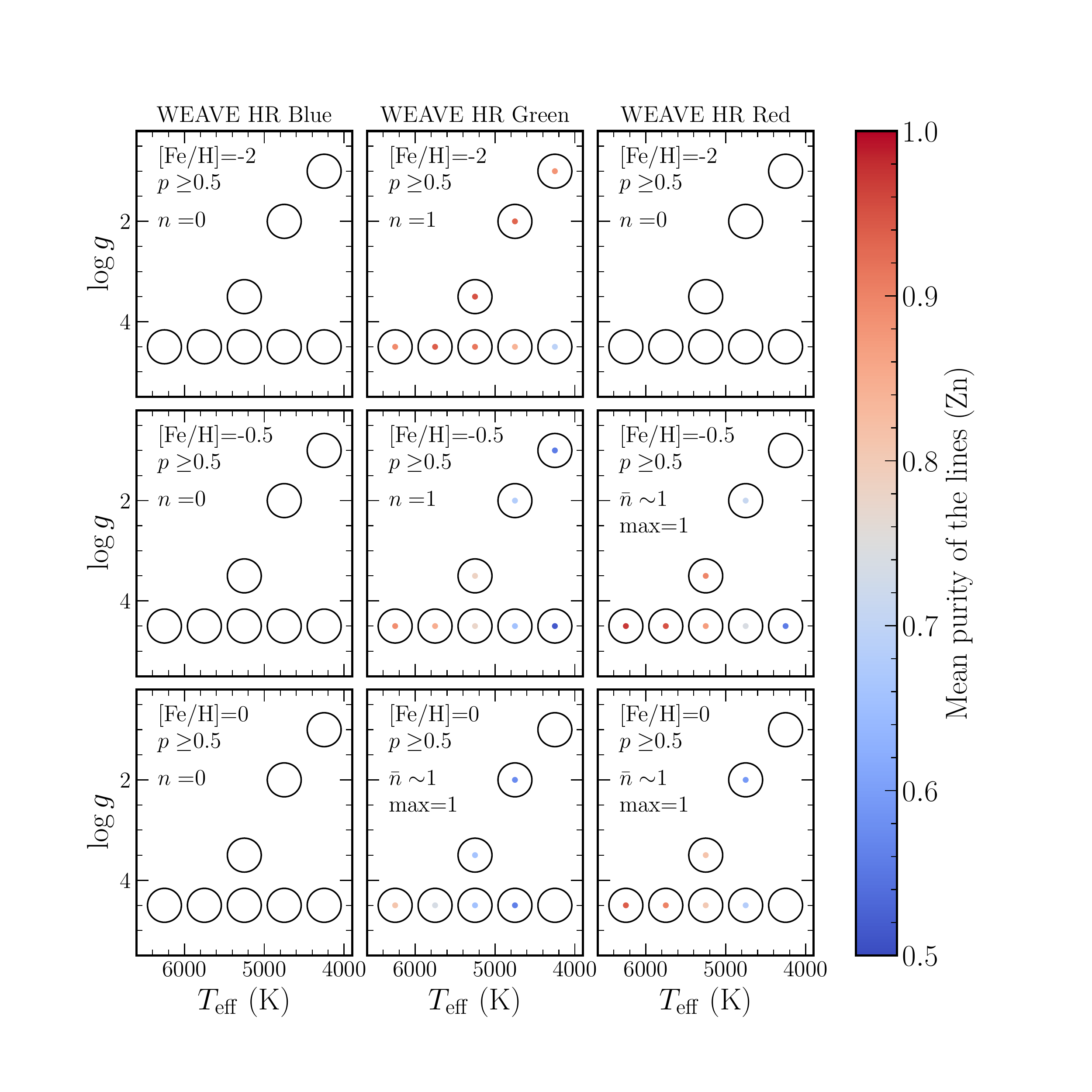}
\end{tabular}
\caption{Same as  Fig.\,\ref{fig:Summary_WEAVE_BGR_alpha_individual1}, for the Fe-peak elements Co, Ni, Cu, Zn.}
\label{fig:Summary_WEAVE_BGR_iron_individual2}
\end{figure*}

\begin{figure*}
   \centering
    \renewcommand{\arraystretch}{0.01} 
\begin{tabular}{cc}
\includegraphics[width=0.49\linewidth]{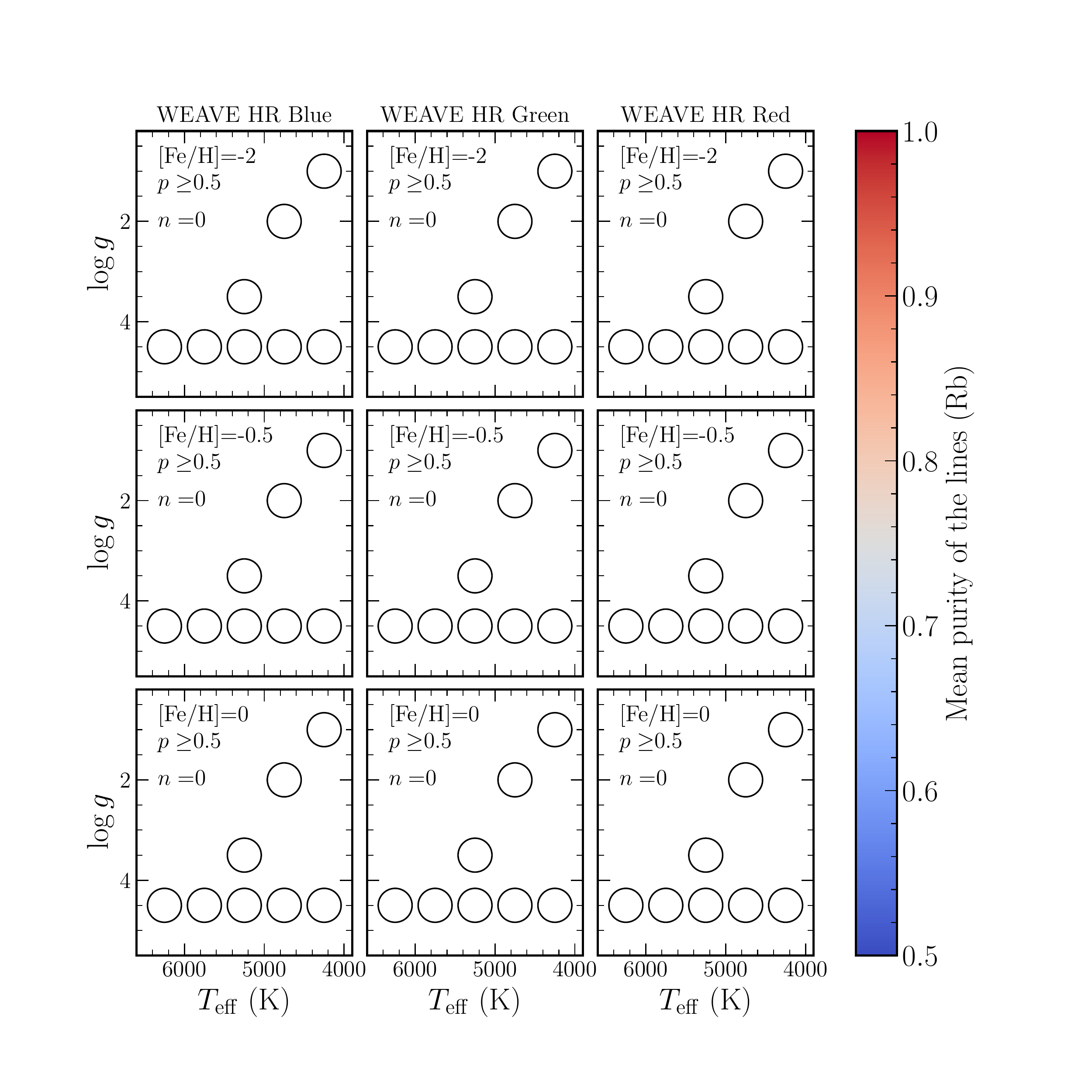}&
\includegraphics[width=0.49\linewidth]{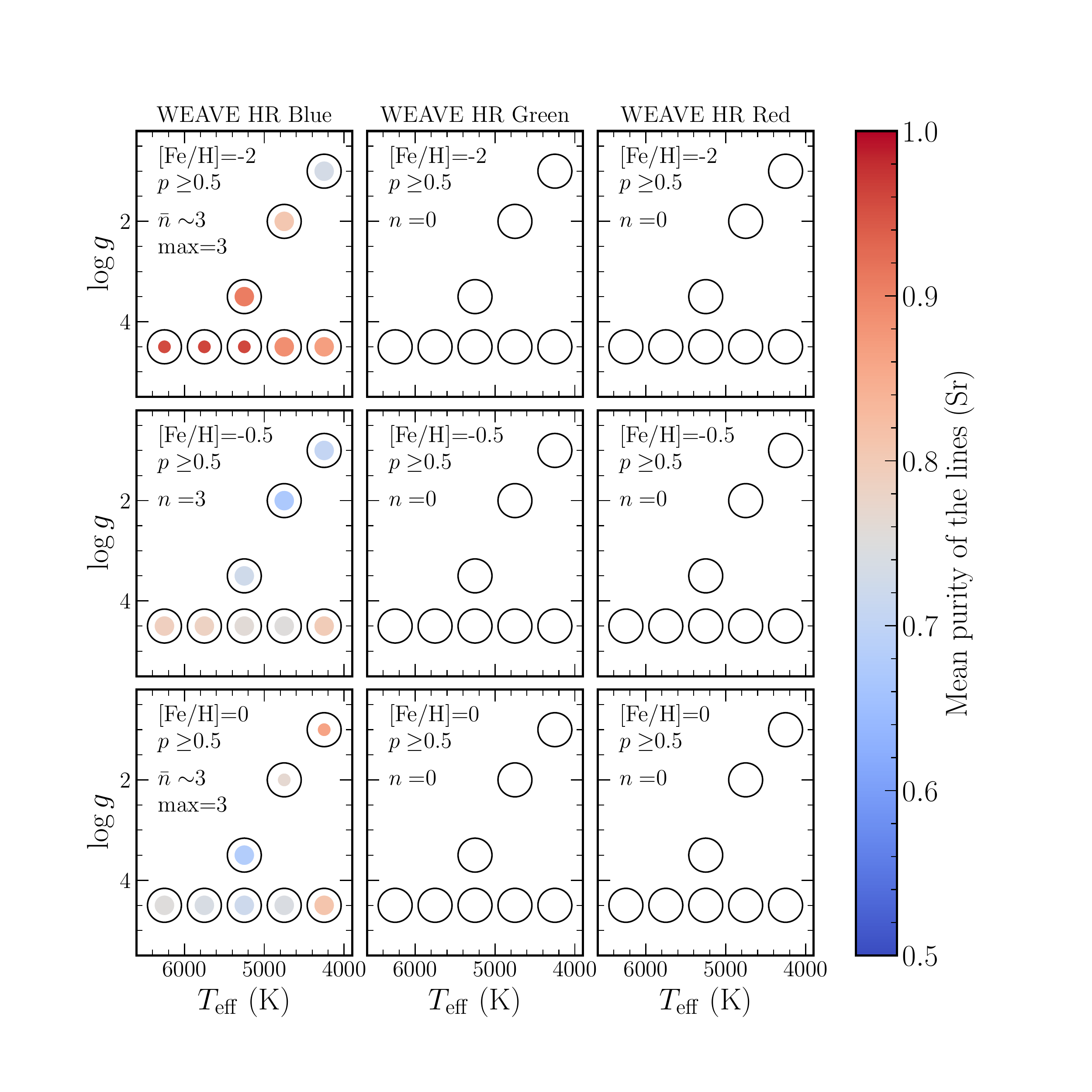}\\
\includegraphics[width=0.49\linewidth]{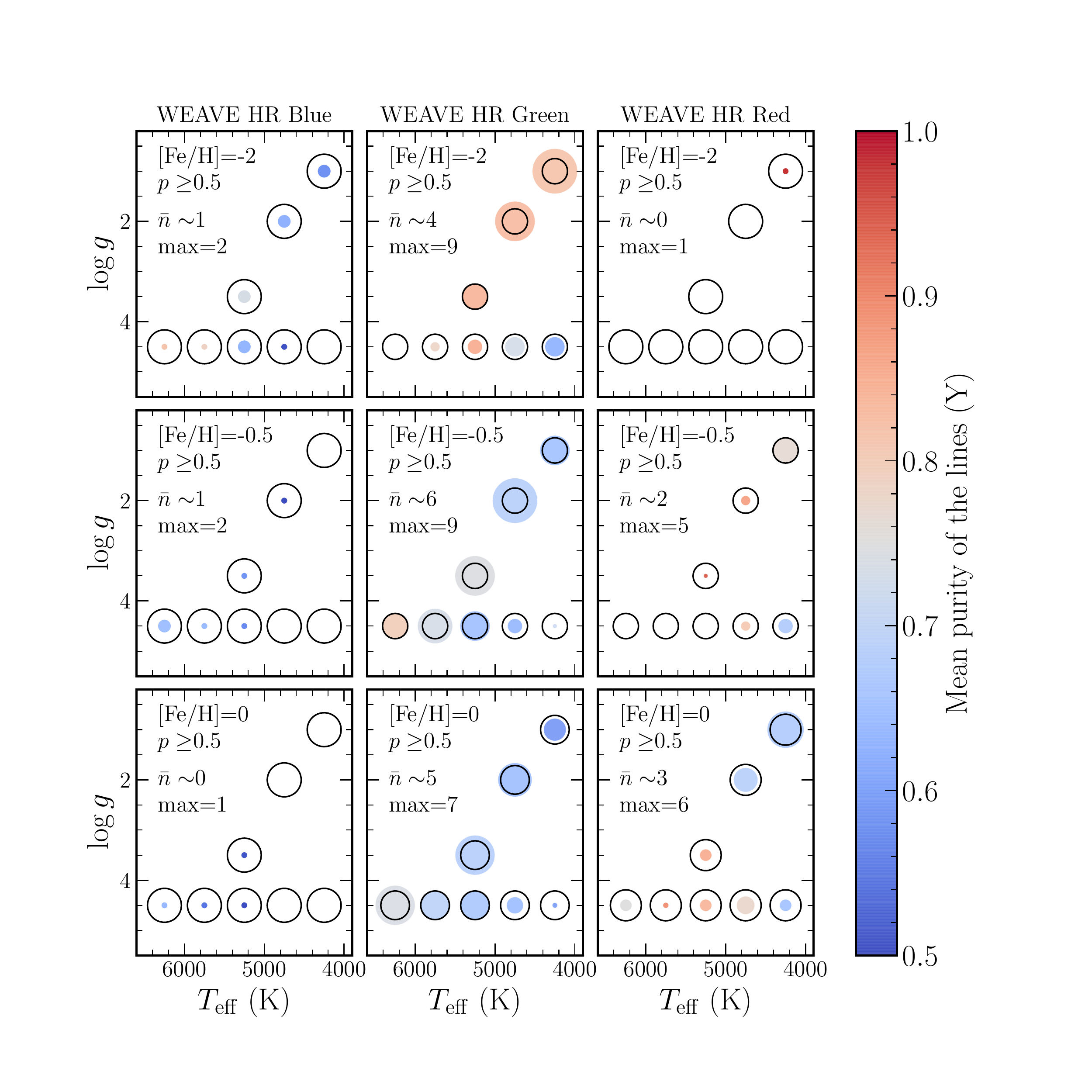}&
\includegraphics[width=0.49\linewidth]{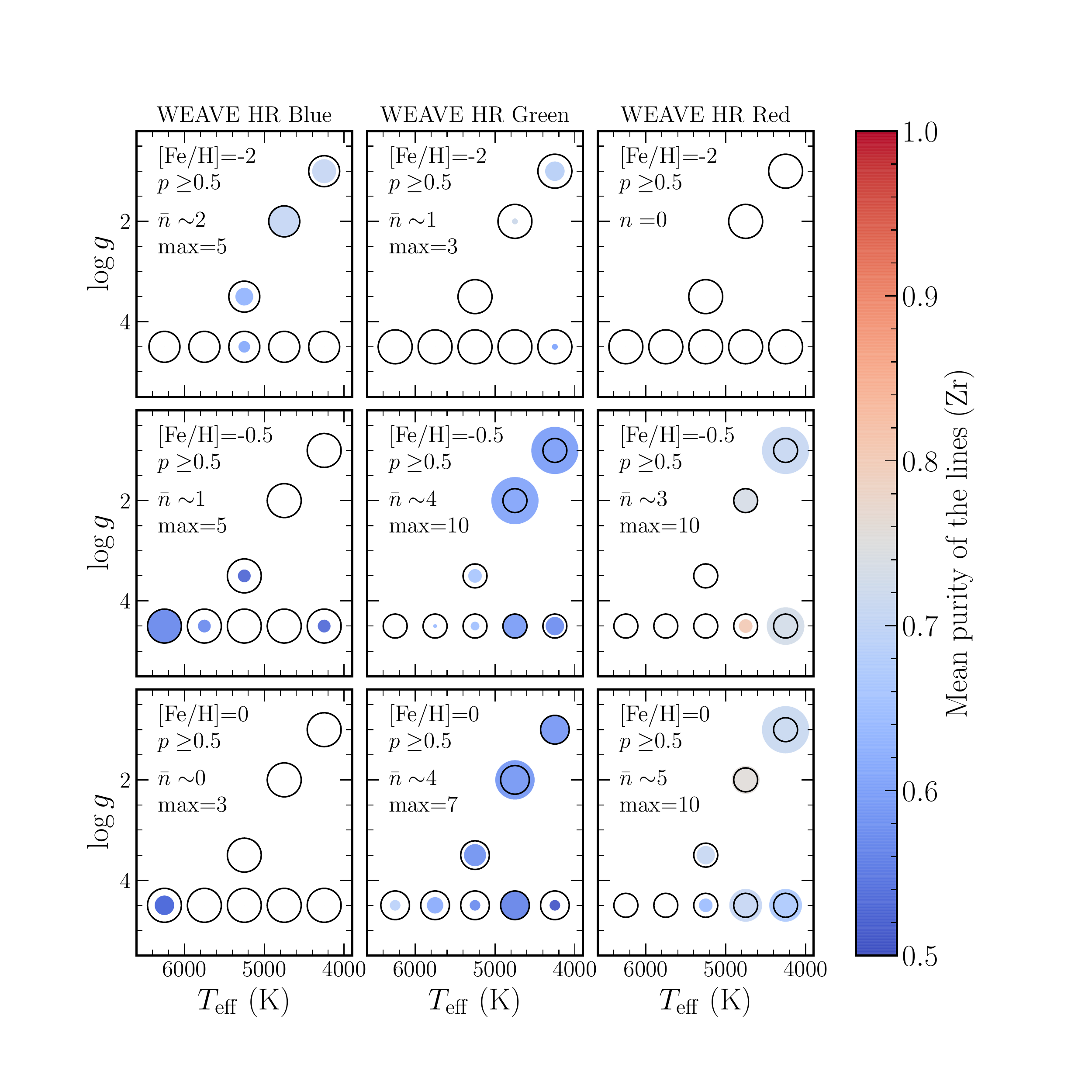}
\end{tabular}
\caption{Same as  Fig.\,\ref{fig:Summary_WEAVE_BGR_alpha_individual1}, for the neutron-capture elements Rb, Sr, Y, Zr.}
\label{fig:Summary_WEAVE_BGR_neutron_individual1}
\end{figure*}

\begin{figure*}
   \centering
    \renewcommand{\arraystretch}{0.01} 
\begin{tabular}{cc}
\includegraphics[width=0.49\linewidth]{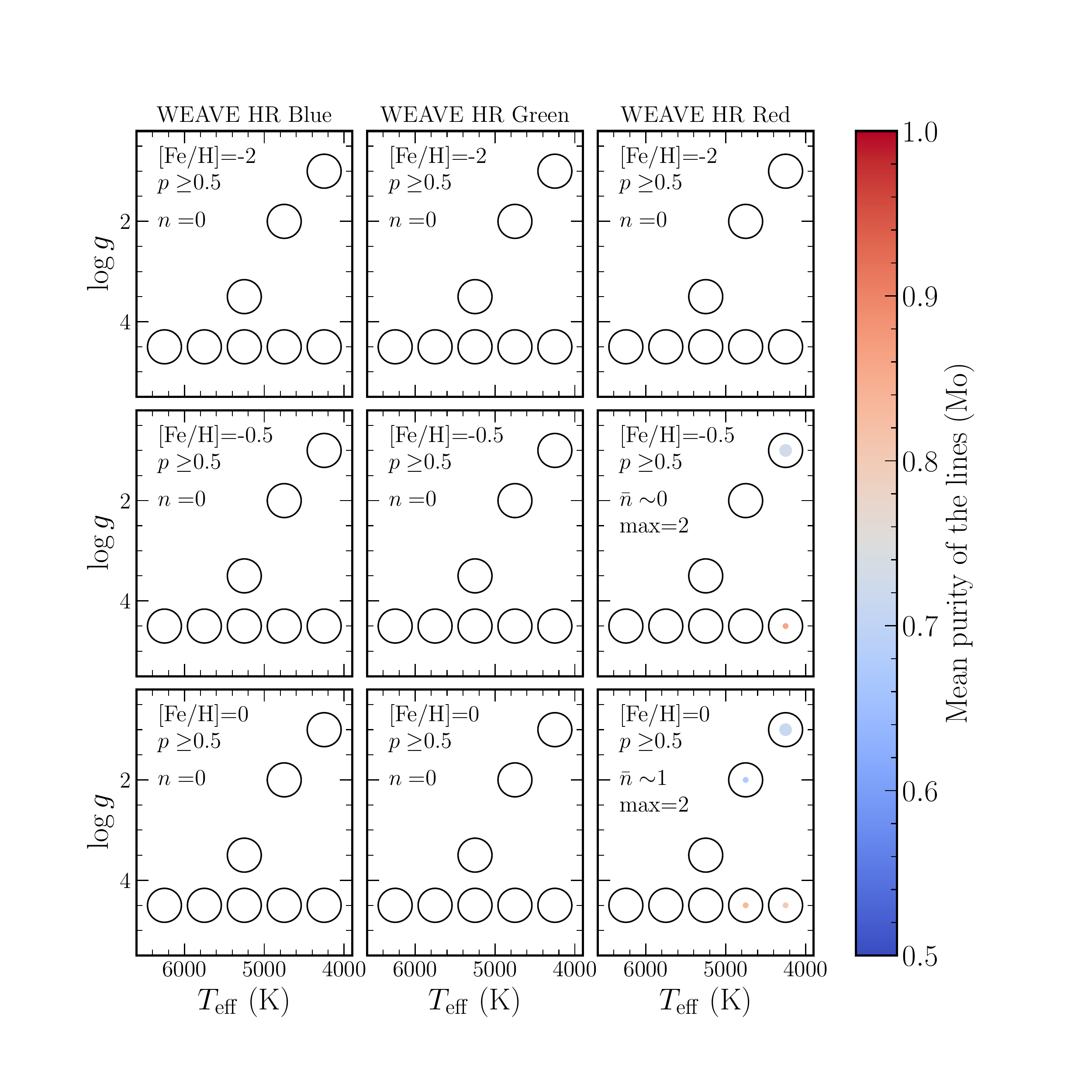}&
\includegraphics[width=0.49\linewidth]{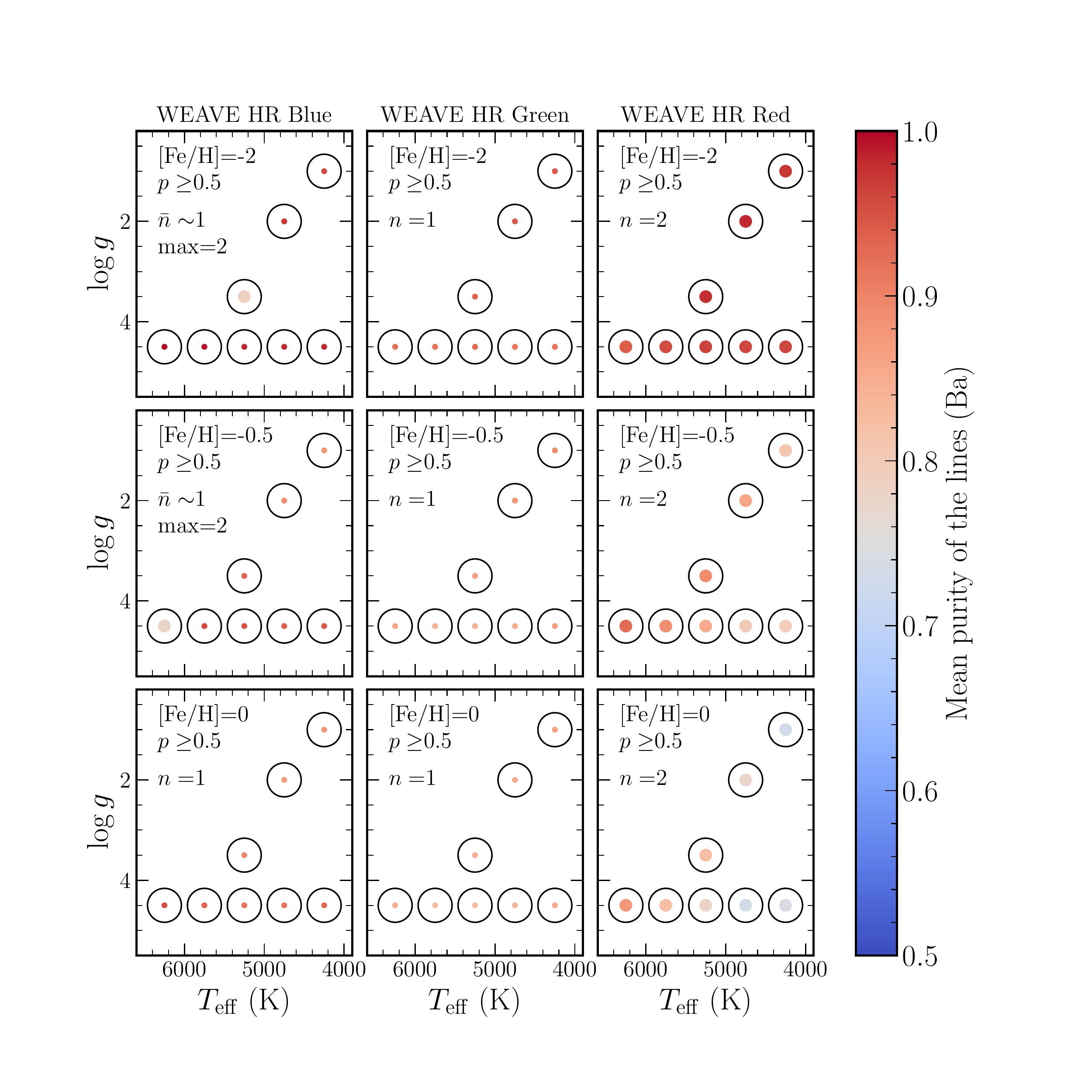}\\
\includegraphics[width=0.49\linewidth]{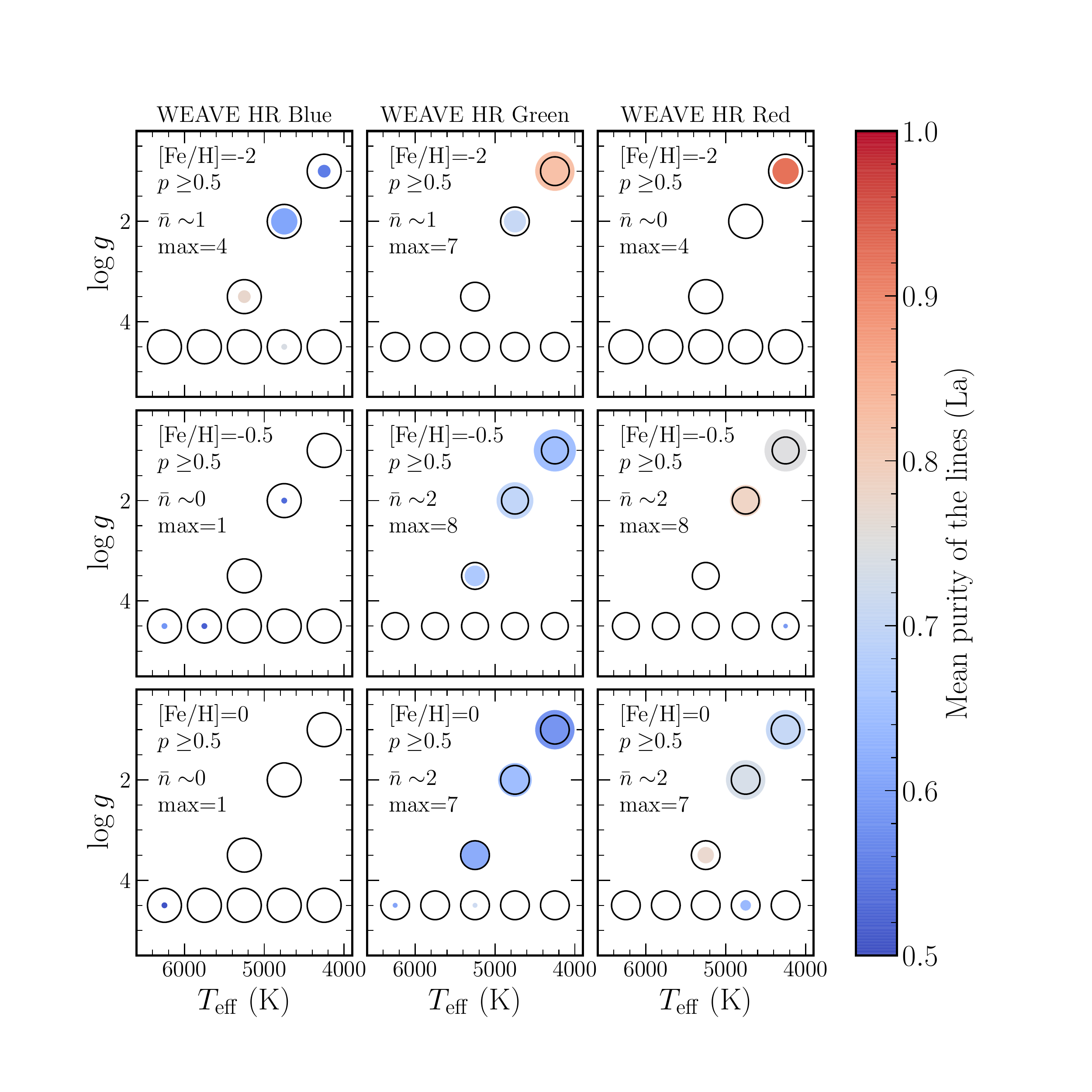}&
\includegraphics[width=0.49\linewidth]{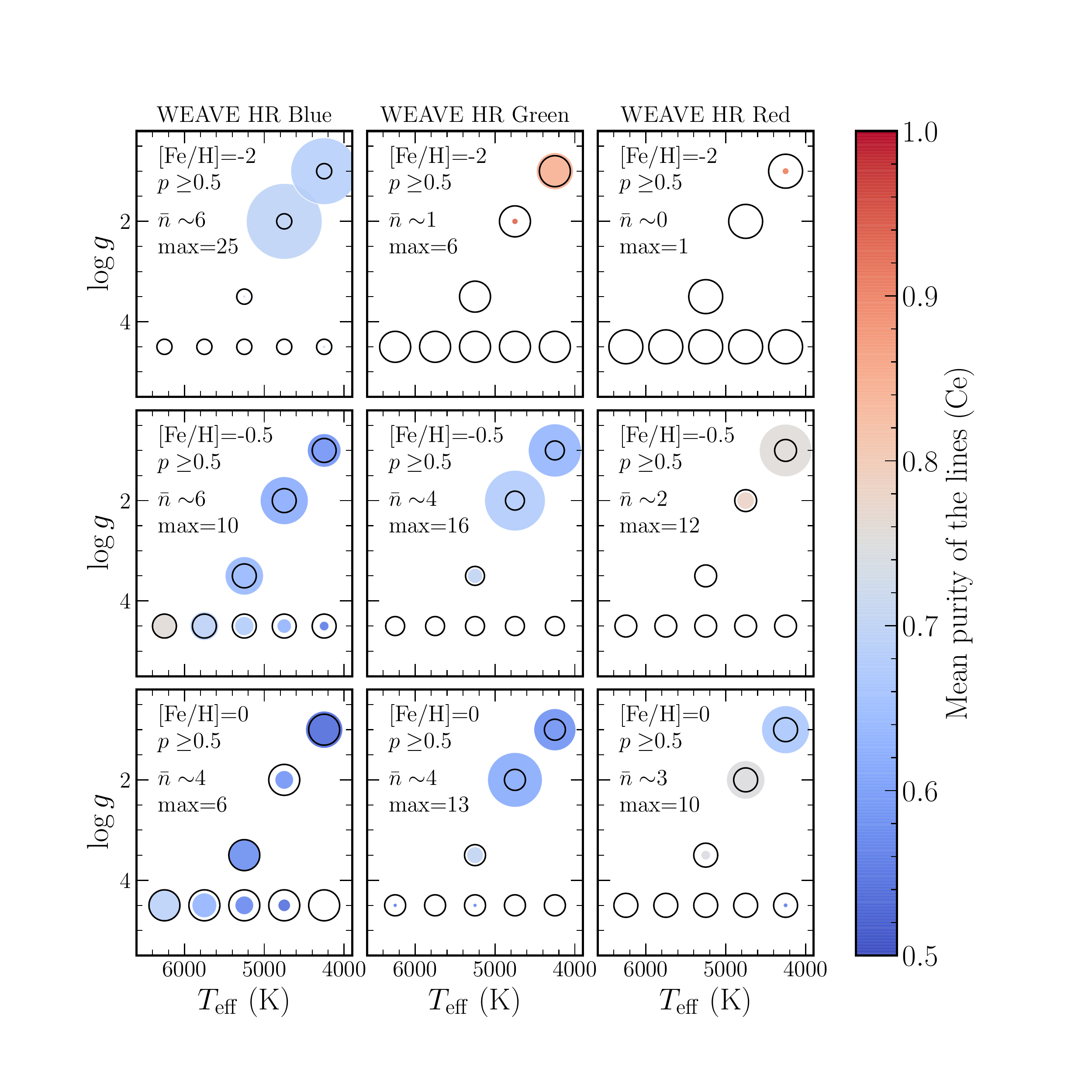}
\end{tabular}
\caption{Same as  Fig.\,\ref{fig:Summary_WEAVE_BGR_alpha_individual1}, for the neutron-capture elements  Mo, Ba, La, Ce.}
\label{fig:Summary_WEAVE_BGR_neutron_individual2}
\end{figure*}

\begin{figure*}
   \centering
    \renewcommand{\arraystretch}{0.01} 
\begin{tabular}{cc}
\includegraphics[width=0.49\linewidth]{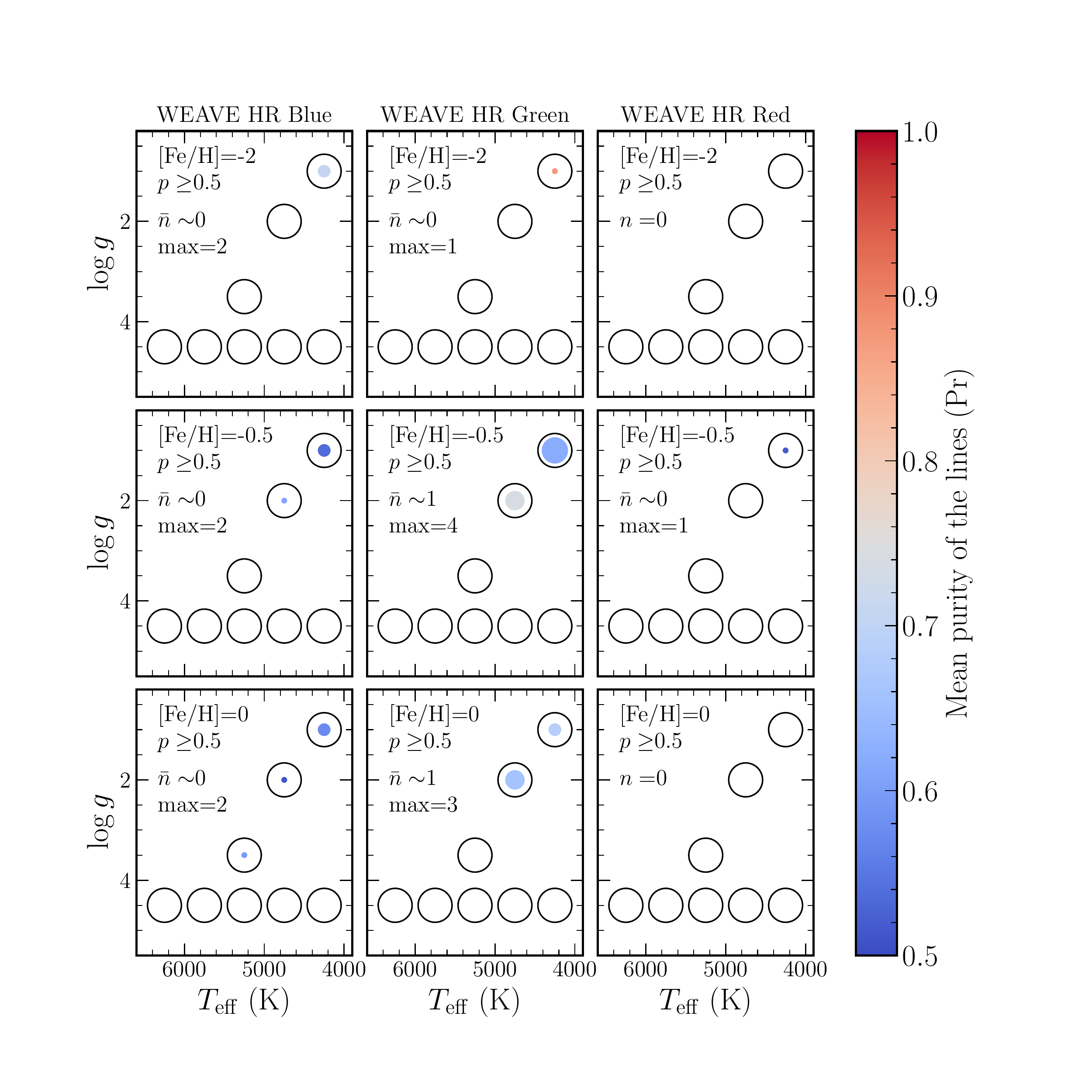}&
\includegraphics[width=0.49\linewidth]{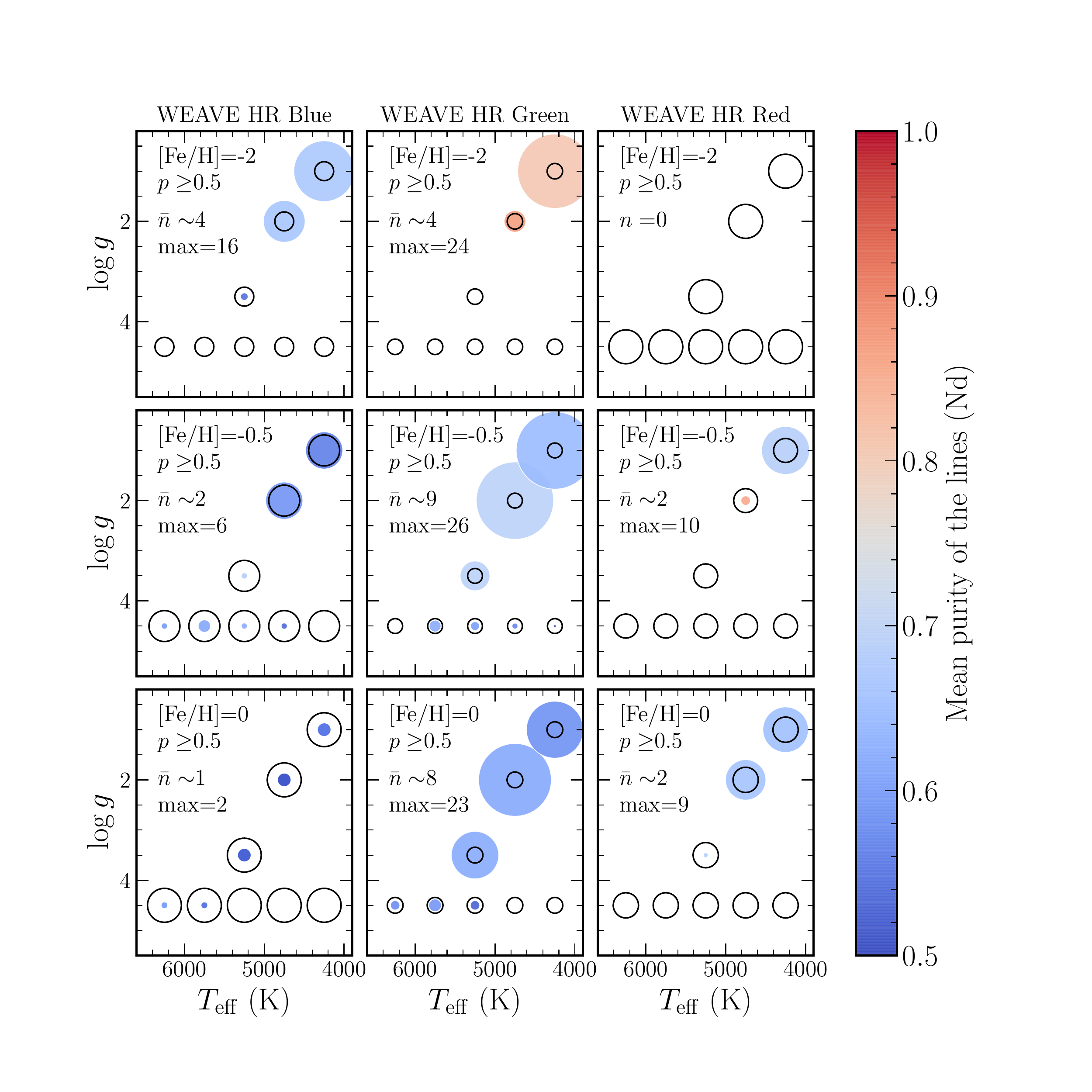}\\
\includegraphics[width=0.49\linewidth]{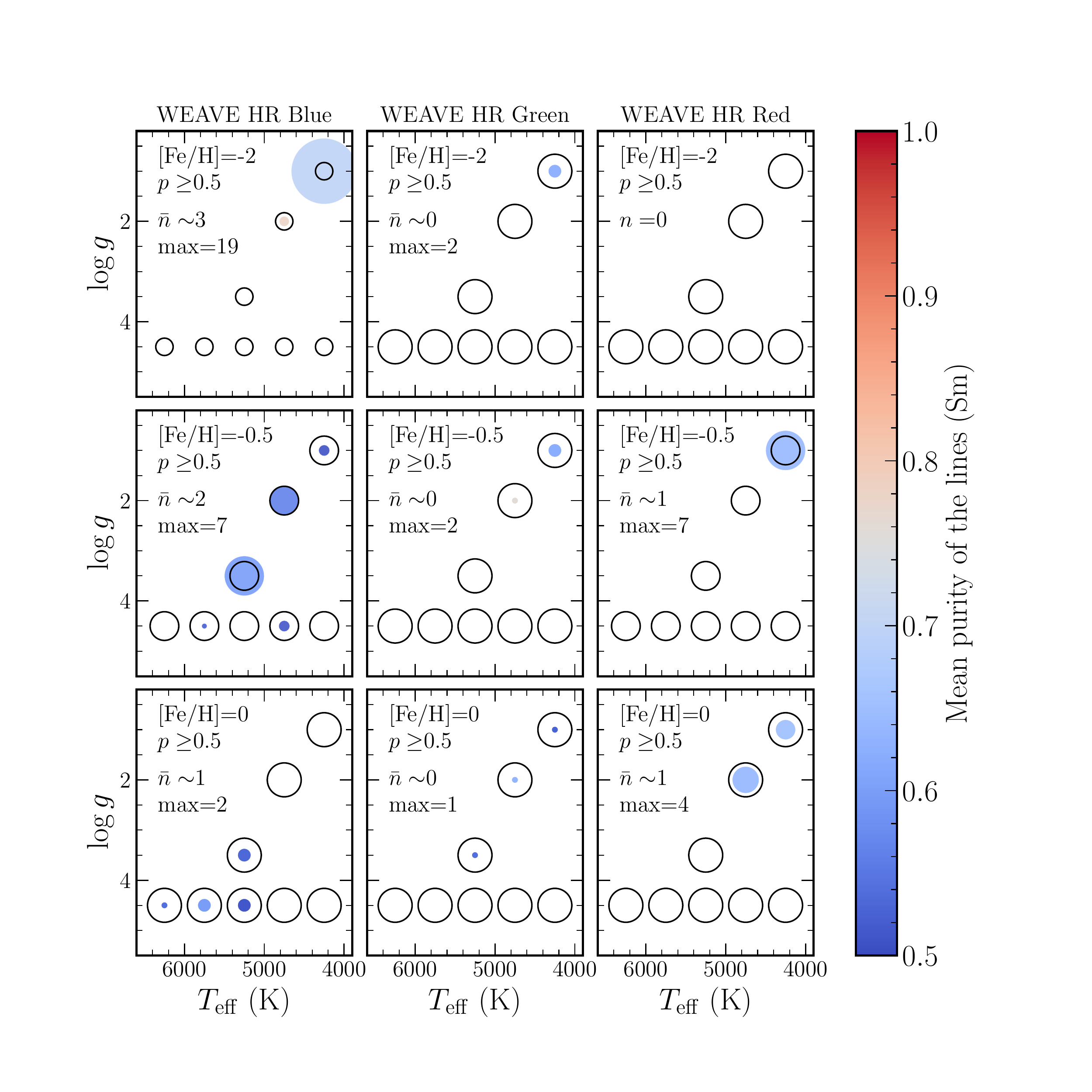}&
\includegraphics[width=0.49\linewidth]{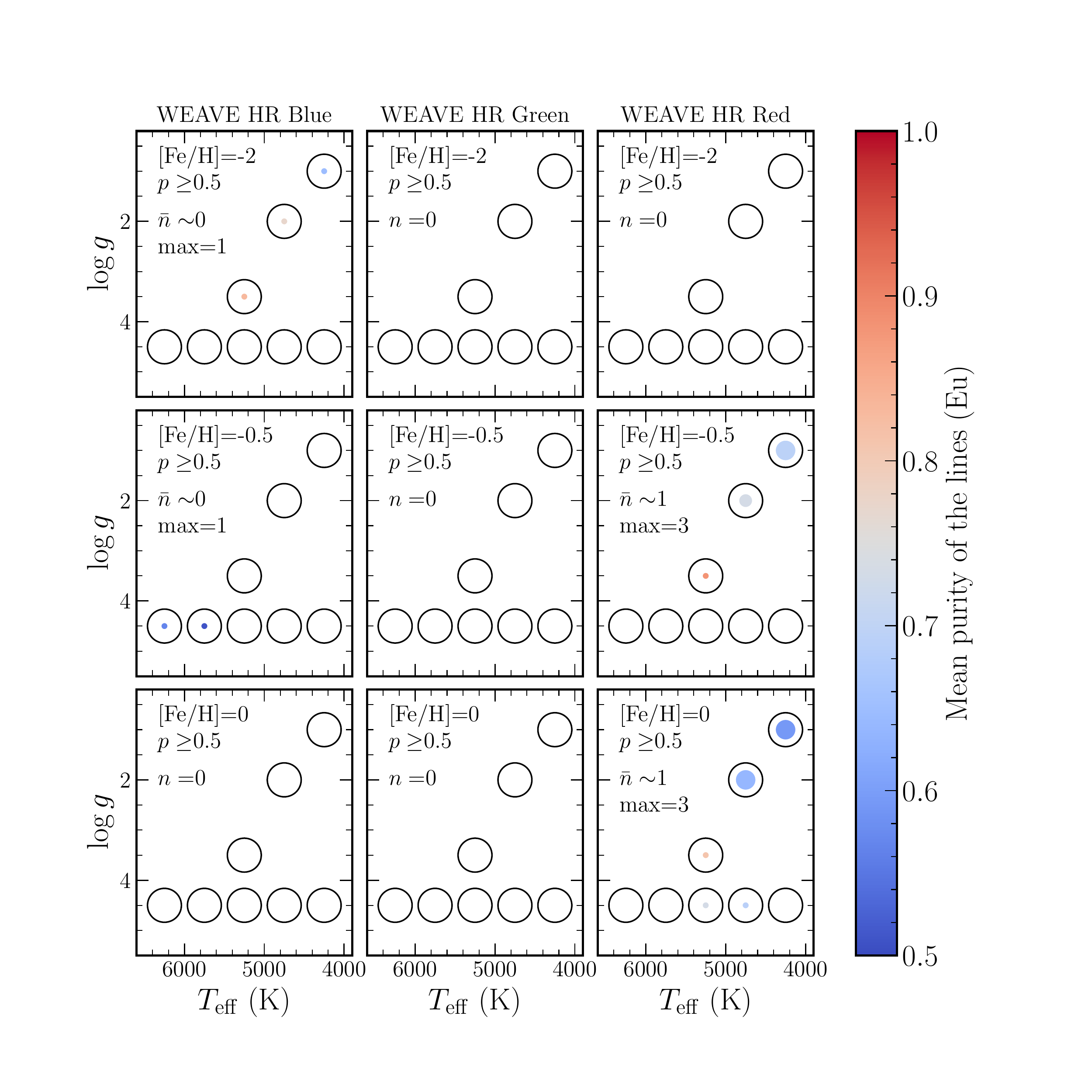}
\end{tabular}
\caption{Same as  Fig.\,\ref{fig:Summary_WEAVE_BGR_alpha_individual1}, for the neutron-capture elements Pr, Nd, Sm, Eu.}
\label{fig:Summary_WEAVE_BGR_neutron_individual3}
\end{figure*}

The plots in this appendix represent the amount of lines detected per element and per combination of \teff-\logg-\meta~ (size of the points), and the mean purity of the lines (colour-code) at each point of the Kiel diagram, for each of WEAVE's HR setup. Figures are separated into  even-Z 
(Fig.\,\ref{fig:Summary_WEAVE_BGR_alpha_individual1} and \ref{fig:Summary_WEAVE_BGR_alpha_individual2}), odd-Z (Figs.\,\ref{fig:Summary_WEAVE_BGR_odd_individual1} and \ref{fig:Summary_WEAVE_BGR_odd_individual2}), Fe-peak (Figs.\,\ref{fig:Summary_WEAVE_BGR_iron_individual1} and \ref{fig:Summary_WEAVE_BGR_iron_individual2}) and neutron-capture elements (Figs.\,\ref{fig:Summary_WEAVE_BGR_neutron_individual1}, \ref{fig:Summary_WEAVE_BGR_neutron_individual2} and \ref{fig:Summary_WEAVE_BGR_neutron_individual3}). The figures are discussed in Sect.\,\ref{sec:weave_HR_application}.

\end{appendix}

\end{document}